\documentclass{article}

\usepackage{amsmath,amssymb}
\usepackage{lmodern}
\usepackage{iftex}
\usepackage{graphicx}  % standard LaTeX graphics tool;
\usepackage{amsmath}   % For getting proper math eqns
\usepackage{amssymb}   % Used for getting various symbols eg. gtrsim, lesssim etc.
\usepackage{bm} % For bold math (esp of lower greek letters)
\usepackage{dcolumn}% Align table columns on decimal point
\usepackage{color}
\usepackage{mathrsfs}
\usepackage{amsfonts}
\usepackage{varioref}
\usepackage[margin=1.6in]{geometry}
\usepackage{slashed}
\usepackage{authblk}
\usepackage{float}
\usepackage{subcaption}
\restylefloat{figure}
\RequirePackage[colorlinks,citecolor=blue,urlcolor=magenta,linkcolor=blue]{hyperref}
\def\p{\partial}
\def\e{\epsilon}

\def\be{\begin{equation}}
\def\ee{\end{equation}}

\labelformat{section}{Section #1} 
\labelformat{subsection}{Section #1} 
\labelformat{subsubsection}{Section #1}
\labelformat{subsubsubsection}{Section #1}
\labelformat{equation}{Eq.~(#1)} 
\labelformat{figure}{Fig.~#1} 
\labelformat{subfigure}{Fig.~\thefigure#1} 
\labelformat{table}{Tab.~#1} 
\labelformat{appendix}{Appendix #1}
\title{\bf IR flow of stochastic power spectra for interacting spectator scalars in the inflationary de Sitter spacetime}
\author[]{Vishal Nath\footnote{vishaln.physics.rs@jadavpuruniversity.in}\ \ \  and  \ Sourav Bhattacharya\footnote{sbhatta.physics@jadavpuruniversity.in}}
\affil[]{Relativity and Cosmology Research Centre, Department of Physics, Jadavpur University,\\ Kolkata 700 032, India. }

\begin{document}
\maketitle
\begin{abstract}
\noindent
In this work we investigate the behaviour  of early inflationary cosmic correlation functions under the flow of the infrared scale for interacting spectator scalar field theories. We start with the derivation of  the one-loop effective potentials for $\phi^4$, $\phi^4$-Yukawa and broken phase $O(N)$ model  in the inflationary de Sitter spacetime, first using the local  Schwinger-DeWitt expansion propagator, and next using the exact de Sitter propagator. The Schwinger-DeWitt expansion being valid in a sub-Hubble length scale,  furnishes suitable framework to determine  the leading curvature corrections to high energy phenomena having  sub-Hubble Compton wavelengths.  These effective potentials  show spontaneous symmetry breaking pattern, and are  next used as  ultraviolet boundary conditions in the Wetterich equation generating the desired IR flow.   The flow is shown to  drive the theory towards symmetry restoration. Using these IR-improved effective potentials, we  next compute the stochastic power spectra and spectral indices for the temporally coincident two point correlation function of the spectator scalar. We show that  the results corresponding to the aforementioned two kind  of propagators are roughly in qualitative agreement with each other, but may differ quantitatively significantly. Both of them show increase in the spectral index in the deep IR for $\phi^4$ and $\phi^4$-Yukawa. The effect of fermionic interaction  has been emphasised. Also for the case of the $O(N)$ scalar, we demonstrate  that the effective potential's shape becomes more and more insensitive to the flow as $N$  gets large. Implication of this on the correlation function has been discussed.
\end{abstract}
\vskip .5cm


\pagebreak
\tableofcontents

%%%%%%%%%%
\section{Introduction}\label{S1}
%%%%%%%%%%%%%
Quantum field theory in curved spacetime deals with fields in fixed, curved backgrounds, where no global Poincarre\'   \ symmetry can be present. Since over the last five decades, there has been an enormous amount of activities in this area of research, making predictions in black hole  physics, particle creation, anomalies, cosmology, and so on, which often has no analogy in the usual Poincarre\'   \ invariant field theory of flat space, see~\cite{Dewitt, Candelas:1975du, Birrell:1982ix, Barvinsky:1985an, Anto, Toms, Hollands:2014eia} and references therein for various pioneering works.  A very  important topic of quantum field theory is the spontaneous breaking of a symmetry of an action, which we will confront in this paper,  generating mass to an initially 
massless field~\cite{Goldstone:1962es, Coleman, Espinosa:1992gq, Andreassen:2014eha, Herring:2024nqg}. We refer our reader to e.g.~\cite{Shore, Panangaden:1981qj, Toms:1983qr, Odintsov, Avramidi:1990je, Shapiro, Parker, Lazzari:2013boa, Markkanen:2018bfx, Toms:2018wpy, Nath:2024doz, Nath:2025vhg} and references therein for various aspects spontaneous symmetry breaking (SSB) and associated phase transitions in curved spacetimes.

It is generally accepted that our very early universe went through a very rapid phase of accelerated expansion with a (nearly) exponential scale factor. This (quasi-)de Sitter primordial inflationary  phase of our early universe is regarded as an appropriate initial condition to address the three puzzles of the standard hot big bang cosmology, namely the horizon problem, the flatness problem and also the problem of non-observability of cosmic relics. The inflationary framework has also been successful in producing nearly scale invariant spectra of various cosmological perturbations as predicted by observations, see e.g.~\cite{Weinberg} and references therein. It turns out  that processes involving massless but non-conformally invariant fields or a light scalar field in the inflationary de Sitter background, undergo rapid quantum fluctuations getting highly amplified at late times. This is essentially a non-perturbative infrared effect associated with redshift due to enormous spacetime expansion, known as the secular effect, see e.g.~\cite{Ford:1984hs, Floratos:1987ek, Tsamis:1996qm, Miao:2006pn, Garbrecht:2006df, Miao:2006pn, Garbrecht:2006df, Seery:2007we, Riotto:2008mv, Beneke:2012kn, Akhmedov:2013vka, Garbrecht:2014dca, Miao:2015oba, Moss:2016uix, Karakaya:2017evp, Mirbabayi:2019qtx, Bhattacharya:2023yhx, Bhattacharya:2026mev} and references therein. See also e.g.~\cite{Boyanovsky:2005px, Onemli:2015tca} and references therein for discussion on the IR effect on cosmological correlation functions.  We refer our reader to~\cite{Fujimoto:1987dp, Serreau:2011fu, Sivasankaran:2020dzp, Prokopec:2011ms, Arai:2013jna, LopezNacir:2014fhi, LopezNacir:2016gzi, Camargo-Molina:2022paw, Bhattacharya:2025jtp, VicenteGarcia-Consuegra:2025two} and references therein for computation of effective action and various associated issues like non-perturbative effects, RG flow, spontaneous symmetry  breaking and Goldstone's theorem etc. 

In this paper we wish to investigate how the IR evolution  affect the two point correlation function of  interacting  spectator scalar field theories for a couple of models  in the inflationary de Sitter background. The IR flow of an effective potential can be generated by the Wetterich equation or the non-perturbative renormalisation group formalism, which progressively integrates out the IR modes~\cite{Wetterich:1989xg, Wetterich:1992yh, Ford:1992mv, Elizalde:1993ee, Morris:1993qb, Berges:2000ew, Pawlowski:2005xe, Delamotte:2007pf, Dupuis:2020fhh, Costeri:2026eqd}. We refer our reader to~\cite{Serreau:2013psa, Serreau:2013eoa, Guilleux:2015pma, FabianGonzalez:2016vgo, Guilleux:2016oqv, Moreau:2018lmz, Mandal:2020umo} and  references therein for application of this formalism in the early inflationary context, many of which focus on  $O(N)$ scalar  in the symmetric phase.  Also, a very well accepted efficient method to study the IR dynamics in the inflationary  background is the stochastic formalism~\cite{Starobinsky:1994bd} (see also e.g.~\cite{Tokuda:2017fdh, Pinol:2018euk, Markkanen:2019kpv, Markkanen:2020bfc, Cable:2023gdz, Bhattacharya:2025eqf} and references therein). In this formalism the non-perturbative IR effect is resummed via a Fokker-Planck equation. We further refer our reader to~\cite{Burgess:2009bs, Prokopec:2017vxx, Hardwick:2019uex} for applications of RG and IR flow methods in the stochastic formalism. \\

\noindent
 Specifically, we wish to  derive in this paper the IR flow of the one loop effective potentials for $\phi^4$, $\phi^4$-Yukawa and the $O(N)$ scalar field theory in the inflationary de Sitter spacetime, \ref{S3}. We will use these effective potentials as the ultraviolet boundary conditions in the flow equation in \ref{S4}.  We next use these IR improved effective potentials to compute the power spectra and spectral indices of temporally coincident two point correlation function. We will show that the spectral index gets enhanced in the deep IR limit. The similarity and differences of results corresponding to different models will be emphasised. Also, in order to derive the effective potentials, we will use  two kinds of propagators, namely, the Schwinger-DeWitt local expansion  using the local  Lorentz invariance~\cite{Toms, Parker} taken up to the quadratic order of spacetime curvature, as well as the exact de Sitter propagator.   It is valid in a neighbourhood small compared to the Hubble horizon, and hence it can furnish a suitable framework to probe the leading curvature correction to high energy processes having Compton wavelengths small compared to the Hubble horizon. (see also~\cite{Nath:2025vhg} for resummation of curvature effects in the same formalism using the 2PI Hartree approximation.) The exact propagator naturally does not put any such restriction. We will also be interested to see how the results originating from these two kind of propagators differ, qualitatively or quantitatively, \ref{S4}.  

 Let us begin with the derivation of the effective potentials in the following section. We will see that they exhibit spontaneous symmetry breaking pattern, which eventually wash away as we move towards deep inside the IR.  We will work with the mostly positive signature of the metric in four spacetime dimensions and will set $c=\hbar=1$ throughout.

%%%%%%%%%%
\section{The one loop effective potentials}\label{S2}
%%%%%%%%%%%%%
In this section we wish to compute the one-loop effective actions for the $\phi^4$, $\phi^4$-Yukawa, and broken phase $O(N)$ model in the de Sitter spacetime background. As we have stated earlier, we will  use the Schwinger-DeWitt  expansion of the propagator using the local Lorentz invariance, {\it as well as}  the exact de Sitter propagator for this purpose.  As we have also discussed, the first kind of propagator, even though not exact, may be considered to be appropriate to describe high energy processes with wavelengths small compared to the Hubble horizon. We will use these effective potentials in the subsequent sections as UV boundary conditions in the Wetterich equation to determine their IR flows. 
%%%%%%%%
\subsection{Local analysis using the Schwinger-DeWitt formalism}\label{ef1}
%%%%

\noindent
We will restrict our computations up to the quadratic order to the spacetime curvature. The tree level action corresponding to the matter sector reads,
\begin{eqnarray}
S= \int d^d x  \sqrt{-g} \left[ -\frac12 g^{\mu\nu} (\nabla_{\mu} \Phi)(\nabla_{\nu} \Phi) -\frac12 m_0^2 \Phi^2 -\frac12 \xi R \Phi^2 - \frac{\lambda \Phi^4}{4!}   + i\overline{\psi}\slashed{\nabla} \psi - M_{0} \overline{\psi} \psi -g \overline{\psi} \psi \Phi \right],
\label{y1}
\end{eqnarray}
where $\slashed{\nabla}= \tilde{\gamma}^{\mu}\nabla_{\mu}$, and $\tilde{\gamma}^{\mu}$ and $\nabla_{\mu}$ are respectively the curved space gamma matrices and  spin covariant derivative. Since we are working in the mostly positive signature of the metric,  the anti-commutation relation for the $\tilde{\gamma}$'s reads
\be
[\tilde{\gamma}_{\mu},\tilde{\gamma}_{\nu}]_+= -2 g_{\mu\nu} \, {\bf I}_{d\times d}.
\label{y2}
\ee

\noindent
Let us first consider the pure scalar field theory, and calculate the one loop effective potential. Up to quadratic order in the spacetime curvature, the renormalised result reads ($\hbar=1$)~\cite{Nath:2024doz} (see~\ref{A} for some detail),
\begin{eqnarray}
&&V_{\rm{eff, Ren}}^{1-{\rm loop}} = \frac12(m^2+\xi R)\phi^2+\frac{\lambda \phi^4}{4!} +\frac{1}{2 (4\pi)^2} \left[\frac{m^{4}}{2}\left( \ln\frac{m^{2}}{4\pi\mu^2}- \frac12 -\psi(2) \right) - \frac{m_0^4}{2} \left(\ln\frac{m_0^{2}}{4\pi\mu^2}-\frac12-\psi(2) \right)\right.\nonumber\\&&\left. +\left(\frac16 -\xi\right)R \left(\psi(2)(m^{2}-m_0^2)-m^2 \ln\frac{m^{2}}{4\pi\mu^2}+m_0^2 \ln\frac{m_0^{2}}{4\pi\mu^2}\right)-\left(\left(\frac16 -\xi\right)^2 R^2 -4f_1\right) \ln \frac{m}{m_0} \right],
\label{v12}
\end{eqnarray}
where $\phi$ is a classical background field, $m^2= m_0^2+\lambda \phi^2/2$ is the effective mass squared, and  we have abbreviated
$$f_1 = \frac12\left(\frac16 -\xi\right)^2 {R^2} +\frac{1}{180}R_{\mu\nu\rho\sigma} R^{\mu\nu\rho\sigma} -\frac{1}{180} R_{\mu\nu} R^{\mu\nu}.$$
We will mention the necessary renormalisation counterterms later in this section. 
\\

\noindent
We now wish to generalise the effective potential by including the Yukawa interaction. The Dirac equation in a curved background reads
\begin{equation}
\left(i\tilde{\gamma}^{\mu}\nabla_{\mu}-M_0\right)\psi =0,
\label{v13}
\end{equation}
where $\tilde{\gamma}^{\mu}(x)=e^{\mu}{}_{a}(x)\gamma^{a}$, and $e^{\mu}{}_{a}(x)$'s are the vierbin, and $\nabla_{\mu}\psi =(\partial_{\mu}-\Gamma_{\mu})\psi$ is the spin covariant derivative, with $\Gamma_{\mu}$ being the spin-connection matrices. Squaring the Dirac operator yields a second-order equation,

\begin{equation}
\left(
g^{\mu\nu}\nabla_{\mu}\nabla_{\nu}
+\frac14R
-M_0^{2}
\right)\psi =0 .
\label{v16}
\end{equation}
Let us define a bispinor $S(x,x')$  as,
\begin{eqnarray}
S(x,x')=\left(
i\tilde{\gamma}^{\mu}\nabla_{\mu}+M_0
\right)\mathcal{G}(x,x'),
\label{v18}
\end{eqnarray}
where $\mathcal{G}(x,x')$ is a scalar Green function satisfying
\begin{equation}
\left(
g^{\mu\nu}\nabla_{\mu}\nabla_{\nu}
+\frac14R
-M_0^{2}
\right)
\mathcal{G}(x,x')
=
-g^{-1/2}(x)\delta(x-x').
\label{v19}
\end{equation}
This implies that $S(x,x')$ is the Green function for the Dirac operator,
\begin{eqnarray}
\left(i\tilde{\gamma}^{\mu}\nabla_{\mu}-M_0
\right)S(x,x') =g^{-1/2}(x)\delta(x-x')\mathbf{1},
\label{v17}
\end{eqnarray}
Similar to the scalar field theory, \ref{A}, one obtains up to quadratic order in the curvature~\cite{Toms}
\begin{eqnarray}
\mathcal{G}(k)=\left[1+(A_{1}+i A_{1\alpha}\partial_k^{\alpha}-A_{1\alpha\beta}\partial_k^{\alpha}\partial_k^{\beta})\left(-\frac{\partial}{\partial M_0^{2}}\right)+A_{2}\left(-\frac{\partial}{\partial M_0^{2}}\right)^{2}\right]\frac{1}{k^{2}+M_0^{2}},
\label{v24}
\end{eqnarray}
with the coefficients,
\begin{equation}
\begin{aligned}
A_1 &= \frac{1}{12}R\,\mathbf{1}, \\[4pt]
A_2 &= \left( \frac{1}{120}R_{;\mu}^{\;\mu} + \frac{1}{288}R^{2} - \frac{1}{180}R_{\mu\nu}R^{\mu\nu} + \frac{1}{180}R_{\mu\nu\sigma\tau}R^{\mu\nu\sigma\tau} \right)\mathbf{1}
      + \frac{1}{48}\,\Sigma_{\alpha\beta}\Sigma_{\gamma\delta}\; R^{\alpha\beta\lambda\tau}R^{\gamma\delta}_{\;\;\lambda\tau},\\[4pt]
A_{1\mu} &= -\frac{1}{24}R_{;\mu}\,\mathbf{1} - \frac{1}{12}\,\Sigma_{\alpha\beta}\; R^{\alpha\beta\;\;\;\lambda}_{\;\;\;\;\;\mu\;\;\;;\lambda}, \\[4pt]
A_{1\mu\nu} &= \frac{1}{30}\left( \frac{1}{4}R_{\mu\nu;\lambda}^{\;\;\;\;\lambda} - \frac{1}{2}R_{;\mu\nu} + \frac{1}{3}R_{\mu\lambda}R^{\lambda}_{\;\nu} - \frac{1}{6}R^{\lambda\tau}R_{\lambda\mu\tau\nu} - \frac{1}{6}R^{\lambda\tau\sigma}_{\;\;\;\;\mu}R_{\lambda\tau\sigma\nu} \right)\mathbf{1}
+ \frac{1}{48}\,\Sigma_{\alpha\beta}\Bigl( R\,R^{\alpha\beta}_{\;\;\;\;\mu\nu} + R^{\alpha\beta\kappa}_{\;\;\;\;\;\mu;\nu\kappa} + R^{\alpha\beta\kappa}_{\;\;\;\;\nu;\kappa\mu} \Bigr)
\\[4pt]&- \frac{1}{96}\,\Sigma_{\alpha\beta}\Sigma_{\gamma\delta}\Bigl( R^{\alpha\beta\lambda}_{\;\;\;\;\mu}R^{\gamma\delta}_{\;\;\;\;\lambda\nu} + R^{\alpha\beta\lambda}_{\;\;\;\;\nu}R^{\gamma\delta}_{\;\;\;\;\lambda\mu} \Bigr),
\label{v25}
\end{aligned}
\end{equation}
where
$$\Sigma_{\alpha\beta} \equiv \frac{1}{4}\left[\gamma_{\alpha},\gamma_{\beta} \right].$$
We have from \ref{v18}
\begin{eqnarray}
S(x,x')= \left(
i\tilde{\gamma}^{\mu}\partial_{\mu}-M_0
\right)\mathcal{G}(x,x') - i\tilde{\gamma}^{\mu}\Gamma_{\mu}(x)\mathcal{G}(x,x').
\label{v26}
\end{eqnarray}
In the coincidence limit, $x\rightarrow x'$, we compute
\begin{eqnarray}
{\rm Tr}S(x,x)= {\rm Tr}\int \frac{d^d k}{(2\pi)^d} \left(-\gamma^{\mu}k_{\mu}+M_0\right)\mathcal{G}(k)
\label{v27}
\end{eqnarray}
Writing now the effective mass in the Yukawa theory as $M=M_{0}+g\phi$, using \ref{v24}, \ref{v25}, the effective potential becomes up to quadratic order in the curvature
\begin{eqnarray}
V_{\rm{eff}}^{\rm{Yuk}} = \int \frac{d^d k}{(2\pi)^d}\left(-2\ln{(k^2+M^2)} +\frac{R}{3(k^2+M^2)}-\frac{f_4}{(k^2+M^2)^2}\right)
\label{v29}
\end{eqnarray}
where we have abbreviated,
 $$f_4 = \frac{R^2}{24}+\frac{R^{\mu\nu}R_{\mu\nu}}{45}-\frac{7R^{\mu\nu\rho\sigma}R_{\mu\nu\rho\sigma}}{360}.$$

\noindent
Let us now discuss the renormalisation of the effective action for the $\phi^4$-Yukawa theory. Collecting the divergent terms from \ref{v29} and \ref{v10}, and introducing the counterterms,
\begin{eqnarray}
&& \delta{\cal L}_{\rm grav} + \delta{\cal L}_{\rm matt} = \left[\delta \Lambda +\delta l R +\delta \alpha_{1} R_{\mu\nu\rho\sigma} R^{\mu\nu\rho\sigma} +\delta \alpha_{2} R_{\mu\nu} R^{\mu\nu} +\delta \alpha_{3} R^2 \right] \nonumber \\ && + \left[\frac12 \delta m^{2}\phi^2  +\frac12 \delta\xi R \phi^2  +\frac{\delta \eta \phi^{3}}{3!} +\frac{\delta \lambda \phi^{4}}{4!} +\delta \tau \phi +\delta \gamma R \phi \right],
\label{v31}
\end{eqnarray}
we obtain the gravitational counterterms
\begin{eqnarray}
\delta \Lambda = \frac{\mu^{-\e}m_0^4}{32 \pi^2 \epsilon}-\frac{2\mu^{-\e}M_0^4}{(4\pi)^2 \epsilon},  \qquad \delta l =\frac{\mu^{-\e}m_0^2\left(\xi  -1/6 \right)}{16 \pi^2 \epsilon}+\frac{2\mu^{-\e}M_0^2}{3(4\pi)^2 \epsilon}, \qquad
\delta\alpha_{1} =\frac{\mu^{-\e}}{16 \pi^2 \epsilon}\frac{1}{90}-\frac{7\mu^{-\e}}{180(4\pi)^2 \epsilon},\nonumber\\
\delta\alpha_{2} =-\frac{\mu^{-\e}}{16 \pi^2 \epsilon}\frac{1}{90}+\frac{2\mu^{-\e}}{45(4\pi)^2 \epsilon}, \qquad
\delta\alpha_{3} =\frac{\mu^{-\e}}{32 \pi^2 \epsilon}\left(\frac{1}{36} + \xi^2 -\frac{\xi}{3} \right)+\frac{\mu^{-\e}}{12(4\pi)^2 \epsilon},\qquad 
\label{v32}
\end{eqnarray}
as well as the matter counterterms
\begin{eqnarray}
&&\delta m^{2} =\frac{\mu^{-\e}m_0^2 \lambda}{16 \pi^2 \epsilon}-\frac{3M_0^2g^2\mu^{-\e}}{(2\pi^2 \epsilon}, \qquad
\delta\xi = \frac{\mu^{-\e}\lambda\left(\xi -1/6  \right)}{16 \pi^2 \epsilon}+\frac{g^2\mu^{-\e}}{12\pi^2 \epsilon},\qquad
\delta\tau = -\frac{gM_0^3\mu^{-\e} }{2\pi^2 \epsilon},  \nonumber\\&&
\delta\gamma =\frac{ g \mu^{-\e}}{12\pi^2 \epsilon}, \qquad 
\delta\eta =-\frac{3g^3M_0\mu^{-\e}}{\pi^2 \epsilon},  \qquad
\delta\lambda = \frac{3\mu^{-\e} \lambda^2}{16 \pi^2 \epsilon}-\frac{3 g^4\mu^{-\e}}{\pi^2 \epsilon}.
\label{v34}
\end{eqnarray}
Combining now \ref{v12} with the UV finite part of \ref{v29}, we obtain the renormalised one loop effective potential for the $\phi^4$-Yukawa theory,
\begin{eqnarray}
&&V_{\rm{eff, Ren}}^{1-{\rm loop}} = \frac12(m^2_0+\xi R)\phi^2+\frac{\lambda \phi^4}{4!} +\frac{1}{4 (4\pi)^2} \left[m^{4} \ln\frac{m^{2}}{4\pi\mu^2}  -2\left(\frac16 -\xi\right)R m^2 \ln\frac{m^{2}}{4\pi\mu^2}-\left(\left(\xi-\frac16\right)^2 R^2 -4f_1\right) \ln \frac{m^2}{m_0^2} \right]\nonumber\\&&-\frac{1}{ (4\pi)^2}\left(g^2\phi^2 +2g\phi M_0\right)\left(\left(M_0+g\phi\right)^2+M_0^2\right)\ln{\frac{(M_0+g\phi)^2}{4\pi\mu^2}}- \frac{1}{8\pi^2}\left(M_0^4 +f_4\right)\ln{\left(1+\frac{g\phi}{M_0}\right)}\nonumber\\&&+\frac{R}{ 3(4\pi)^2}\left( M_0+ g\phi\right)^2\ln{\frac{(M_0+g\phi)^2}{4\pi\mu^2}}+\phi\left[\frac{4gM_0^3}{(4\pi)^2}\ln\frac{M_0^2}{4\pi\mu^2} +\frac{g(M_0^4+f_4)}{8\pi^2M_0} -\frac{2gRM_0}{3(4\pi)^2}\ln\frac{M_0^2e}{4\pi\mu^2}\right]\nonumber\\&&-\frac{\phi^2}{2}\left[\frac{\lambda}{4(4\pi)^2}\left(2m_0^2\ln\frac{m_0^2}{4\pi\mu^2}+m_0^2+2\left(\xi-1/6\right)R\ln\frac{m_0^2 e}{4\pi\mu^2}
-\frac{\left(\xi-1/6\right)^2R^2-4f_1}{m_0^2}\right)\right.\nonumber\\&&\left. -\left(\frac{12g^2M_0^2}{(4\pi)^2}\ln\frac{M_0^2 e^{\frac43}}{4\pi\mu^2}
-\frac{g^2(M_0^4+f_4)}{8\pi^2M_0^2}-\frac{2g^2R}{3(4\pi)^2}\ln\frac{M_0^2 e^3}{4\pi\mu^2}\right)\right]\nonumber\\&& +\frac{\phi^3}{6}\left[\frac{24g^3M_0}{(4\pi)^2}\ln\frac{M_0^2 e^2}{4\pi\mu^2} +\frac{g^3(M_0^4+f_4)}{4\pi^2M_0^3}-\frac{4g^3R}{3(4\pi)^2M_0}\right] -\frac{\phi^4}{24}\left[\frac{3\lambda^2}{2(4\pi)^2}\left(\ln\frac{m_0^2 e^{\frac32}}{4\pi\mu^2}+\frac{\left(\xi-1/6\right)R}{m_0^2}\right.\right.
\nonumber\\
&&\left.\left. +\frac{\left(\xi-1/6\right)^2R^2-4f_1}{2m_0^4}\right) -\left(\frac{24g^4}{(4\pi)^2}\ln\frac{M_0^2 e^{\frac{14}{3}}}{4\pi\mu^2}+\frac{3g^4(M_0^4+f_4)}{4\pi^2M_0^4} -\frac{g^4R}{12\pi^2M_0^2}\right)\right]\ + \ {\cal O}(R^3), 
\label{v36}
\end{eqnarray}
where we have abbreviated $m^2=m_0^2+\lambda \phi^2/2$ as earlier, $f_1$ is defined below  \ref{v12}, $f_4$  below \ref{v29}, and we have further added some  {\it finite} counterterms  so that \ref{v36} satisfies the following renormalisation conditions 
$$\frac{\p V_{\rm{eff, Ren}}^{1-{\rm loop}}}{\p\phi}\Bigg\vert_{\phi=0}= 0= \frac{\p^3 V_{\rm{eff, Ren}}^{1-{\rm loop}}}{\p\phi^3}\Bigg\vert_{\phi=0}, \quad \frac{\p^2 V_{\rm{eff, Ren}}^{1-{\rm loop}}}{\p\phi^2}\Bigg\vert_{\phi=0}= m_0^2 +\xi R, \quad \frac{\p^4 V_{\rm{eff, Ren}}^{1-{\rm loop}}}{\p\phi^4}\Bigg\vert_{\phi=0}= \lambda.$$
\noindent
As such, \ref{v36} is valid in any arbitrary curved background. For the de Sitter space we are particularly interested in, substituting appropriate values for the curvature invariants yields the desired result. We will  use the effective potential shown in \ref{v36} for the Wetterich flow, in the {\it small effective mass} limit in \ref{S3}. Also in \ref{v36}, we will take $\lambda \gtrsim 4 g^2$, so that the potential remains bounded from below.

%%%%%%%%%%%%%
\subsection{Effective potential using the exact de Sitter propagator}\label{ef2}
%%%%%%%%%%%%

\noindent
We now wish to derive the one loop effective potential using the exact de Sitter propagator. The de Sitter metric appropriate to describe an inflationary background reads 
\begin{eqnarray}
ds^2 = -dt^2 + a^2(t) d{\vec x}^2= a^2(\eta) \left(-d\eta^2 +  d{\vec x}^2\right)
\label{v37}
\end{eqnarray}
where $a(t)= e^{Ht}$ or $a(\eta)=-1/H\eta$, and $H=\sqrt{\Lambda/3}$ is the de Sitter Hubble constant.

Let us consider the $\phi^4$-theory first. Evaluation of such potential via computing the eigenvalues and summing them numerically was reported long back in~\cite{Fujimoto:1987dp}. The de Sitter propagator for the free scalar field theory in $d$-spacetime dimensions reads, e.g.~\cite{Birrell:1982ix},
\begin{eqnarray}
iG(x,x') =\frac{H^{d-2}}{(4\pi)^{d/2}}\frac{\Gamma\!\left(\frac{d-1}{2}+\nu\right)\Gamma\!\left(\frac{d-1}{2}-\nu\right)}{\Gamma\!\left(\frac{d}{2}\right)}{}_2F_1
\left(\frac{d-1}{2}+\nu,\frac{d-1}{2}-\nu;\frac{d}{2};1+\frac{y}{4}\right),
\label{v38}
\end{eqnarray}
where
 $$y(x,x')=a(\eta) a(\eta') \left[-(\eta-\eta')^2+|\mathbf{x}-\mathbf{x}'|^2\right]$$
is the de Sitter invariant distance function, and $\nu =\sqrt{9/4 - m^2/H^2}$.

For the pure scalar field  part of \ref{y1}, the effective potential is given by, 
\begin{eqnarray}
\frac12\int dm^2 \, iG(x,x),
\label{v39}
\end{eqnarray}
where $m^2 =m_0^2+\lambda \phi^2/2 +\xi R$ is the effective mass squared. Note that in  the Schwinger–DeWitt formalism, $\xi R$ is kept explicitly as part of the curvature expansion series, whereas in the exact de Sitter,  the same has been included into the definition of the effective mass parameter. We have in the coincidence limit
\begin{eqnarray}
iG(x,x) = \left(m_0^2+\frac{\lambda \phi^2}{2} +\left(\xi -\frac16\right)12H^2\right)f_d+f_{\rm{fin}}
\label{v40}
\end{eqnarray}
where $f_d = -1/8\pi^2 \epsilon$, and the UV finite part reads
\begin{eqnarray}
f_{\rm{fin}} = \frac{1}{16\pi^2}\left(m^2 - 2H^2\right)
\left[\psi\!\left(\frac12+\nu\right)+\psi\!\left(\frac12-\nu\right)-1+\gamma+\ln\frac{H^2}{4\pi\mu^2}\right]+\frac{H^2}{16\pi^2}.
\label{v41}
\end{eqnarray}
The renormalisation of the above effective potential can be achieved as of the preceding section, and  
the final result  reads
\begin{eqnarray}
&&V_{\rm{eff,scalar}}^{\rm{exact}}=\frac12\int dm^2 f_{\rm{fin}} = -\frac{H^4}{8\pi^2}\Bigg[ \nu\left(\frac14-\nu^2\right)
\ln\frac{\Gamma\!\left(1/2+\nu\right)}{\Gamma\!\left(1/2-\nu\right)}
\nonumber\\&& -\left(\frac14-3\nu^2\right)
\left(\psi^{(-2)}\!\left(1/2+\nu\right)+\psi^{(-2)}\!\left(1/2-\nu\right)\right)-6\nu
\left(\psi^{(-3)}\!\left(1/2+\nu\right)-
\psi^{(-3)}\!\left(1/2-\nu\right)\right)\nonumber\\&&
+6\left(\psi^{(-4)}\!\left(1/2+\nu\right)+
\psi^{(-4)}\!\left(1/2-\nu\right)\right)\Bigg]+\frac{1}{16\pi^2}\left(\frac12 m^4-2H^2m^2\right)\left(\gamma -1+\ln\!\frac{H^2}{4\pi\mu^2}\right)+\frac{H^2m^2}{16\pi^2}
\label{v42}
\end{eqnarray}
In particular, in the  small effective mass limit $m^2/H^2 \ll  1$, the effective potential becomes
\begin{eqnarray}
\frac{V_{\rm{eff,scalar}}^{\rm{exact}}}{H^4} \simeq  \frac{3}{16\pi^2}\ln\!\frac{m^2}{H^2} -\frac{13 m^2}{96\pi^2 H^2}\ln\!\frac{m^2}{H^2}+ \frac{m^2}{32\pi^2 H^2}\left[2\gamma -\frac{11}{3}- 2\ln\frac{H^2}{4\pi\mu^2}\right] + \frac{m^4}{64\pi^2 H^4}\ln\!\frac{m^2}{H^2}+ \frac{m^4}{64\pi^2 H^4}\left[\gamma-\frac{3}{2}+ \ln\frac{H^2}{4\pi\mu^2}\right]\nonumber\\ + \mathcal{O}\!\left(\frac{m^6}{H^6}\right)
\label{v43}
\end{eqnarray}

\noindent
 Let us now come to the $\phi^4$-Yukawa theory. For the Yukawa theory in the absence of any quantum scalar fluctuations, the computation of the one loop effective potential can be seen in~\cite{Miao:2006pn}. Since the effective action requires the propagator in the coincidence limit, it will be sufficient for our present purpose to consider its small separation expansion,  which reads, 
\begin{eqnarray}
 && iS(x,x') 
 = \frac{\Gamma(1-\e/2)}{2^2\pi^{2-\e/2} (aa')^{3/2-\e/2}}\left[i \slashed{\p}+a M \right]\frac{1}{\Delta x^{2-\e}} + \frac{M H^{2-\e}}{2^{2-\e/2}\pi^{2-\e/2}} \frac{\Gamma \left(d/2+ i M/H \right) \Gamma \left(d/2- iM/H \right)}{\Gamma (1+ iM/H)\Gamma (1- iM/H) }\left(-\frac{2}{\e}+\gamma-1 \right) \times {\bf I}_{d\times d}\nonumber\\ &&
 +{\cal O}(y). \label{v44}
\end{eqnarray}
In the coincidence limit we have
\begin{eqnarray}
 {\rm Tr} iS(x,x) =  \frac{{M }H^{2-\e}}{2^{-\e/2}\pi^{2-\e/2}} \left[\left(-\frac{2}{\e}+\frac32+\gamma \right)+ \left(-\frac{2}{\e}-\frac12+\gamma \right)\frac{M^2}{H^2}+ \left(1+\frac{M^2}{H^2}\right) \left(\psi\left(1+iM/H\right)+ \psi\left(1-iM/H\right) \right) \right]. %\nonumber\\&&
 \label{v45}
\end{eqnarray}
We have
\begin{eqnarray}
-\int dM \, iS(x,x)\vert_{\rm fin.} = -\left[ \left( -\frac{2}{\e}+\frac32 +\gamma \right) \frac{M^2 H^{2-\e}}{2^{1-\e/2}\pi^{2-\e/2}} +   \left( -\frac{2}{\e}-\frac32 +\gamma \right) \frac{M^4 H^{-\e}}{2^{2-\e/2}\pi^{2-\e/2}} + \int dM F_{\rm fin}   \right],
\label{v46}
\end{eqnarray}
where as earlier the fermion effective mass is defined as $M= M_0+g\phi$, and 
\begin{eqnarray}
F_{\rm fin}= \frac{MH^2}{\pi^2} \left( 1+ \frac{M^2}{H^2}\right)\left[\psi\left(1+ \frac{iM}{H}\right)+ \psi\left(1- \frac{iM}{H} \right)\right]. 
\label{v48}
\end{eqnarray}
Taking the small mass limit $M/H \ll 1$, limit and using the formula 
$$\psi\left(1+ ix\right)+ \psi\left(1- ix\right) = -2\gamma -2\zeta(3)x^2+2\zeta(5)x^4-2\zeta(7)x^6+2\zeta(9)x^8-\cdots,$$
we have
\begin{eqnarray}
-\int dM\, F_{\text{fin}} = -\frac{H^4}{\pi^2}\Bigg[- \frac{\gamma M^2}{H^2}+\frac{\zeta(3)-\gamma}{2}\frac{M^4}{H^4}+\frac{\zeta(3)-\zeta(5)}{3}\frac{M^6}{H^6}+\frac{\zeta(7)-\zeta(5)}{4}\frac{M^8}{H^8}+\mathcal{O}\!\left(\frac{M^{10}}{H^{10}}\right)\Bigg].
\label{v49}
\end{eqnarray}
Combining the above result with \ref{v43}, we have the one loop effective potential for the $\phi^4$-Yukawa theory in the de Sitter background in the {\it small mass limit},
\begin{eqnarray}
&&\frac{V_{\rm{eff}}^{\rm{exact}}}{H^4} = \frac{(m_0^2+12\xi H^2)\phi^2}{2H^4}+\frac{\lambda \phi^4}{4!H^4}+\frac{1}{16\pi^2}\left(3 -\frac{13 m^2}{6 H^2} + \frac{m^4}{4H^4}\right)\ln\!\frac{m^2}{H^2}\nonumber\\&& -\frac{\lambda\phi^2}{32\pi^2} \left[\frac{3}{(m_0^2+12\xi H^2)}-\frac{13}{6H^2}\ln\frac{(m_0^2+12\xi H^2)e}{H^2} +\frac{(m_0^2+12\xi H^2)}{2H^4}\ln\frac{(m_0^2+12\xi H^2)e^{\frac12}}{H^2}\right]\nonumber\\&&+\frac{\lambda^2\phi^4}{384\pi^2}\left[\frac{9}{(m_0^2+12\xi H^2)^2}+\frac{13}{2H^2(m_0^2+12\xi H^2)}-\frac{3}{2H^4}\ln\frac{(m_0^2+12\xi H^2)e^{\frac32}}{H^2})\right]\nonumber\\&& -\frac{(g\phi)^5}{\pi^2H^6}\Bigg[\frac{\zeta(3)-\zeta(5)}{3} \left(6M_0+g\phi \right)-\frac{\zeta(5)-\zeta(7)}{4H^2}\left(56M_0^3 +28M_0^2g\phi +8M_0(g\phi)^2 +(g\phi)^3\right)\Bigg] + {\rm subleading ~terms},\nonumber\\
\label{v50}
\end{eqnarray}
where we recall that $m^2=m_0^2 +\xi R + \lambda \phi^2/2$ ($R= 12 H^2$), and $M=M_0+g\phi$, and we have used finite counterterms as earlier so that \ref{v50} satisfies the renormalisation condition appearing below \ref{v36}. \\

\noindent
We next wish to extend the computation of the one loop effective potential to an $O(N)$ symmetric model of the scalar field theory. Due to the $O(N)$ symmetry, fermions will be discarded in the following computations. As we mentioned earlier, the effective action for this model in the symmetric phase was computed earlier in the de Sitter spacetime  in~\cite{Serreau:2011fu, Guilleux:2015pma}. We will compute the same below in the broken phase. Derivation of $O(N)$  non-perturbative effective potential  in the 2PI Hartree approximation can be seen in~\cite{Lazzari:2013boa, Nath:2025vhg, Arai:2013jna,   LopezNacir:2014fhi, LopezNacir:2016gzi}, including the effect of negative rest mass squared. Our motivation behind this study is to see the effect of the number of species $N$ in deep IR.  \\

%%%%%%%%%%%%%
\subsection{Effective potential for $O(N)$ model in the broken phase in de Sitter}\label{ef3}
%%%%%%%%%%%%

\noindent
The action for the $O(N)$ symmetric scalar field theory reads
\begin{eqnarray}
S[\Phi_i]=-\! \int \sqrt{-g} d^dx \left[ \frac{1}{2} \Phi^i (-\Box +m_0^2) \Phi^i+\frac{1}{2}\xi R \Phi^i\Phi^i+\frac{\lambda}{4!N} (\Phi^i\Phi^i)^2 \right],
\label{v50'}
\end{eqnarray}
where $i= 1, 2, \cdots N$. We make the decomposition of $\Phi^i$ into classical background ($\phi^i$) and quantum fluctuation,
$\Phi^i = \phi^i +\varphi^i$. In the broken phase only one field has a classical background while the rest $(N-1)$ are purely quantum. We take  $\phi^1=\phi$,  $\phi^i=0$ for all $i= 2, 3, \cdots N-1$ and rename the quantum  fluctuations as $\varphi^1=\sigma$, $\varphi^i=\pi^i$ ($ i=2, 3, \cdots N-1$).\\

\noindent
Let us first compute the effective potential in the local Schwinger-DeWitt formalism,  as earlier, up to the quadratic order in the spacetime curvature. The relevant mass matrix reads
\begin{eqnarray}
m_{ij}^2=m_0^2\delta_{ij}+\frac{\lambda}{3N}\left[\Phi^i\Phi^j +\frac12\Phi_k^2\delta_{ij}\right].
\label{v52}
\end{eqnarray}
Accordingly, we have for the radial mode, $i=j=1$ and $\Phi^1=\phi$ (say), 
\begin{eqnarray}
m_\sigma^2= m_0^2+\frac{\lambda \phi^2}{2N},
\label{v55}
\end{eqnarray}
and for the  transverse modes, $i=j=2,\ldots,N$  ($\phi^i=0$), 
\begin{eqnarray}
m_\pi^2=m_0^2+\frac{\lambda \phi^2}{6N}.
\label{v56}
\end{eqnarray}
The one loop effective potential is obtained as,
\begin{eqnarray}
V_{\rm eff}(\phi)= \frac{1}{2}(m_0^2+\xi R)\phi^2 +\frac{\lambda\phi^4}{4!N} +\frac{1}{2}\int d m_\sigma^2\,iG_\sigma(x,x)+\frac{N-1}{2}\int d m_\pi^2\,iG_\pi(x,x),
\label{v57'}
\end{eqnarray} 
where,
\begin{eqnarray}
iG_{\sigma,\pi}(x,x)= m_{\sigma ,\pi}^2f_d+\left(\xi -\frac16 \right)R f_d+f_{\rm{fin}}^{\sigma,\pi}.
\label{v58}
\end{eqnarray}
$f_d = -\mu^{-\epsilon}/(8\pi^2 \epsilon)$ as earlier and
\begin{eqnarray}
&&f_{\rm{fin}}^{\sigma,\pi} =  \frac{m_{\sigma ,\pi}^2}{(4\pi)^2} \left(\ln\frac{m_{\sigma ,\pi}^2}{4\pi\mu^2} -\psi(2) \right)+\frac{\left(\xi-1/6\right)R}{(4\pi)^2}\left(\ln\frac{m_{\sigma ,\pi}^2}{4\pi\mu^2} -\psi(1)\right)+\frac{ f_1}{(4\pi)^2m_{\sigma ,\pi}^2}
\label{v59}
\end{eqnarray}
%%\cite{Nath:2025vhg}
After  imposing the renormalisation conditions 
$$ \left.\frac{\partial^2 V_{\rm eff,N}}{\partial\phi^2}\right|_{\phi=0}= m_0^2 +\xi R, \quad \left.\frac{\partial^4 V_{\rm eff,N}}{\partial\phi^4}\right|_{\phi=0}= \frac{\lambda}{N},$$
we obtain the renormalised result
\begin{eqnarray}
&&V_{\text{eff, N}}(\phi) =\frac{1}{2}(m_0^2 + \xi R)\phi^2
+ \frac{\lambda \phi^4}{24N} + \frac{1}{4(4\pi)^2}
\Bigg[m_\sigma^4  + 2(\xi - \tfrac{1}{6})R\, m_\sigma^2
+ 2 f_1\Bigg]\ln \frac{m_\sigma^2}{4\pi \mu^2}\nonumber\\
&& + \frac{N-1}{4(4\pi)^2}\Bigg[m_\pi^4+ 2(\xi - 1/6)R\, m_\pi^2 + 2 f_1\Bigg] \ln \frac{m_\pi^2}{4\pi \mu^2}-\frac{\lambda(N+2)\phi^2}{24N(4\pi)^2}\Bigg[2m_0^2\ln \frac{m_\pi^2}{4\pi \mu^2}+m_0^2\nonumber\\&& +2R\left(\xi-\frac16\right)\ln\frac{m_0^2 e}{4\pi\mu^2}+\frac{2f_1}{m_0^2}\Bigg]-\frac{\lambda^2(N+8)\phi^4}{144N^2(4\pi)^2}\Bigg[\ln\frac{m_0^2 e^{\frac32}}{4\pi\mu^2}+\frac{\left(\xi-1/6\right)R}{m_0^2}-\frac{f_1}{m_0^4}\Bigg]
\label{v61}
\end{eqnarray}

\bigskip

\noindent
Let us now come to the case of the  exact de Sitter spacetime. We need to use the exact propagator of \ref{v40}, and  the masses of the fields read in place of \ref{v55}, \ref{v56},
\begin{eqnarray}
&& m_{\sigma}^2= m_0^2 +12\xi H^2 +\frac{ \lambda \phi^2}{2N},\quad
m_{\pi}^2= m_0^2 +12\xi H^2+\frac{ \lambda \phi^2}{6N},
\label{v62}
\end{eqnarray}
We will take both these masses to be small compared to the Hubble rate. We refer our reader 
to~\cite{Arai:2013jna, LopezNacir:2014fhi, LopezNacir:2016gzi} for  ${\cal O}(N)$ model effective action  using the Hartree approximation . We have the one loop effective potential
\begin{eqnarray}
&&\frac{V_{\text{eff, exactN}}(\phi)}{H^4} =\frac{(m_0^2 + 12\xi H^2)\phi^2}{2H^4} + \frac{\lambda\phi^4}{24N H^4}+\frac{1}{16\pi^2}\left(3 -\frac{13 m_{\sigma}^2}{6 H^2}+ \frac{m_{\sigma}^4}{4H^4}\right)\ln\!\frac{m_{\sigma}^2}{H^2}  +\frac{(N-1)}{16\pi^2}\left(3 -\frac{13 m_{\pi}^2}{6 H^2} + \frac{m_{\pi}^4}{4 H^4}\right)\ln\!\frac{m_{\pi}^2}{H^2}\nonumber\\&& -\frac{(N+2)\lambda\phi^2}{1152\pi^2N(m_0^2+12\xi H^2)H^4}
\left[36H^4-26H^2(m_0^2+12\xi H^2)\ln\frac{(m_0^2+12\xi H^2)e}{H^2}\right.\nonumber\\&&\left. +6(m_0^2+12\xi H^2)^2\ln\frac{(m_0^2+12\xi H^2)e^{\frac12}}{H^2}\right]+\frac{(N+8)\lambda^2\phi^4}{13824\pi^2N^2H^4(m_0^2+12\xi H^2)^2}\left[36H^4+26H^2(m_0^2+12\xi H^2)\right.\nonumber\\&&\left. -6(m_0^2+12\xi H^2)^2\ln\frac{(m_0^2+12\xi H^2)e^{\frac32}}{H^2}\right],
\label{v65}
\end{eqnarray}
where $m_{\sigma}^2$ and $m_{\pi}^2$ are given in  \ref{v62}. The results shown in \ref{v61}, \ref{v65}, along with those derived in \ref{ef1}, \ref{ef2} completes our discussion on the one loop effective potentials in the Schwinger-DeWitt expansion formalism up to quadratic order in the curvature, as well as  in the exact de Sitter background. We now wish to determine below the IR flow of these potentials using the Wetterich equation. The above effective potentials will serve as the initial condition representing the UV limit in the flow equation.
%%%%%%%%%%
\section{Wetterich flow of the effective potential}\label{S3}
%%%%%%%%%%

\noindent
In this section, we derive the IR flow of the effective potential for a scalar field in de Sitter space using the non-perturbative renormalisation group or the Wetterich formalism. The main objective is to understand how quantum fluctuations generated during inflation modify the effective theory as infrared modes are progressively integrated out. Starting from the Wetterich equation, the flow of the effective action is evaluated for constant field configurations, reducing the problem to the running of the effective potential $V_\kappa(\phi)$, where $\kappa$ is the scale specifying the IR evolution, with $\kappa=1$ corresponds to the UV, and $\kappa \to 0$ to the deepest IR.  We will see below restoration of the broken symmetry. The detail of  this formalism can be seen in the original references~\cite{Wetterich:1992yh} (also e.g.~\cite{Serreau:2011fu, Guilleux:2015pma} and references therein). \\

\noindent
By ignoring the variation of the background field, we begin with the Wetterich equation~\cite{Wetterich:1992yh}, 
\begin{eqnarray}
\dot{V}_\kappa[\phi]=\frac12 {\rm{Tr}}{\left(\dot{R_{\kappa}}(x,y)G_{\kappa}(y,x)\right)}, 
\label{w1}
\end{eqnarray}
where $V_{\kappa}$ is an effective potential evolving under the flow of an IR scale $\kappa$,  $R_\kappa$ is an IR regulator, $G_\kappa$ is the propagator evolving under the flow, and $y$ is integrated out. We also have,  $\dot{V}_\kappa[\phi] = \kappa\partial_{\kappa}V_\kappa[\phi]$, and 
\begin{eqnarray}
iG_\kappa^{-1}(x,y)=\frac{1}{\sqrt{g(x)g(y)}}\frac{\delta^2\Gamma_\kappa[\phi]}{\delta\phi(y)\delta\phi(x)}
+R_\kappa(x,y),
\label{w2}
\end{eqnarray}
where $\Gamma_\kappa[\phi]$ is the effective action. $\kappa \to 1$ stands for the UV, whereas $\kappa \to 0$ for the deepest IR. The effective potentials we have derived in the preceding section corresponds to $\kappa=1$, serving as the initial condition for the Wetterich equation. Following~\cite{Serreau:2011fu, Guilleux:2015pma}, we now introduce the spatial momentum space owing to the spatial homogeneity of de Sitter to write
\begin{eqnarray}
R_\kappa(x,y) = R_\kappa(\eta,\eta',|\vec x-\vec y|)
= \int \frac{d^{3}\vec{k}}{(2\pi)^{3}}e^{i\vec{k}\cdot(\vec x-\vec y)}
\tilde R_\kappa(\eta,\eta',k),
\label{w3}
\end{eqnarray}
where $k=|\vec{k}|$. Redefining now the field as $\phi(x)\rightarrow a(\eta)\phi(x)$, and introducing the physical momentum operator, $p=k/a$,  one has~\cite{Serreau:2011fu},
\begin{eqnarray}
\tilde R_\kappa(\eta,\eta',k)=(H^2\eta\eta')^{3}
k^3\hat R_\kappa(p,p'),
\label{w6}
\end{eqnarray}
where
\begin{eqnarray}
\hat R_\kappa(p,p')=H\,\frac{\delta(p-p')}{p^2}R_\kappa(p).
\label{w7}
\end{eqnarray}
$R_\kappa(p)$ should be such that it  suppresses the modes with $p\ensuremath{<}\kappa$. Likewise, we have for the propagator,
\begin{eqnarray}
\tilde G_\kappa(\eta,\eta',k) = \frac{H^2\eta\eta'}{k} \hat G_\kappa(p,p').
\label{w9}
\end{eqnarray}
In this spatial momentum space description, the flow equation  for a constant field configuration can be written as~\cite{Guilleux:2015pma},
\begin{eqnarray}
\dot V_\kappa(\phi)
=\int \frac{d^{3}\vec{p}}{(2\pi)^{3}}\frac{\dot R_\kappa(p)}{2p}
\hat F_\kappa(p,p),
\label{w12}
\end{eqnarray}
where $F_\kappa(p,p')$ is the momentum space representation of the statistical function 
\begin{eqnarray}
G_\kappa(x,x')=\langle T\phi(x)\phi(x')\rangle =
F_\kappa(x,x')-\frac{i}{2}\mathrm{sgn}(\eta-\eta')
\rho_\kappa(x,x'),
\label{w13}
\end{eqnarray}
with
$$ F_\kappa(x,x')=\frac{1}{2}\langle\{\phi(x),\phi(x')\}\rangle, \quad \rho_\kappa(x,x')=i\langle[\phi(x),\phi(x')]\rangle.$$
Using the above ingredients, one derives the equation satisfied by the flow improved Green function
\begin{eqnarray}
\left(\partial_p^2 +1-\frac{\nu_\kappa^2-1/4}{p^2}-
\frac{R_\kappa(p)}{Z_\kappa H^2p^2}\right)\hat G_\kappa(p,p')
=i\delta(p-p').
\label{w18}
\end{eqnarray}
where $Z_\kappa$ is the scale-dependent field strength renormalisation factor,  $\eta_\kappa =-\dot{Z}_\kappa/Z_\kappa$, and 
$$\nu_\kappa =\sqrt{\frac{9}{4}-\frac{V_\kappa''(\phi)}
{Z_\kappa H^2}}$$
In the absence of field strength renormalisation, which will be our case, we have $Z_\kappa=1$ and $\eta_\kappa=0$. 

The mode function $u_\kappa(p)$ satisfies, 
\begin{eqnarray}
&& \left(\partial_p^2 +1-\frac{\nu_\kappa^2-1/4}{p^2}\right)
u_\kappa(p)=0,\qquad(p\ge\kappa)\nonumber\\ &&
\left(\partial_p^2 +1-\frac{\kappa^2}{H^2p^2}-\frac{\bar\nu_\kappa^2-1/4}{p^2}\right)u_\kappa(p)=0,
\qquad(p\ensuremath{<}\kappa),
\label{w22}
\end{eqnarray}
where $\bar\nu_\kappa =\sqrt{\nu_\kappa^2 -\kappa^2/H^2}$. The solutions are given by  respectively
\begin{eqnarray}
&& u_\kappa(p)=\sqrt{\frac{\pi p}{4H}}e^{i\gamma_\kappa}H_{\nu_\kappa}^{(1)}\left(\frac{p}{H}\right), \qquad 
u_\kappa(p)=\sqrt{\frac{\pi p}{4H}}e^{i\gamma_\kappa}\left[c_\kappa^{-}\left(\frac{\kappa}{p}\right)^{\bar\nu_\kappa}
+c_\kappa^{+}\left(\frac{p}{\kappa}\right)^{\bar\nu_\kappa}
\right].
\label{w25}
\end{eqnarray}
Substituting these solutions into \ref{w12}, the flow of the effective potential becomes
\begin{eqnarray}
\dot V_\kappa(\phi)= \frac{1}{(2\pi)^{2}}
\int_0^\kappa dp\,p\Big[(2-\eta_\kappa)\kappa^2
+\eta_\kappa p^2\Big]|u_\kappa(p)|^2.
\label{w26}
\end{eqnarray}
Finally, assuming
\begin{eqnarray}
|V_\kappa''(\phi)|\ll1,
\label{w29}
\end{eqnarray}
which corresponds to the effective mass (such as $m_0^2+\lambda \phi^2/2$) small compared to $H^2$, assuming $\kappa$ to be small compared to unity. Recalling now  that we do not require any field strength renormalisation for our present purpose $(\eta_\kappa=0)$, \ref{w26}  simplifies to
\begin{eqnarray}
\dot V_\kappa(\phi)=\frac{8\pi\kappa^2H^{4}}{V_\kappa''(\phi)+\kappa^2},
\label{w30}
\end{eqnarray}
As we have emphasised earlier, we wish to solve the above equation for $V_\kappa(\phi)$ by treating \ref{v36}, \ref{v50}, \ref{v61} and \ref{v65} as $\kappa \to 1$.\\

\begin{figure}[H]
\begin{center}
\end{center}
\centering
\begin{subfigure}{.4\textwidth}
  \centering
  \includegraphics[scale=0.40]{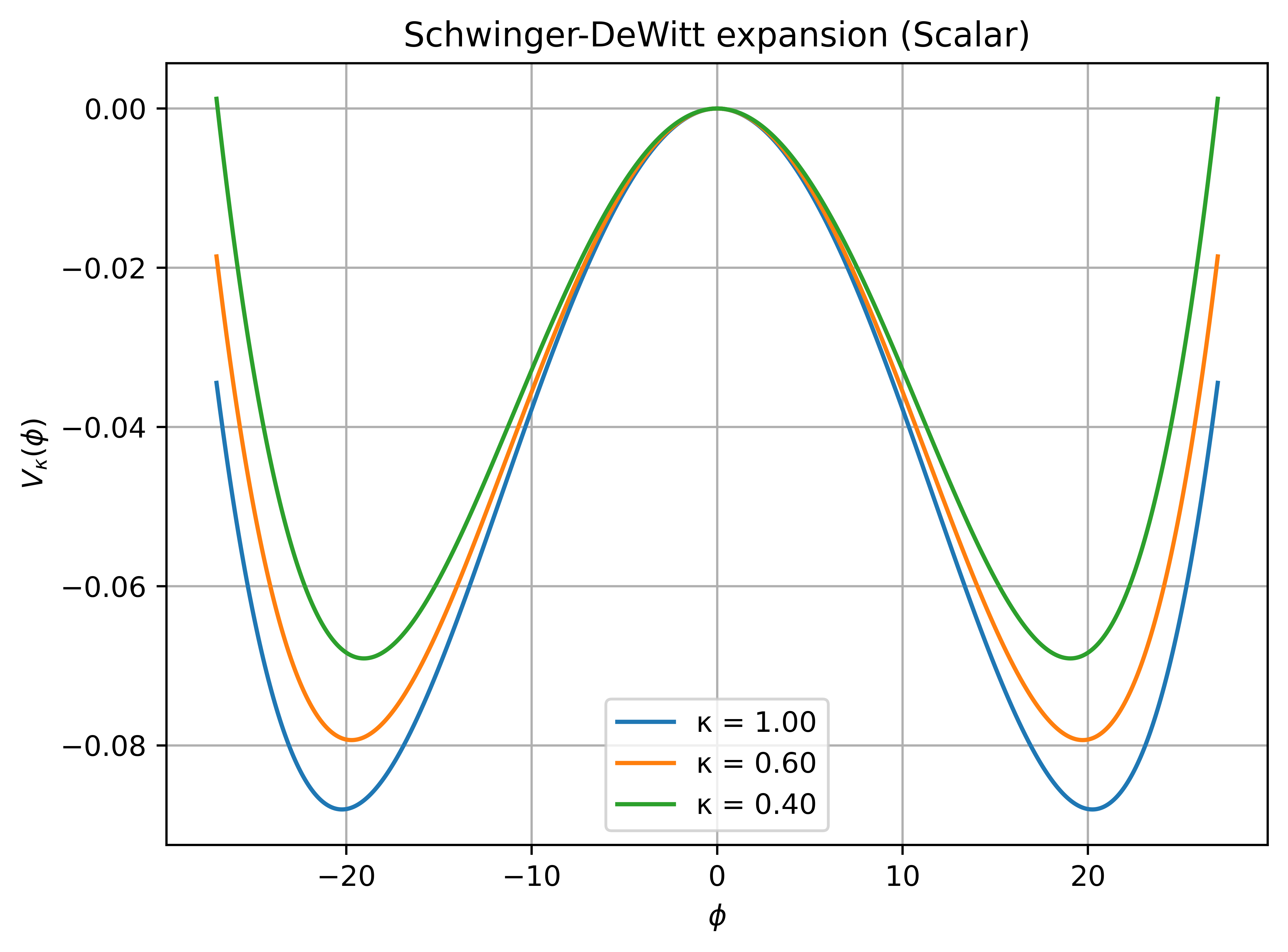}
  \caption{$\lambda\phi^4$ interaction}
  \label{figsub1}
\end{subfigure}%
\begin{subfigure}{.6\textwidth}
  \centering
  \includegraphics[scale=0.40]{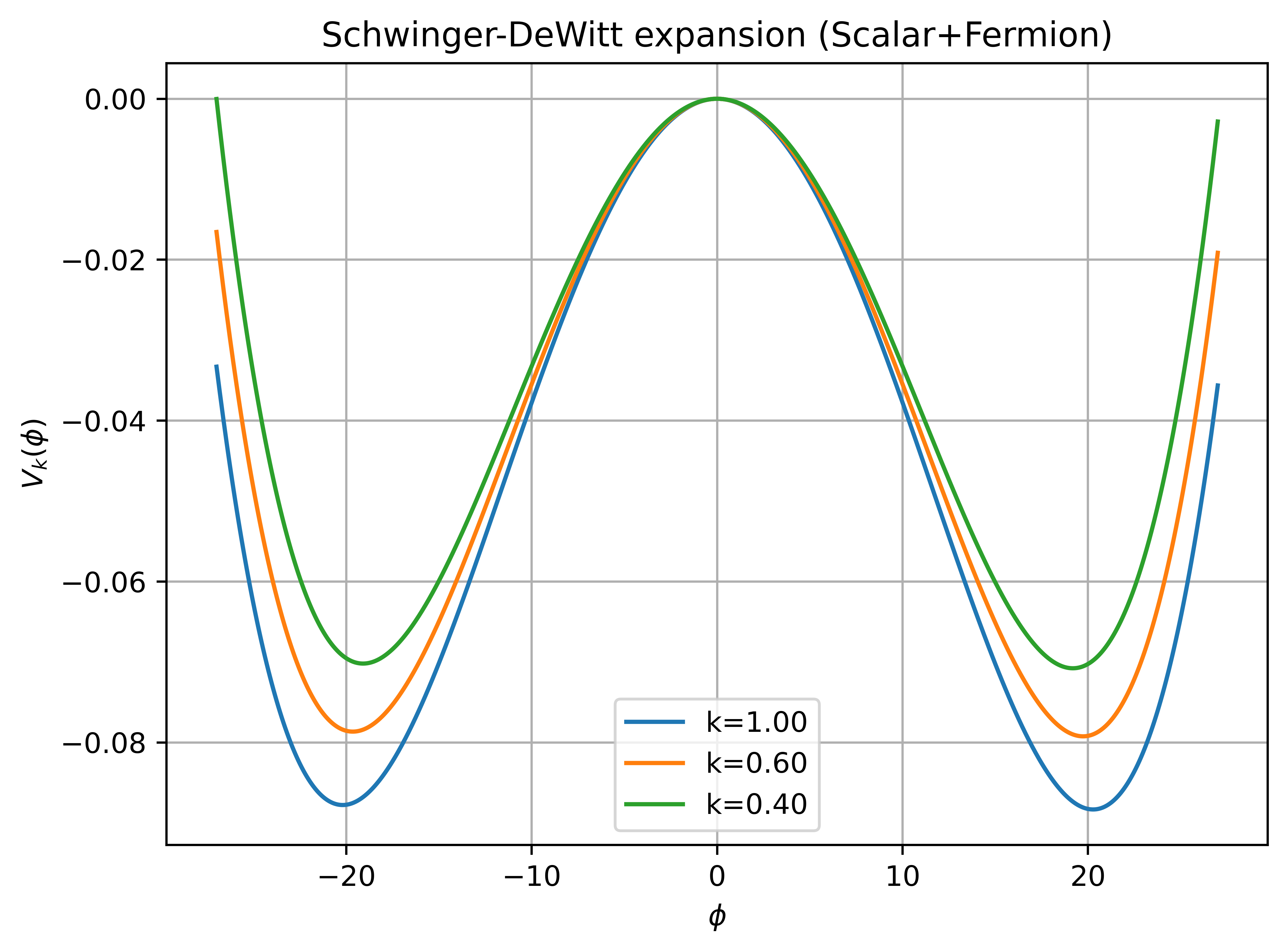}
  \caption{$\lambda\phi^4 +\rm{yukawa}$ interaction}
  \label{figsub2}
\end{subfigure}
\caption{{\it \small IR flow~\ref{w30} of the Schwinger-DeWitt effective potential~\ref{w33},  with  $\lambda=10^{-5},\  \xi = -0.0134, \bar{m}_0=0.4$. The right hand side plots include the Yukawa interaction with coupling $g=2\times 10^{-3}$ and $\bar{M_0} = 0.5 $. The blue curves correspond to the standard one loop effective potential. Thus moving towards the IR restores the symmetry. Note also the asymmetry introduced by the Yukawa coupling.  } }
\label{fig1}
\end{figure}

\begin{figure}[H]
\begin{center}
\end{center}
\centering
\begin{subfigure}{.4\textwidth}
  \centering
  \includegraphics[scale=0.4]{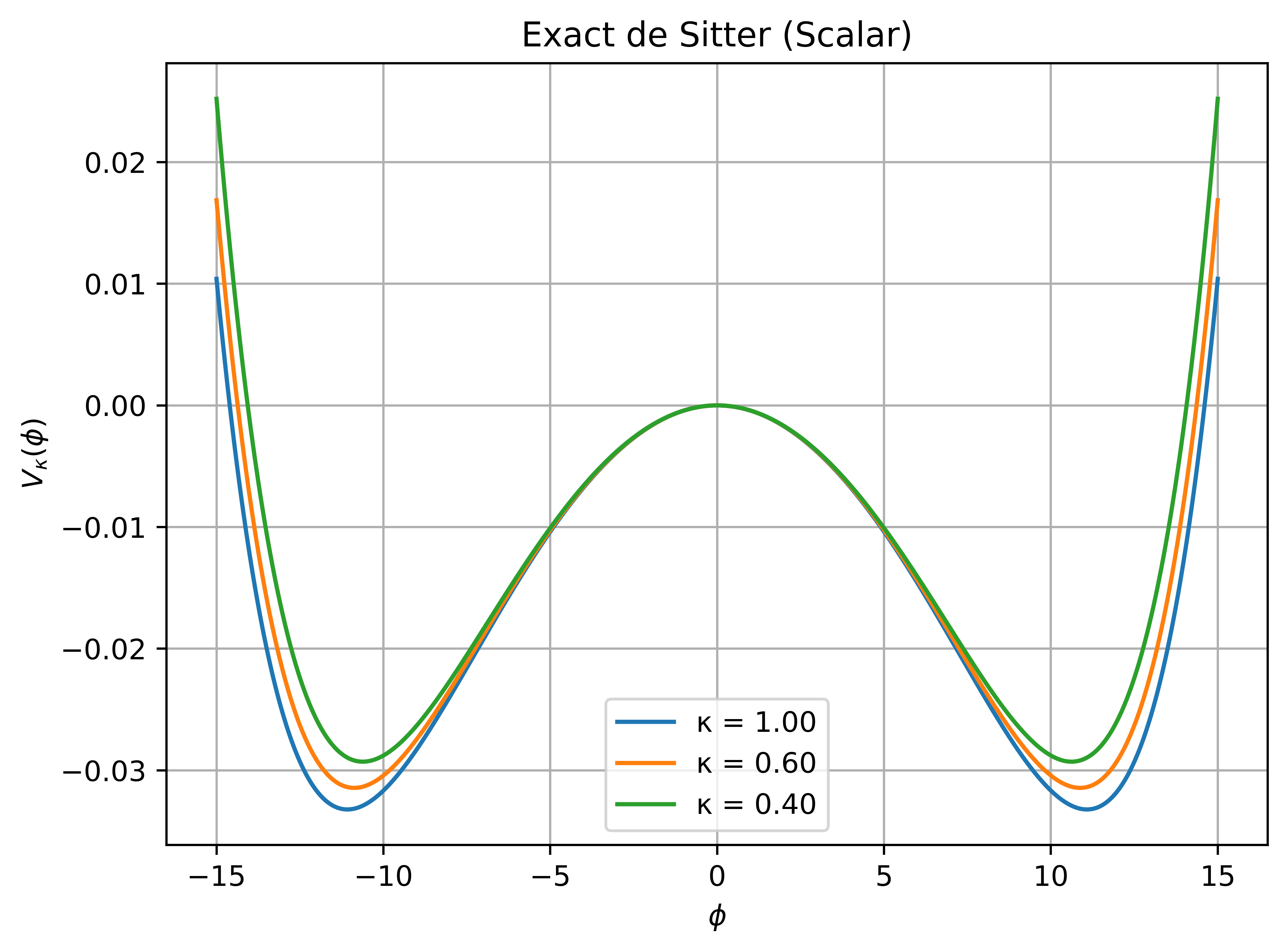}
  \caption{$\lambda\phi^4$ interaction}
  \label{figsub3}
\end{subfigure}%
\begin{subfigure}{.6\textwidth}
  \centering
  \includegraphics[scale=0.4]{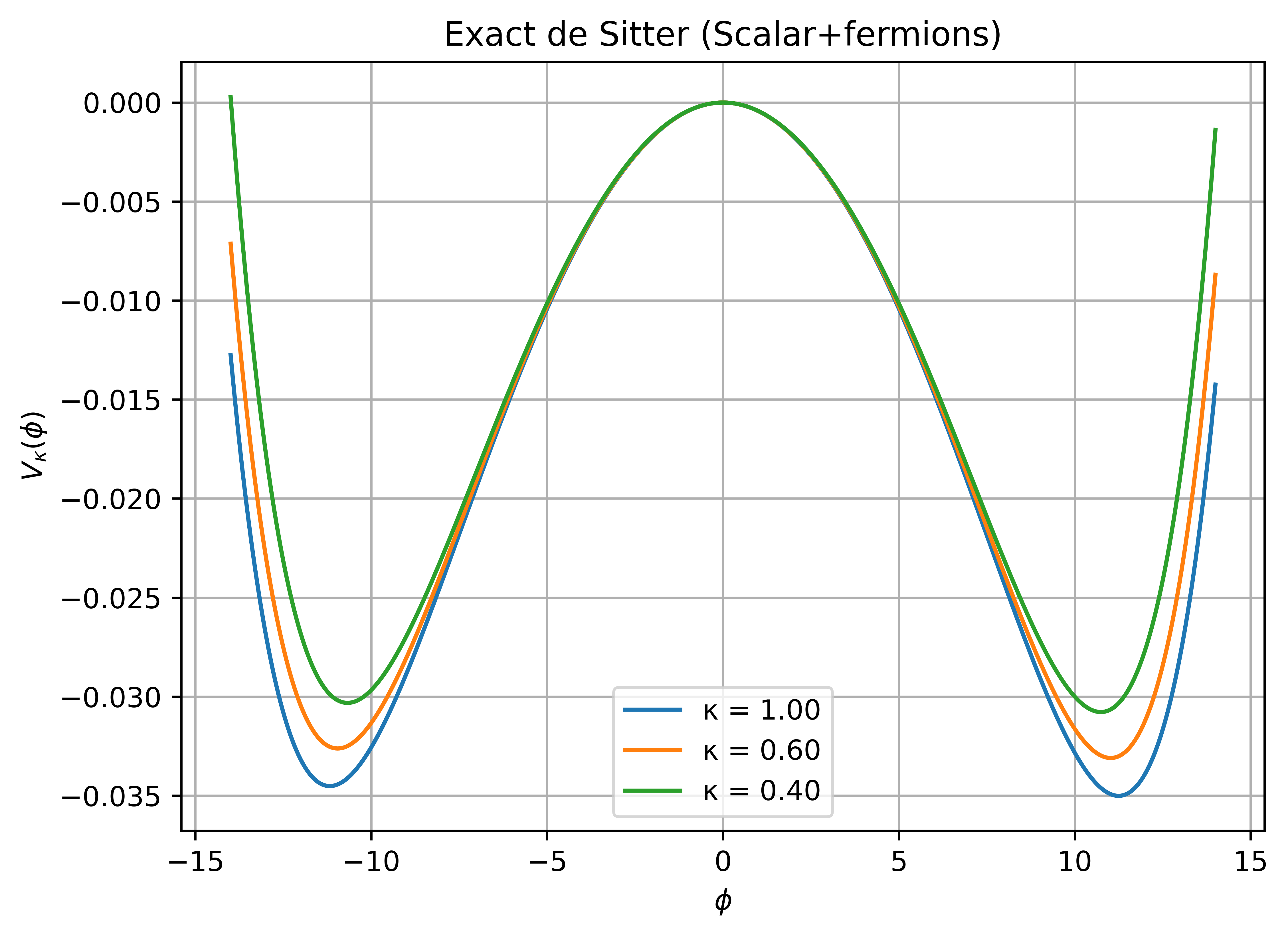}
  \caption{$\lambda\phi^4 +\rm{yukawa}$ interaction}
  \label{figsub4}
\end{subfigure}
\caption{{\it \small IR flow~\ref{w30} of the effective potential derived in the exact de Sitter spacetime~\ref{v50} (the blue curve above), with numerical values borrowed from \ref{fig1}.  Both these plots show qualitatively similar behaviour as of \ref{fig1}, however, there are quantitative differences. See main text for discussion.}}
\label{fig2}
\end{figure}

\noindent
Let us begin with the one loop effective action for the $\phi^4$-Yukawa theory derived using the Schwinger-DeWitt formalism, \ref{v36}. Writing  $R = 12H^2$, $R^{\mu\nu}R_{\mu\nu} =36 H^4$ and $R^{\mu\nu\rho\sigma}R_{\mu\nu\rho\sigma} = 24H^4$, for the de Sitter spacetime,  we expand for the sake of computational convenience  \ref{v36} in the small mass limit ($(m_0^2+\lambda \phi^2)/H^2 \ll 1$) as,
\begin{eqnarray}
&&\frac{V_{\rm eff,Ren}^{1-{\rm loop}}}{H^4}\Bigg\vert_{\kappa=1}=\frac12\left(\bar m_0^2+12\xi\right)\bar\phi^2+\frac{\lambda}{24}\bar\phi^4-\frac{g^5}{720\pi^2\bar M_0^5}\left(18\bar M_0^4-12\bar M_0^2+126\right)\bar\phi^5+\left[\frac{\lambda^3}{1536\pi^2\bar m_0^6}\left\{\bar m_0^4+12\left(\frac16-\xi\right)\bar m_0^2\right.\right.
\nonumber\\
&&\left.\left. -72\left(\frac16-\xi\right)^2+\frac4{15}\right\}+\frac{g^6}{1440\pi^2\bar M_0^6}\left(6\bar M_0^4-12\bar M_0^2+190\right)\right]\bar\phi^6-\frac{g^7}{2520\pi^2\bar M_0^7}\left(3\bar M_0^4-12\bar M_0^2+285\right)\bar\phi^7\nonumber\\
&&+\left[\frac{\lambda^4}{4096\pi^2\bar m_0^8}\left\{72\left(\frac16-\xi\right)^2-\frac4{15}-\bar m_0^4-24\left(\frac16-\xi\right)\bar m_0^2\right\}+\frac{g^8}{20160\pi^2\bar M_0^8}\left(9\bar M_0^4-60\bar M_0^2+1995\right)\right]\bar\phi^8 +\cdots.
\label{w33}
\end{eqnarray}
where a `bar' over any quantity represents scaling with respect to  $H$. Note that small effective  mass also means $\lambda \phi^2/H^2 \ll 1$. In addition, we have also  assumed $\lambda \bar{\phi}^2/\bar{m}_0^2$ to be small as well. Keeping these all in mind,   we have plotted the IR flow of this effective potential  in \ref{figsub1},  \ref{figsub2}. The first set corresponds to pure $\lambda \phi^4$ theory whereas the second includes the Yukawa interaction. All these plots show spontaneous symmetry breaking (SSB) pattern, but diminishing as $\kappa$-value decreases, i.e. as we move towards the IR. Note also that the Yukawa interaction, as expected, introduces an asymmetry in the SSB pattern, as can be seen by observing the two minima of the second plot of \ref{fig1}.  
\noindent Likewise, for the effective potential using the  exact de Sitter propagator \ref{v50}, we have depicted the IR flow  in \ref{figsub3}, \ref{figsub4}. The results obtained in \ref{fig1}, \ref{fig2}   exhibit the same qualitative behaviour, namely the gradual restoration of symmetry as the IR scale $\kappa$ decreases. However as can be seen, the restoration proceeds more slowly when the exact de Sitter propagator is employed. This should correspond to the fact that the Schwinger--DeWitt propagator is based on a local curvature expansion and provides only an approximate description of the spacetime geometry in a small, sub-Hubble region. In contrast, the exact de Sitter propagator fully incorporates the global geometry and the super-Hubble modes, allowing the IR flow to account for the complete long-wavelength dynamics. As a consequence, the evolution of the effective potential becomes more gradual, delaying the disappearance of the SSB minima and hence the onset of the symmetric phase. 

The symmetry restoration phenomenon described above is in qualitative agreement with that of~\cite{Bhattacharya:2025jtp}, where similar conclusion was achieved via the two loop 2PI formalism, using the Hartree-like approximation. 
\begin{figure}[H]
\begin{center}
\includegraphics[scale=0.4]{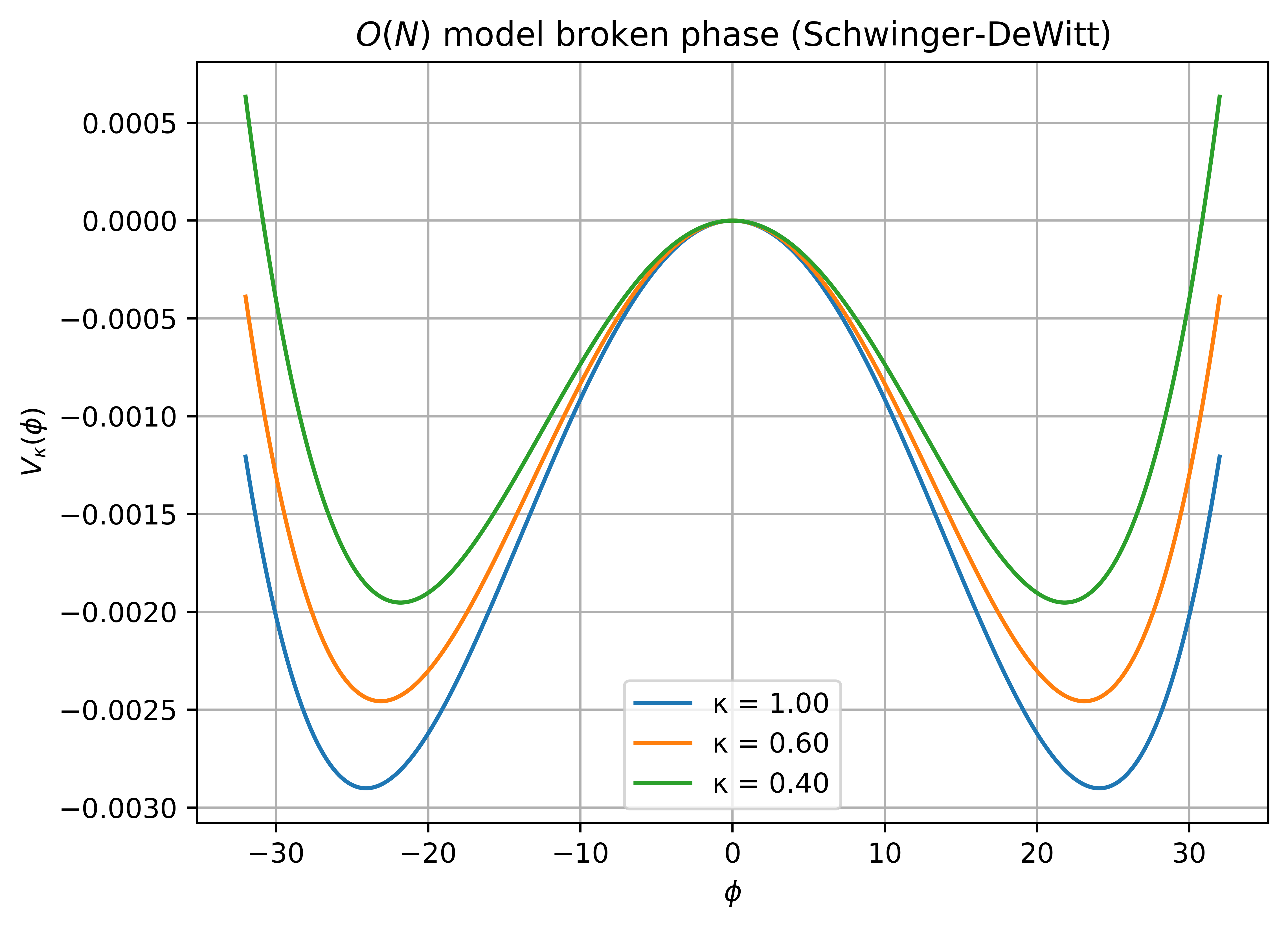}
\includegraphics[scale=0.4]{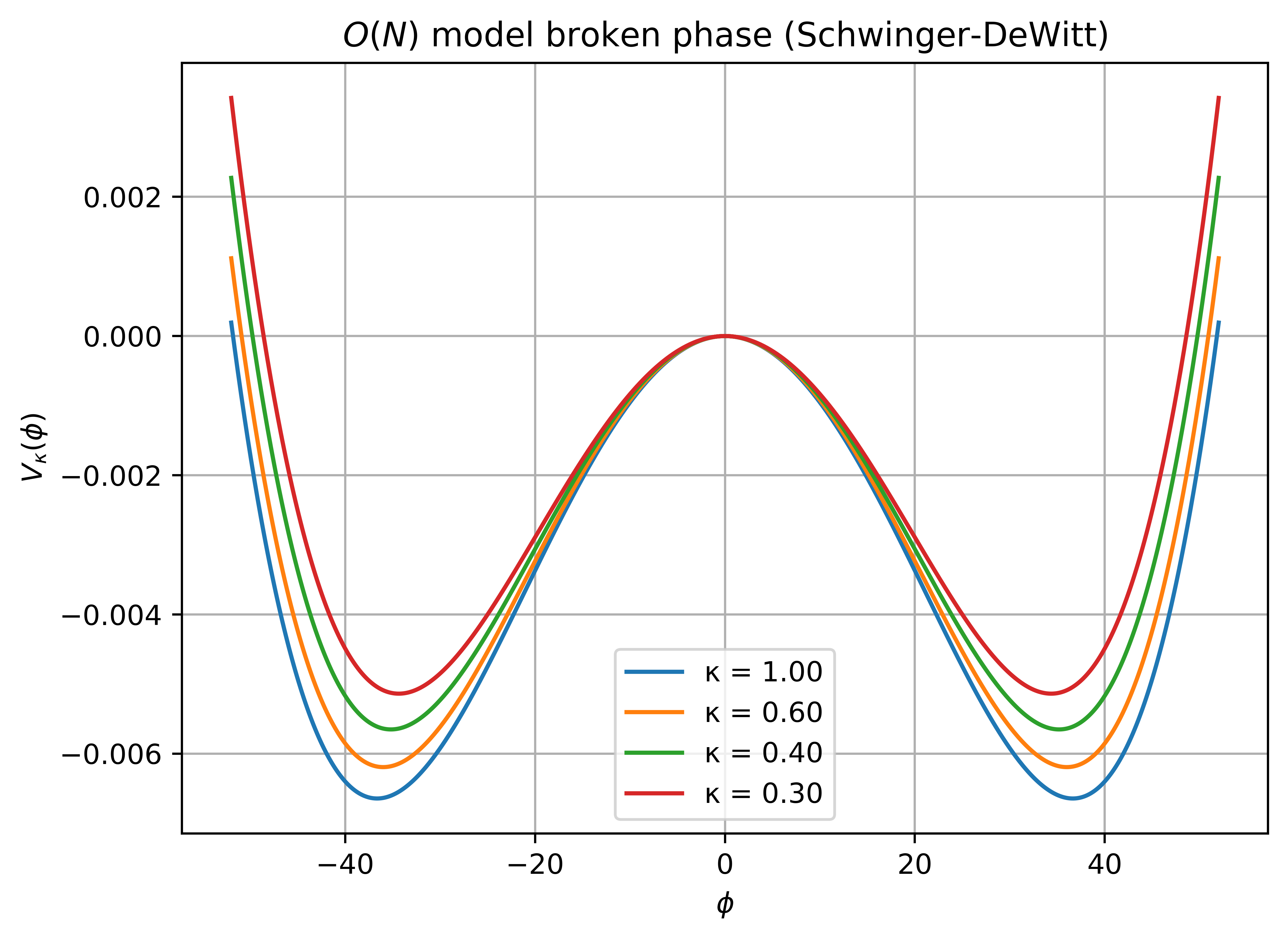}
\includegraphics[scale=0.4]{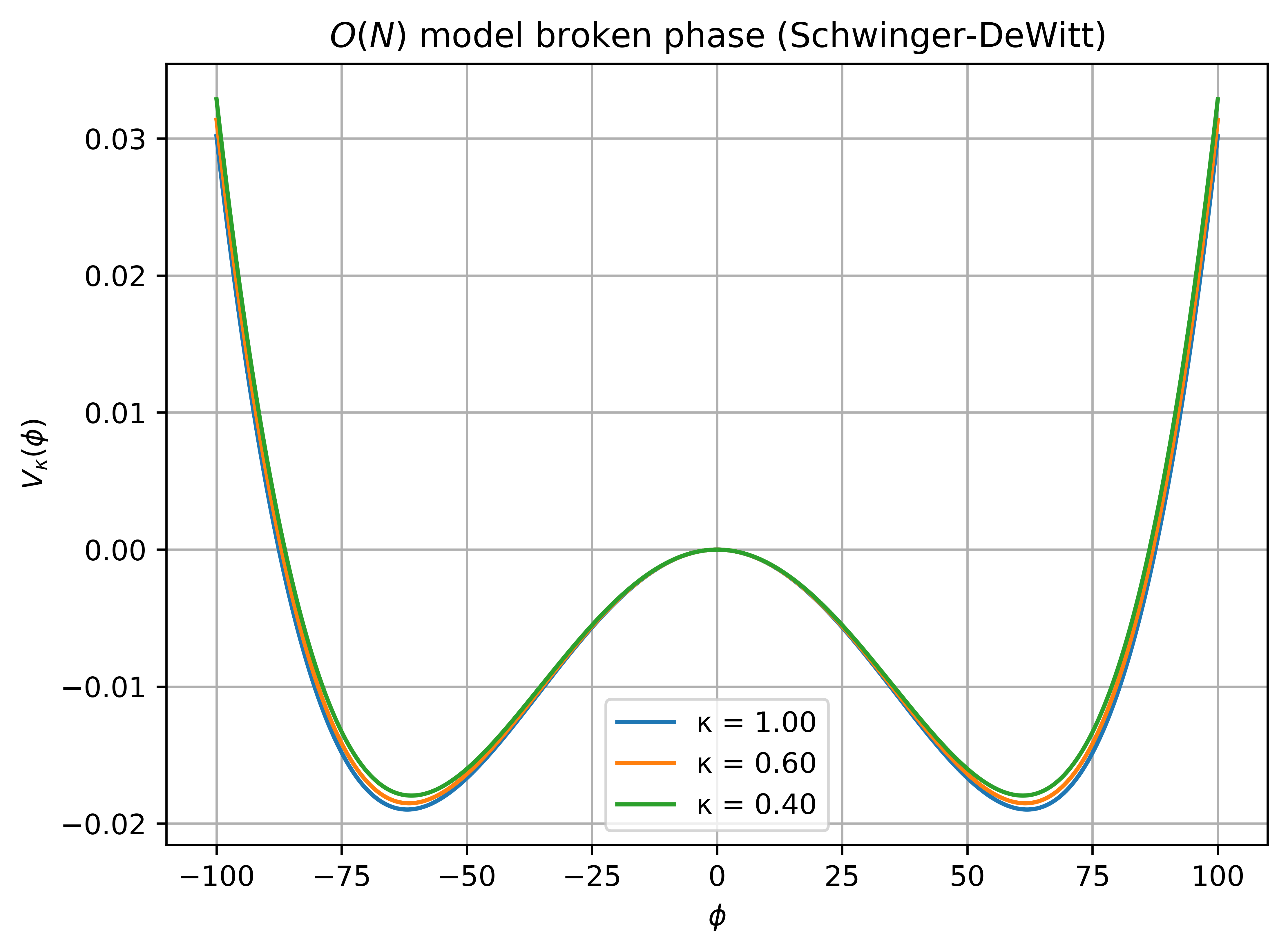}
\includegraphics[scale=0.4]{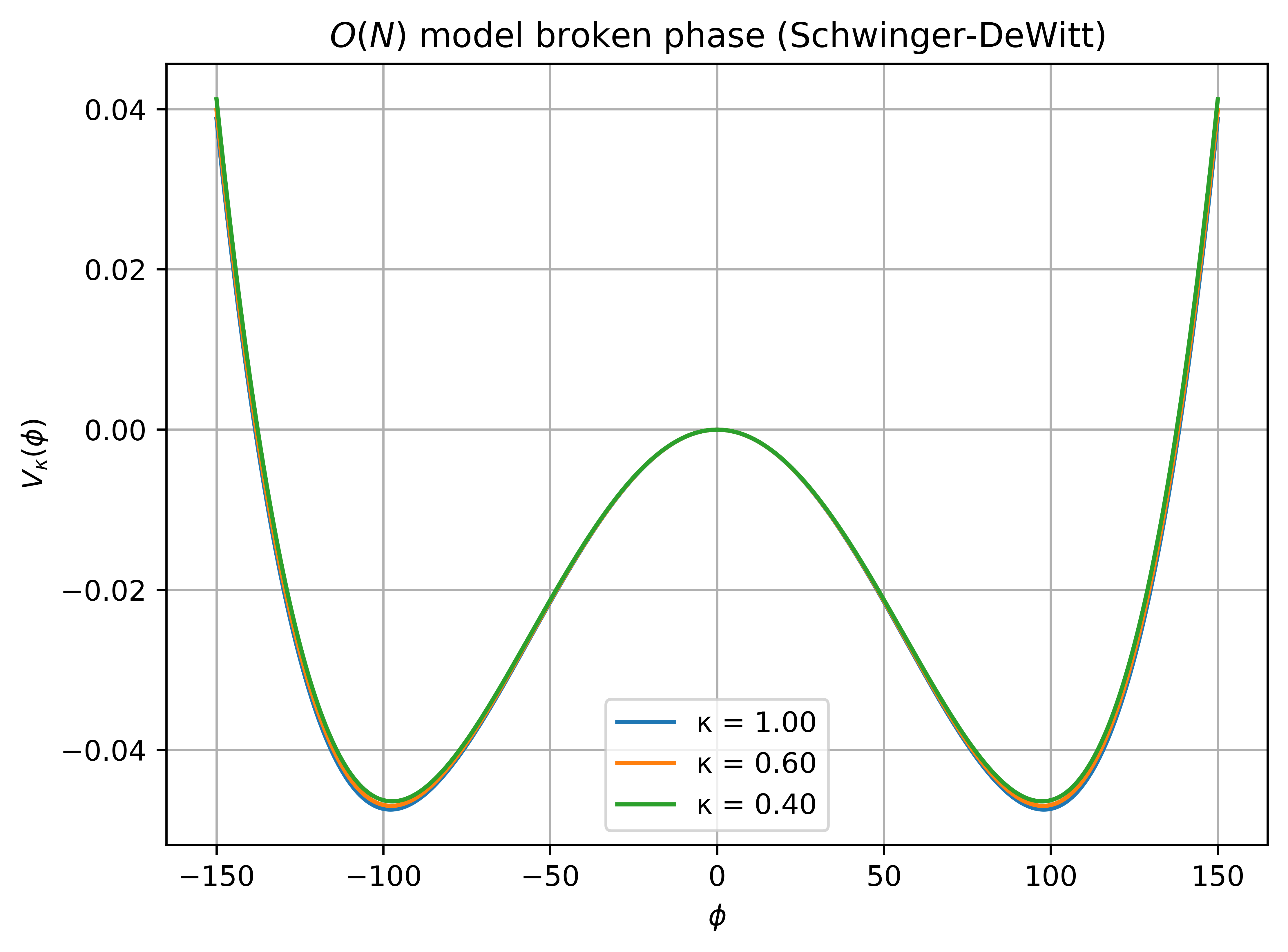}
\caption{ \it \small IR flow~\ref{w30} of the $O(N)$ effective potential~\ref{v61}  for $N = 3, \ 7, \ 20, \ 50$ respectively,  with numerical values $\lambda=10^{-5}, \ \xi = -4.3126\times 10^{-3},\  \bar{m_0}=0.1$.  Symmetry restoration with $\kappa$ decreasing ceases for large N-values.}
\label{fig3}
\end{center}
\end{figure}

\begin{figure}[H]
\begin{center}
\includegraphics[scale=0.4]{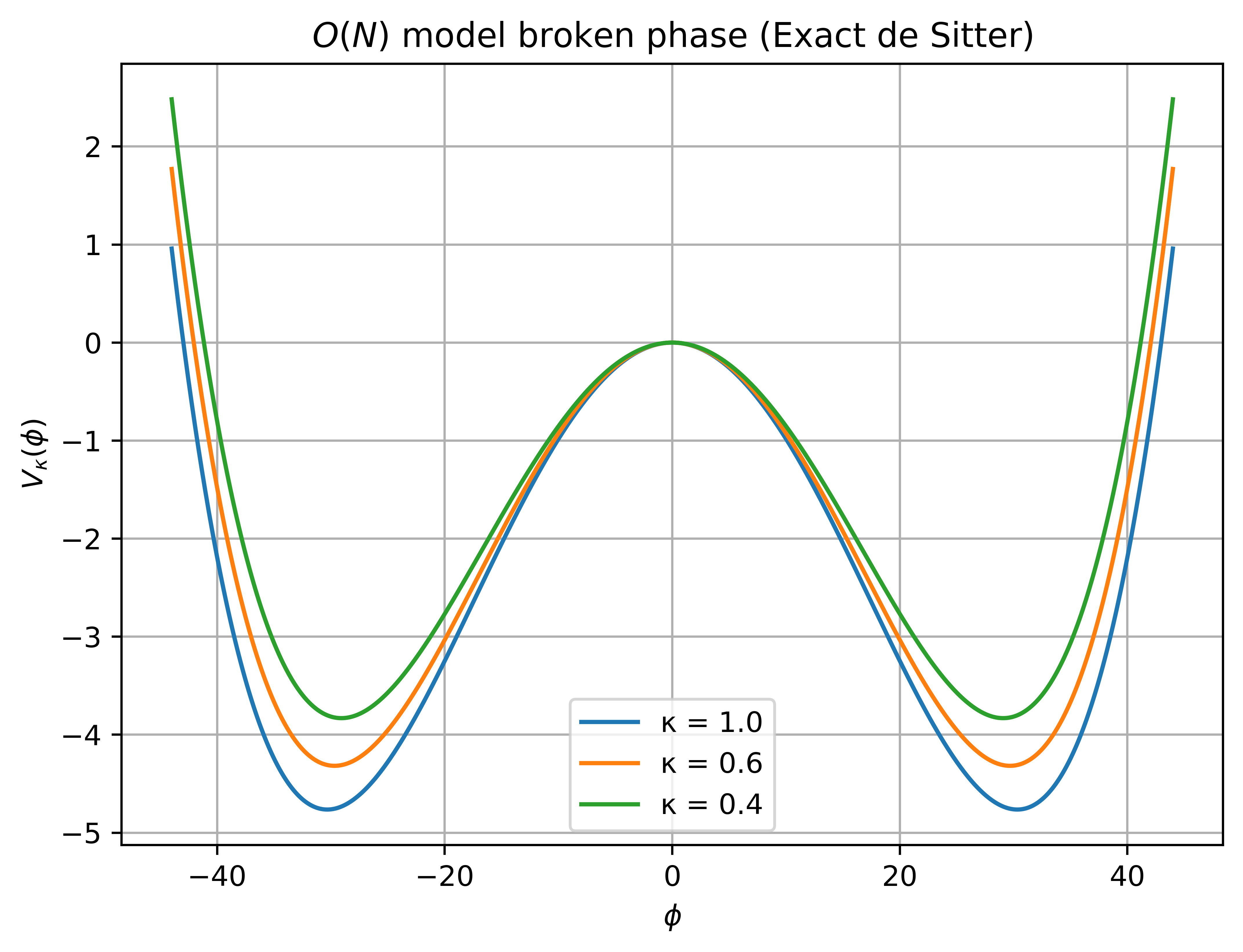}
\includegraphics[scale=0.4]{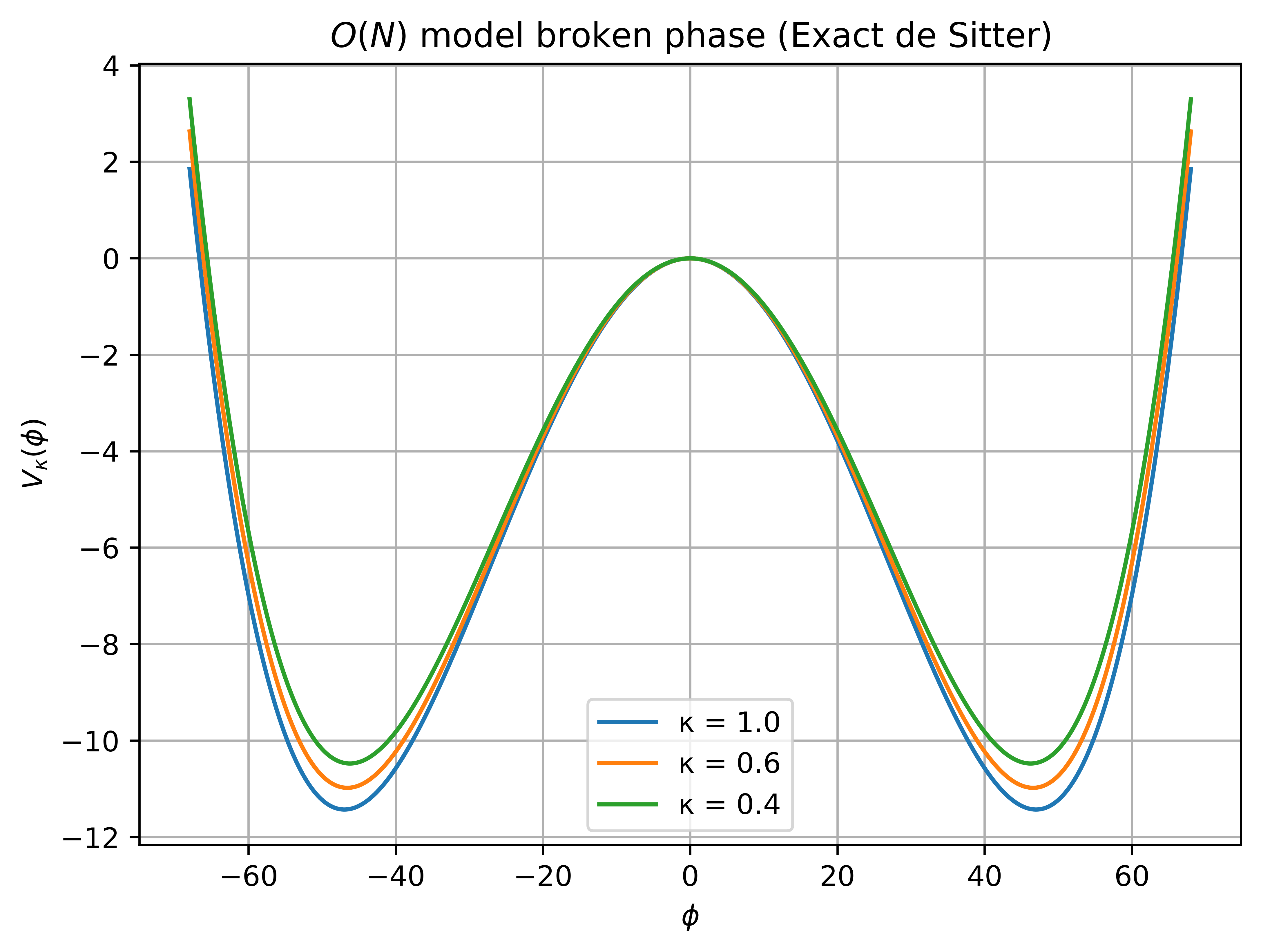}
\includegraphics[scale=0.4]{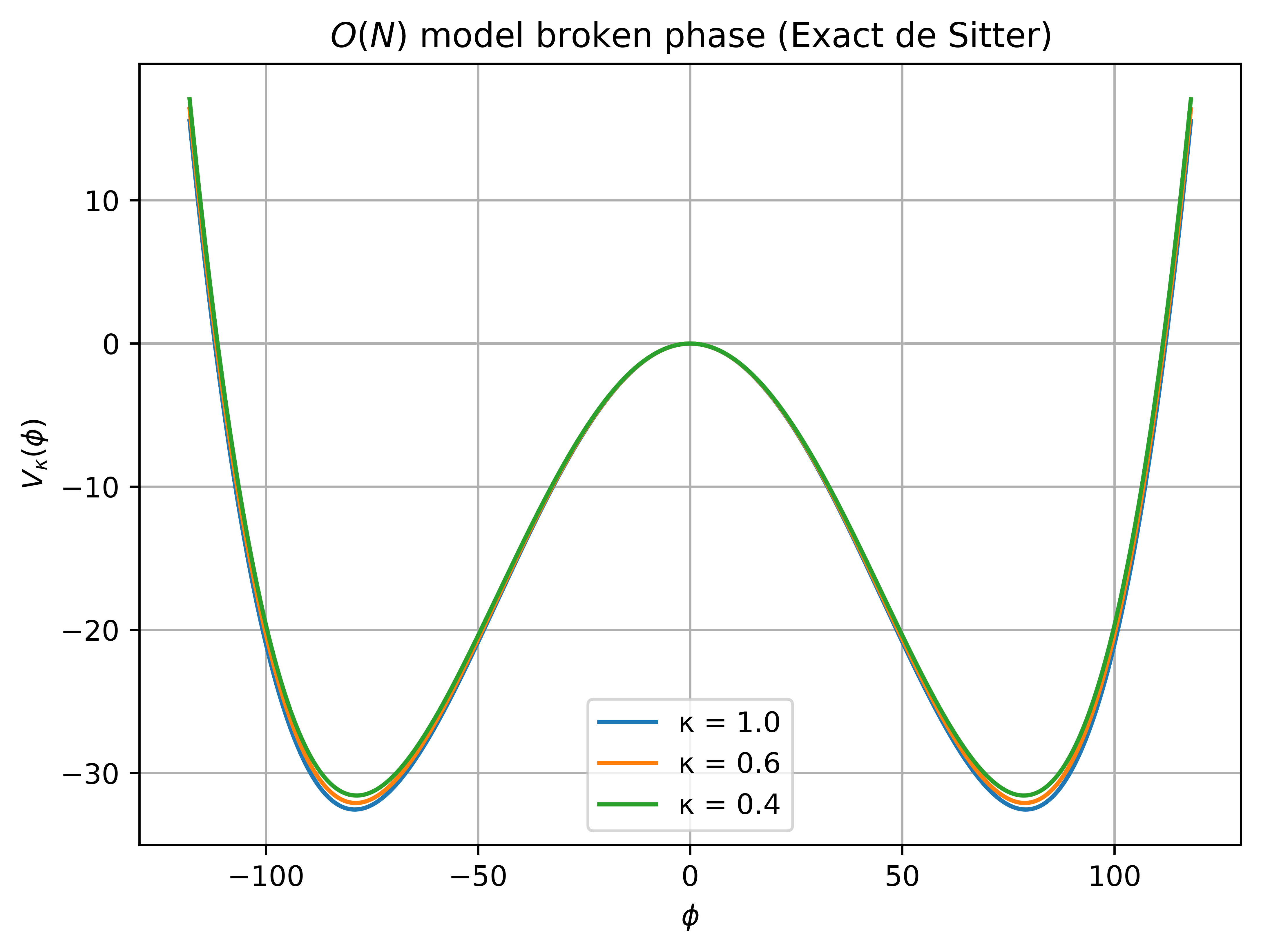}
\includegraphics[scale=0.4]{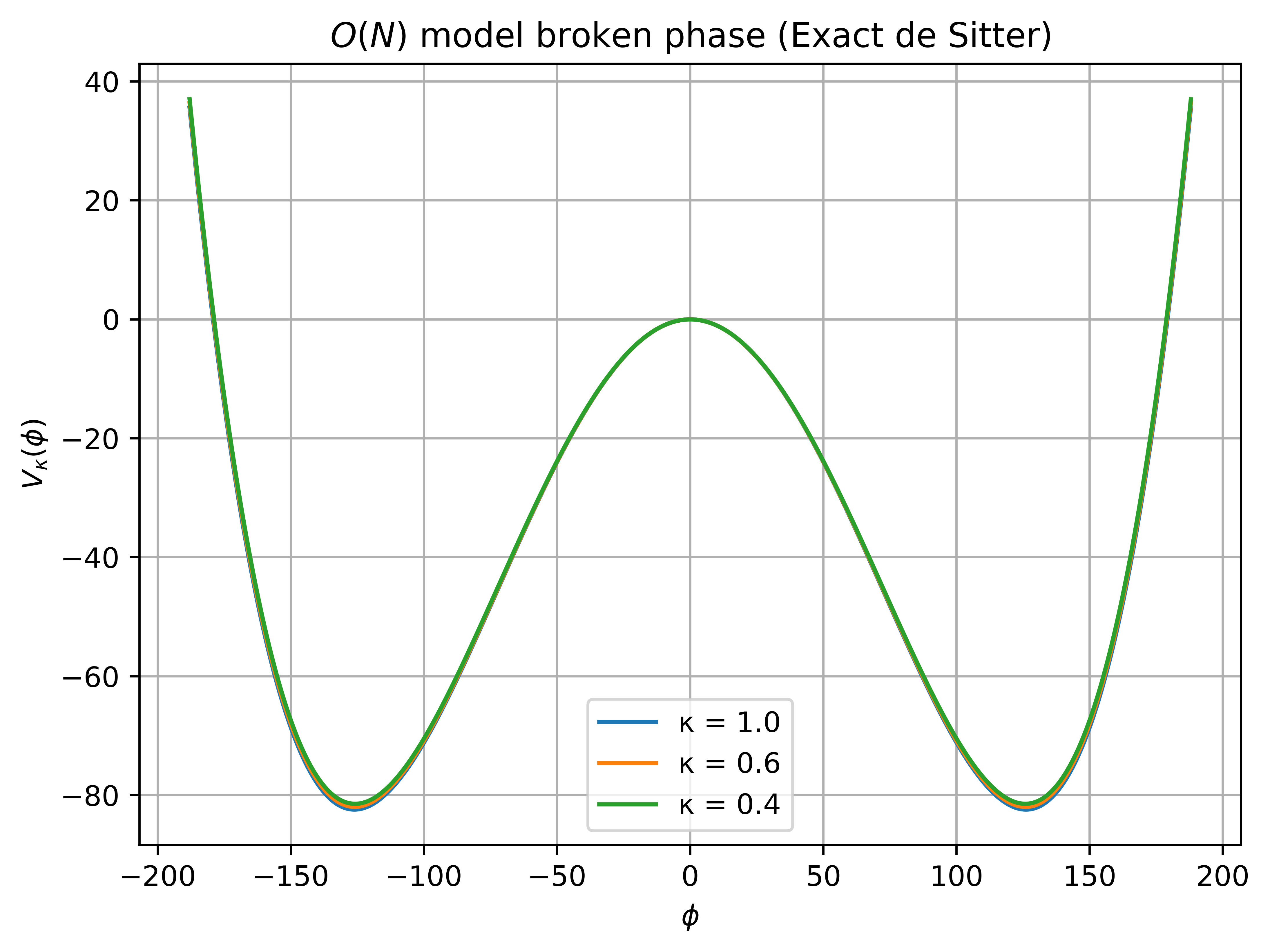}
\caption{\it \small IR flow~\ref{w30} of the $O(N)$ effective potential~\ref{v65} derived using the exact de Sitter propagator,  for $N=3,7,20,50$ respectively,  with numerical values borrowed from \ref{fig3}.  It shows similar qualitative behaviour as of \ref{fig3}.}
\label{fig4}
\end{center}
\end{figure}

Finally  for the $O(N)$ effective potentials~\ref{v61}, \ref{v65}, the IR flow  have been depicted in \ref{fig3}, \ref{fig4}. We note in both these plots that as  $N$ increases, the contribution from the $N-1$ transverse modes becomes dominant.  The flow approaches a nontrivial large-$N$ regime in which the radial mode is strongly suppressed while the collective transverse modes' fluctuations govern the dynamics. This shows significant qualitative large-$N$ modification of the phase structure compared to that of the small-$N\geq 1$ cases.

Having now obtained the IR flows of various one loop effective potentials in de Sitter, we now wish to apply them to determine the IR behaviour of the stochastic two point correlation function and the associated spectral index.

%%%%%%%%%%%%%
\section{IR-improved stochastic power spectrum in the de Sitter spacetime}\label{S4}
%%%%%%%%%%%%%

\noindent
The purpose of the  present section is to study how the IR effects modify the temporally coincident  stochastic  two point correlation function of a spectator scalar field in the inflationary de Sitter spacetime. Physically this might correspond to the evolution of an observable quantity  due to the expansion of the spacetime, which stretches the wavelength towards deep IR and hence effectively reduces the energy scale associated with the process.  We will use the stochastic spectral expansion method for our computation~\cite{Starobinsky:1994bd} (see also~\cite{Markkanen:2019kpv, Markkanen:2020bfc, Bhattacharya:2025eqf} and references therein).

 For each RG scale $\kappa$, we define the dimensionless potential,
\begin{eqnarray}
{\cal V}_\kappa(\phi)=\frac{4\pi^2}{3H^4}V_\kappa(\phi).
\label{w77}
\end{eqnarray}
The corresponding equilibrium one point probability distribution function reads
\begin{eqnarray}
\rho_{\rm eq}^{\kappa}(\phi)=N_\kappa e^{-2 {\cal V}_\kappa(\phi)},
\label{w78}
\end{eqnarray}
where $N_\kappa$ is a normalisation constant. The stochastic eigenvalue equation takes the form
\begin{eqnarray}
\left[-\frac12\frac{\partial^2}{\partial\phi^2}
+W_\kappa(\phi)\right]\psi_n^{(\kappa)}(\phi)=\frac{4\pi^2E_n^{(\kappa)}}{H^3}\psi_n^{(\kappa)}(\phi),
\label{w81}
\end{eqnarray}
where
\begin{eqnarray}
W_\kappa(\phi) =\frac12\left[\left({\cal V}_\kappa'(\phi)\right)^2
-{\cal V}_\kappa''(\phi)\right],
\label{w79}
\end{eqnarray}
and $E_n^{(\kappa)}$ and $\psi_n^{(\kappa)}(\phi)$ respectively denote the stochastic eigenvalues and eigenfunctions. The eigenfunctions are orthonormal and complete
\begin{eqnarray}
\int d\phi\,\psi_n^{(\kappa)}(\phi)\psi_m^{(\kappa)}(\phi)
=\delta_{nm} \quad {\rm and} \quad 
\sum_n\psi_n^{(\kappa)}(\phi)\psi_n^{(\kappa)}(\phi')=
\delta(\phi-\phi').
\label{w83}
\end{eqnarray}
The probability distribution function admits the spectral expansion
\begin{eqnarray}
\rho_\kappa(\phi,t)=e^{-{\cal V}_\kappa(\phi)}\sum_n a_n^{(\kappa)}
\psi_n^{(\kappa)}(\phi)e^{-E_n^{(\kappa)}t},
\label{w84}
\end{eqnarray}
where the coefficients  $a_n^{(\kappa)}$ are determined by the initial conditions. In order that a late time equilibrium state exists, we must have for the lowest eigenvalue
\begin{eqnarray}
E_0^{(\kappa)}=0,
\label{w85}
\end{eqnarray}
with corresponding eigenfunction
\begin{eqnarray}
\psi_0^{(\kappa)}(\phi)\sim  e^{-{\cal V}_\kappa(\phi)}.
\label{w86}
\end{eqnarray}
Probability distributions at two different times are related through an evolution kernel
\begin{eqnarray}
\rho_\kappa(\phi,t+\Delta t)=\int d\phi_0\,U_\kappa(\Delta t;\phi,\phi_0)\rho_\kappa(\phi_0,t),
\label{w87}
\end{eqnarray}
with the initial condition
\begin{eqnarray}
U_\kappa(0;\phi,\phi_0)=\delta(\phi-\phi_0).
\label{w88}
\end{eqnarray}
Using the completeness relation for the eigenfunctions, the kernel can be expanded as
\begin{eqnarray}
U_\kappa(t;\phi,\phi_0)=\sum_n e^{-E_n^{(\kappa)}t}\psi_n^{(\kappa)}(\phi)\psi_n^{(\kappa)}(\phi_0).
\label{w89}
\end{eqnarray}
For a general observable $f(\phi)$, the two-point correlation function becomes
\begin{eqnarray}
\xi_f^{(2,\kappa)}(t)=\sum_n\left(f_n^{(\kappa)}\right)^2
e^{-E_n^{(\kappa)}t},
\label{w90}
\end{eqnarray}
where the spectral coefficients are
\begin{eqnarray}
f_n^{(\kappa)}=\int d\phi\,\psi_0^{(\kappa)}(\phi)f(\phi)
\psi_n^{(\kappa)}(\phi).
\label{w91}
\end{eqnarray}

%%%%%%%%%%%%%%%%
\subsection{RG-Improved Power Spectrum}\label{S5}
%%%%%%%%%%%%%%%%

\noindent
For the scalar-field observable $f(\phi)=\phi(t,\vec{x})\phi(t,\vec{x}')$, the equal-time spatial correlation function reads
\begin{eqnarray}
\xi_\kappa^{(2)}(\mathbf{x},\mathbf{x}')=\sum_n \left(
f_n^{(\kappa)}\right)^2(a(t) rH)^{-2E_n^{(\kappa)}/H},
\label{w92}
\end{eqnarray}
The power spectrum is defined as the Fourier transform of the two-point correlation function,
\begin{eqnarray}
P_\kappa(k)=\frac{k^3}{2\pi^2}\int d^3\vec{r}\,e^{i\mathbf{k}\cdot \vec{r}}\xi_\kappa^{(2)}(r),
\label{w94}
\end{eqnarray}
where we have abbreviated, $\vec{r}=\vec{x}-\vec{x}'$. Substituting the spectral representation of the correlation function yields 
\begin{eqnarray}
P_\kappa(k)=\frac{2}{\pi}\sum_n\left(f_n^{(\kappa)}
\right)^2\Gamma\left(2-\frac{2E_n^{(\kappa)}}{H}\right)
\sin\left(\frac{\pi E_n^{(\kappa)}}{H}\right)\left(\frac{k}{Ha}
\right)^{2E_n^{(\kappa)}/H}.
\label{w95}
\end{eqnarray}
Taking  $k/H\ll 1$, the power spectrum can be shown to behave at the leading order
\begin{eqnarray}
P_\kappa(k)\simeq \left(f_n^{(\kappa)}\right)^2\left(n_s^{(\kappa)}-1\right) \left(\frac{k}{Ha}
\right)^{n_s^{(\kappa)}-1},
\label{w96}
\end{eqnarray}
which allows the RG-dependent spectral index to be identified as
\begin{eqnarray}
n_s^{(\kappa)}=1+\frac{2E_1^{(\kappa)}}{H}+\cdots.
\label{w97}
\end{eqnarray}
The spectrum will be evaluated at the customary pivot scale $k=0.05\,{\rm Mpc}^{-1}$~\cite{Markkanen:2019kpv}.
For each RG scale \(\kappa\), the flowed potential \(V_\kappa(\phi)\) is obtained from the IR evolution described in the preceding section. The stochastic potential \(W_\kappa(\phi)\) of \ref{w79} is then also constructed.  The spectral coefficients are then evaluated by solving the stochastic eigenvalue equation, and finally the IR-improved power spectrum and the spectral index are determined. We will do this numerically by using a discretised field-space lattice system. For example using \ref{fig2}, the IR evolution of the first few eigenvalues corresponding to the $\phi^4$-Yukawa effective potential derived using the exact de Sitter propagator has been listed in \ref{T1}.
\begin{table}[ht]
\centering
\begin{tabular}{cccc}
\hline
$\kappa$ & $E_1$ & $E_2$ & $E_3$ \\
\hline
1.0000 & 0.00315695 & 0.00559442 & 0.01863609 \\
0.9190 & 0.00305927 & 0.00547275 & 0.01859288 \\
0.8278 & 0.00291041 & 0.00528916 & 0.01853048 \\
0.7367 & 0.00269774 & 0.00503075 & 0.01844874 \\
0.6456 & 0.00238026 & 0.00465376 & 0.01834397 \\
0.5646 & 0.00195236 & 0.00416381 & 0.01823994 \\
0.4734 & 0.00121816 & 0.00338743 & 0.01820160 \\
0.3823 & 0.00032525 & 0.00272636 & 0.01881890 \\
0.2911 & 0.00000008 & 0.00439880 & 0.02366539 \\
0.2000 & 0.00000065 & 0.00874488 & 0.03299571 \\
\hline
\end{tabular}
\caption{\small \it RG flow of the lowest stochastic eigenvalues $E_n$ obtained from the RG-improved effective potential. The eigenvalues remain nearly constant throughout most of the flow and exhibit a noticeable increase in the deep infrared regime. We have taken the same numerical values as of \ref{fig1},  $\lambda=10^{-5},\  \xi = -0.0134, \bar{m}_0=0.4$, $g=2\times 10^{-3}$ and $\bar{M_0} = 0.5 $.}
\label{T1}
\end{table}

\noindent
Let us now consider the  IR evolutions corresponding to the one loop effective potential derived using  both the Schwinger-DeWitt and exact de Sitter propagators, \ref{fig1}, \ref{fig2}, \ref{fig3} and \ref{fig4}.
\begin{figure}[H]
\begin{center}
\includegraphics[scale=0.34]{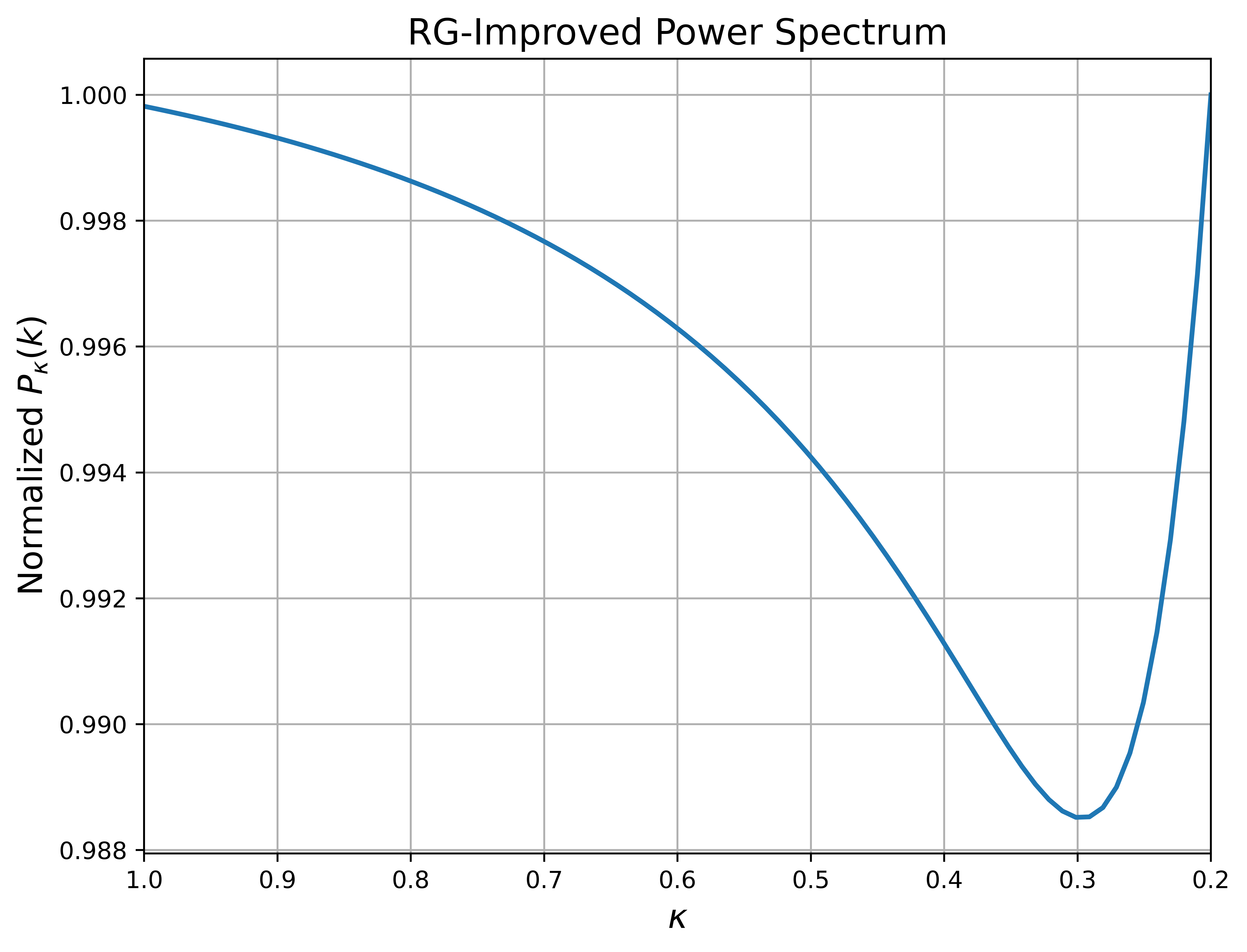}
\includegraphics[scale=0.34]{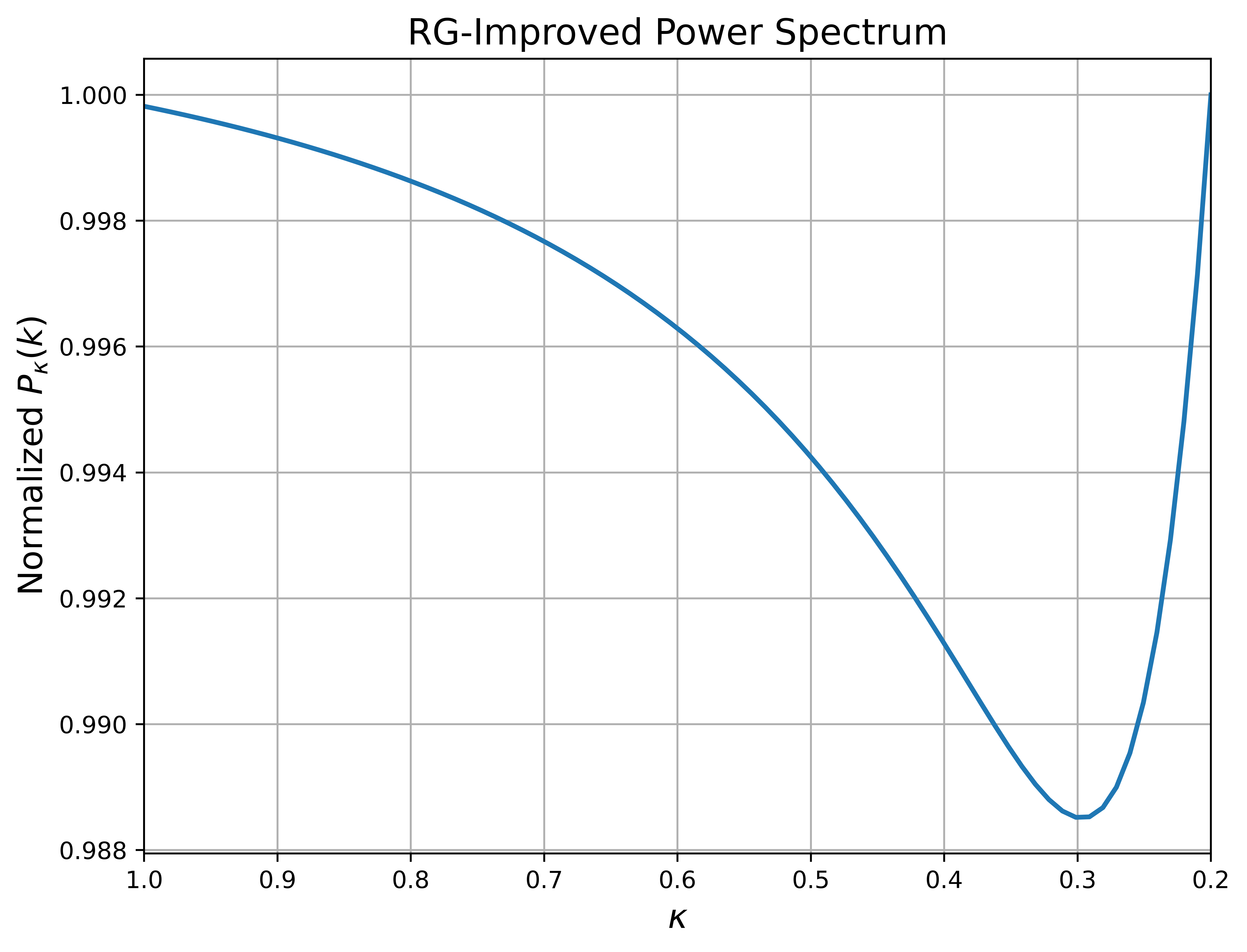}
\includegraphics[scale=0.34]{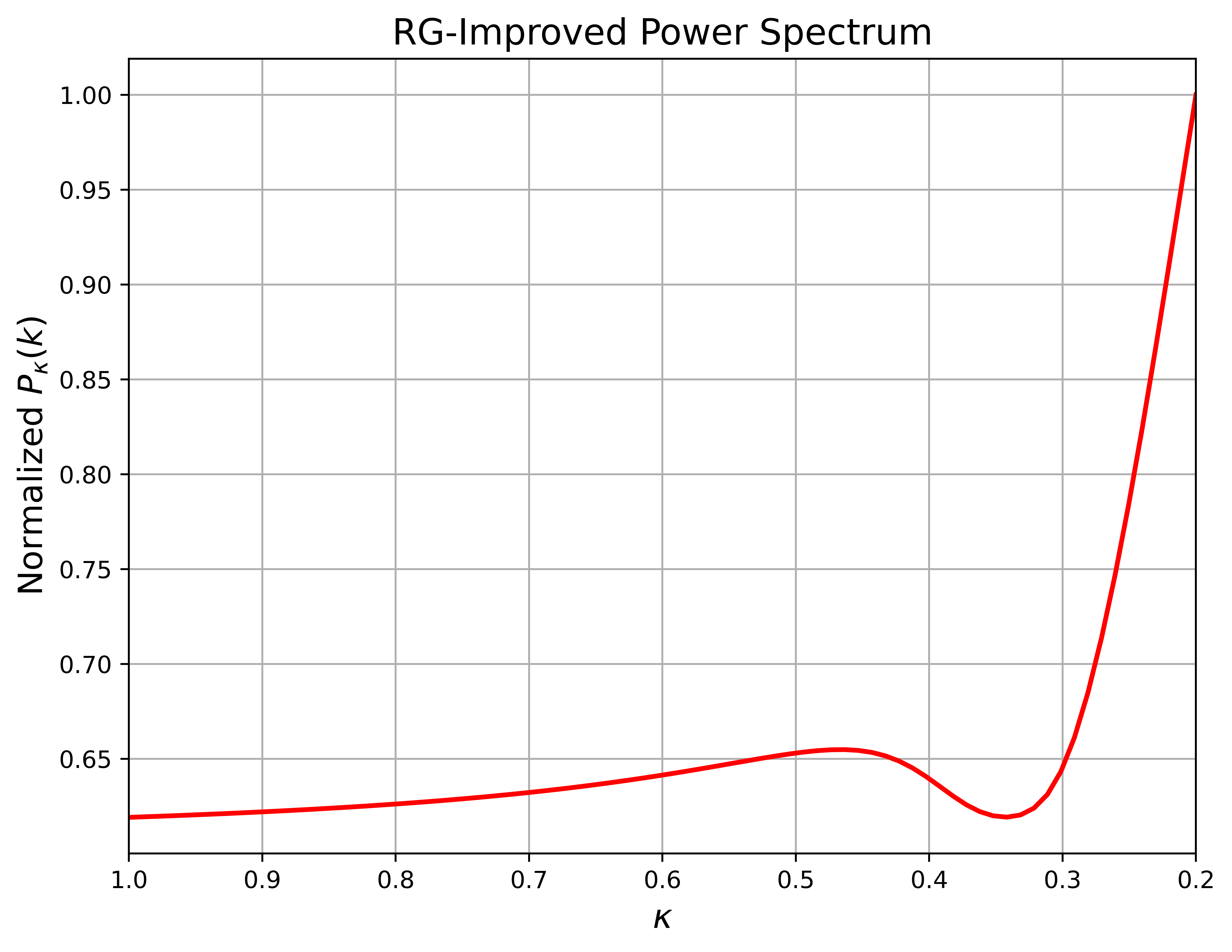}
\includegraphics[scale=0.34]{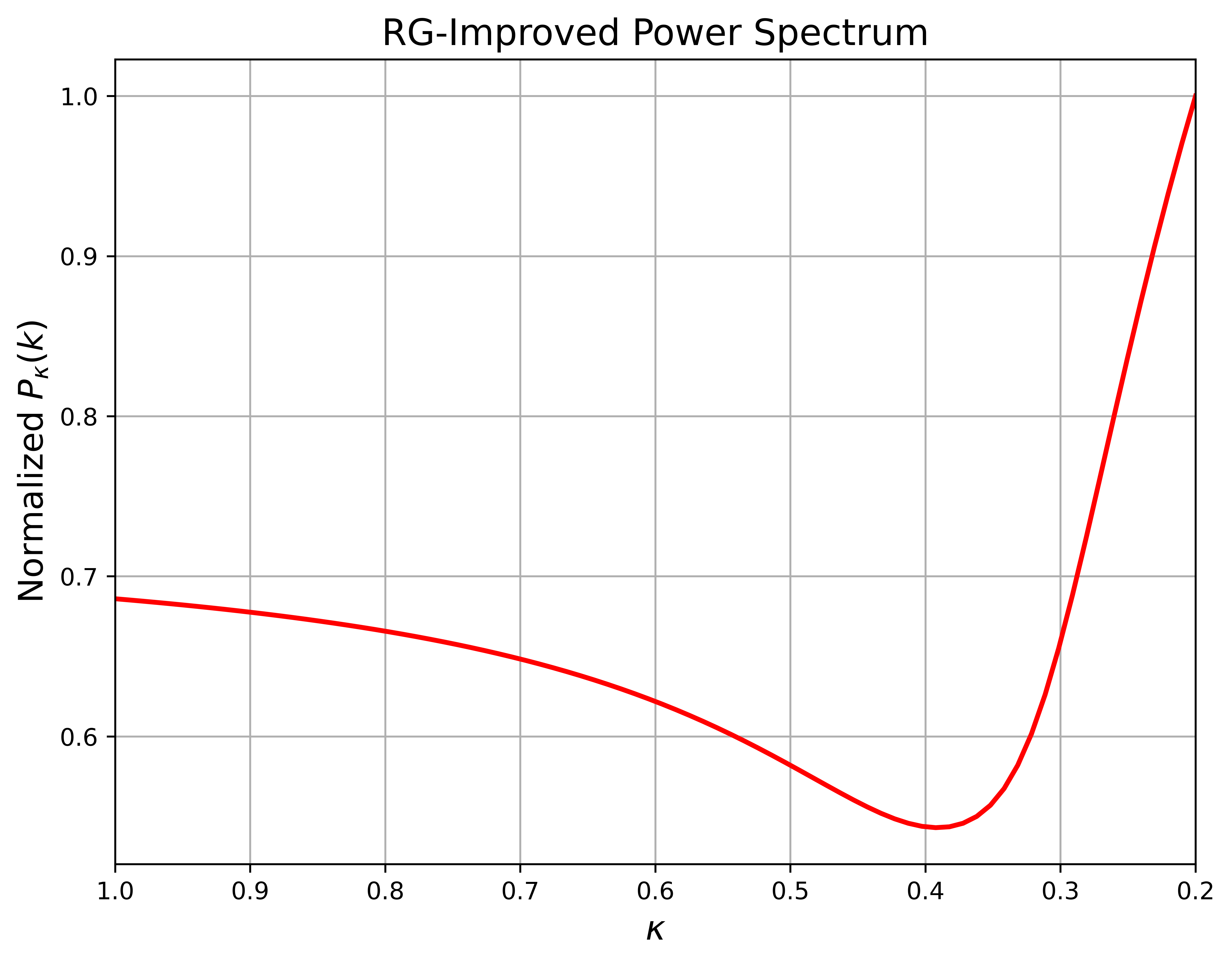}
\caption{  \it \small IR flow of the power spectra corresponding to the $\phi^4$ (left) and $\phi^4$-Yukawa (right) effective potentials, for both Schwinger-DeWitt  (top row) and exact de Sitter cases (bottom row). The  numerical values have been borrowed from~\ref{fig1}.  In both cases the power spectrum decreases to reach a minimum and then it increases monotonically as we move towards IR.}
\label{fig5}
\end{center}
\end{figure}
\begin{figure}[H]
\begin{center}
\includegraphics[scale=0.25]{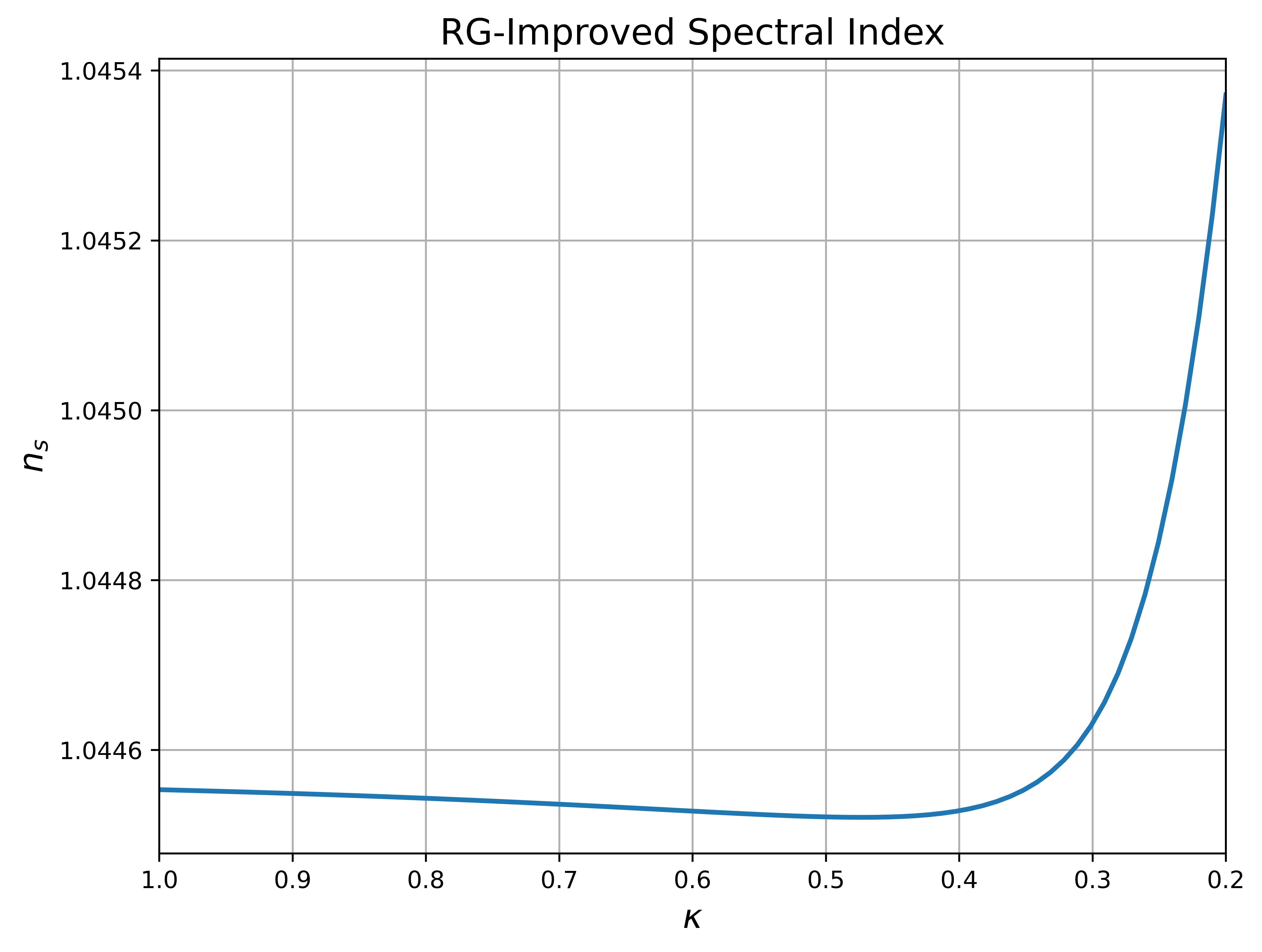}
\includegraphics[scale=0.25]{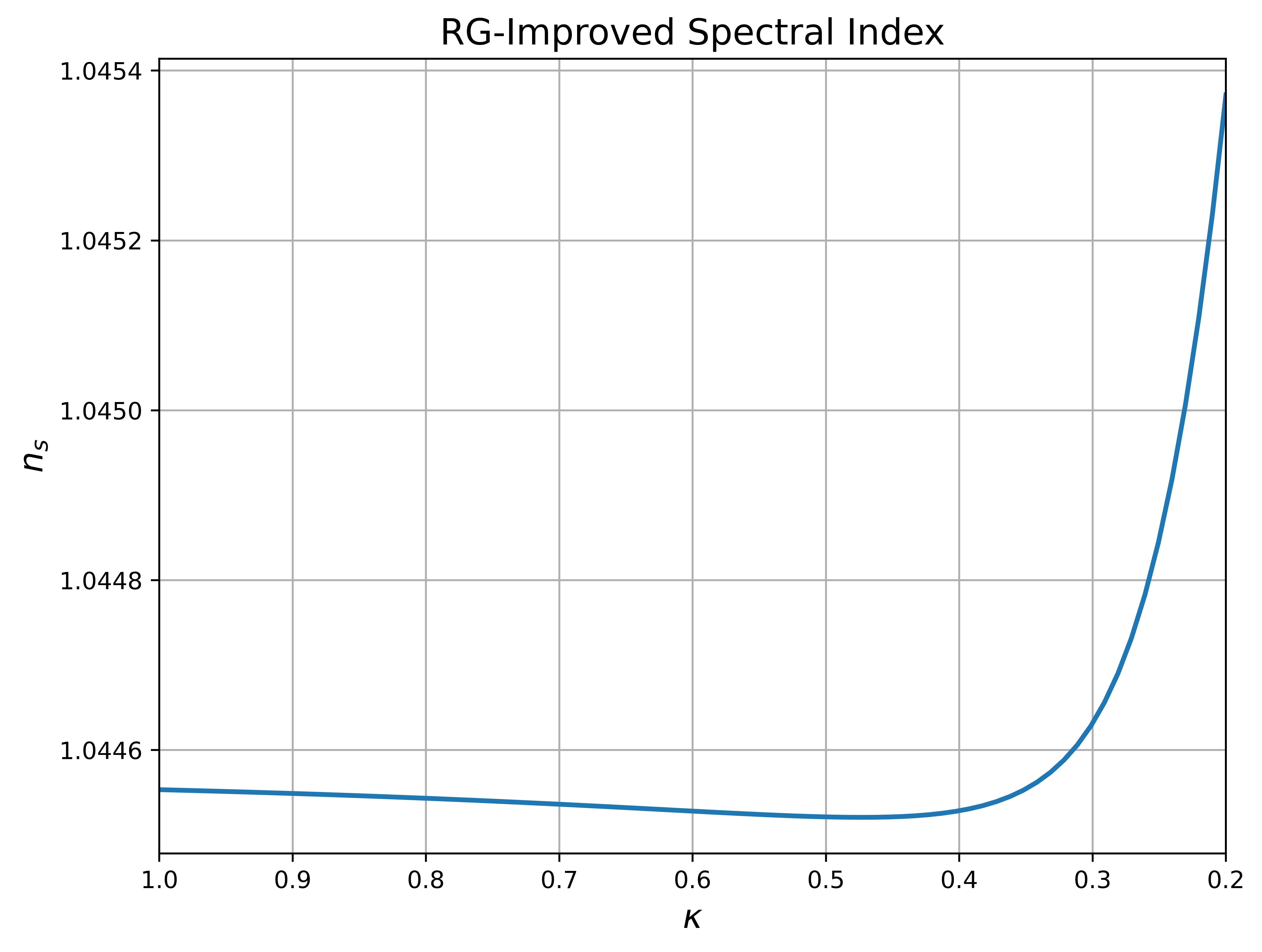}\nonumber\\
\includegraphics[scale=0.25]{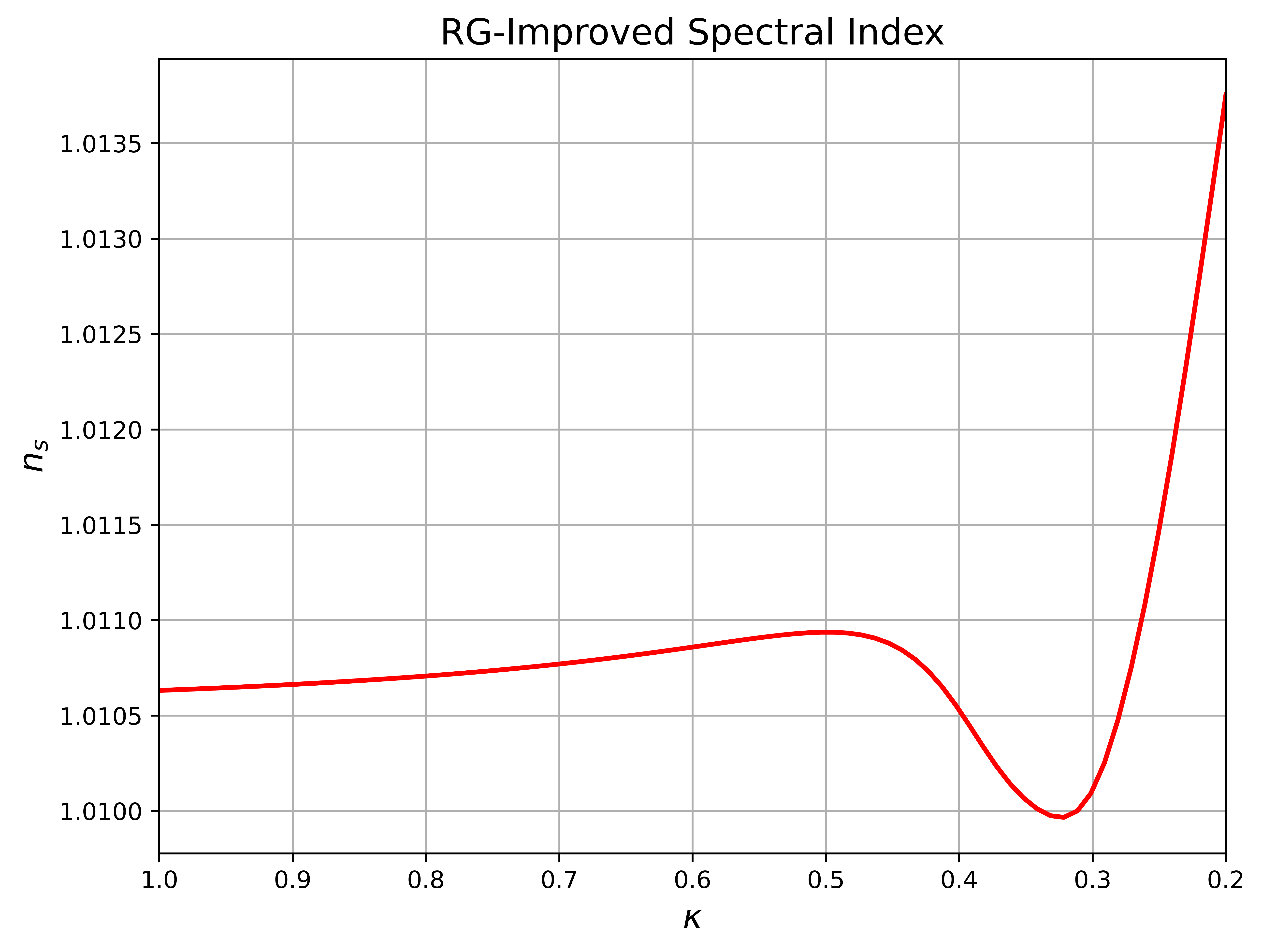}
\includegraphics[scale=0.25]{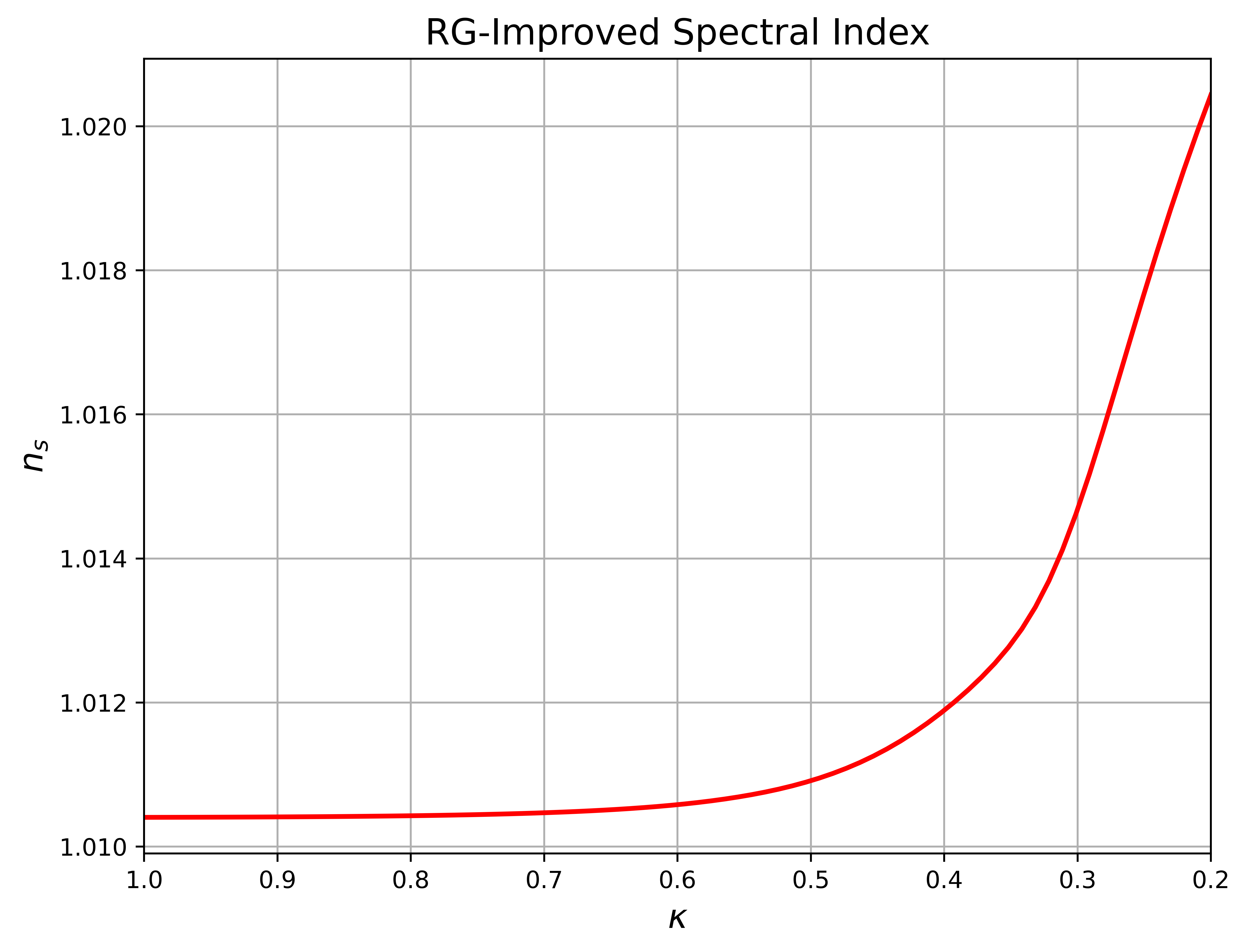}
\caption{  \it \small IR flow of the  spectral indices corresponding to \ref{fig5} corresponding to the $\phi^4$ (left) and $\phi^4$-Yukawa (right) effective potentials, for both Schwinger-DeWitt  (top row) and exact de Sitter cases (bottom row). }
\label{fig6}
\end{center}
\end{figure}
\ref{fig5} and \ref{fig6} depict the IR flow of the power spectra and spectral indices for the $\phi^4$ and $\phi^4$-Yukawa theory. Likewise  \ref{fig7}, \ref{fig8} depict the same for the $O(N)$ model.
\begin{figure}[H]
\begin{center}
\includegraphics[scale=0.23]{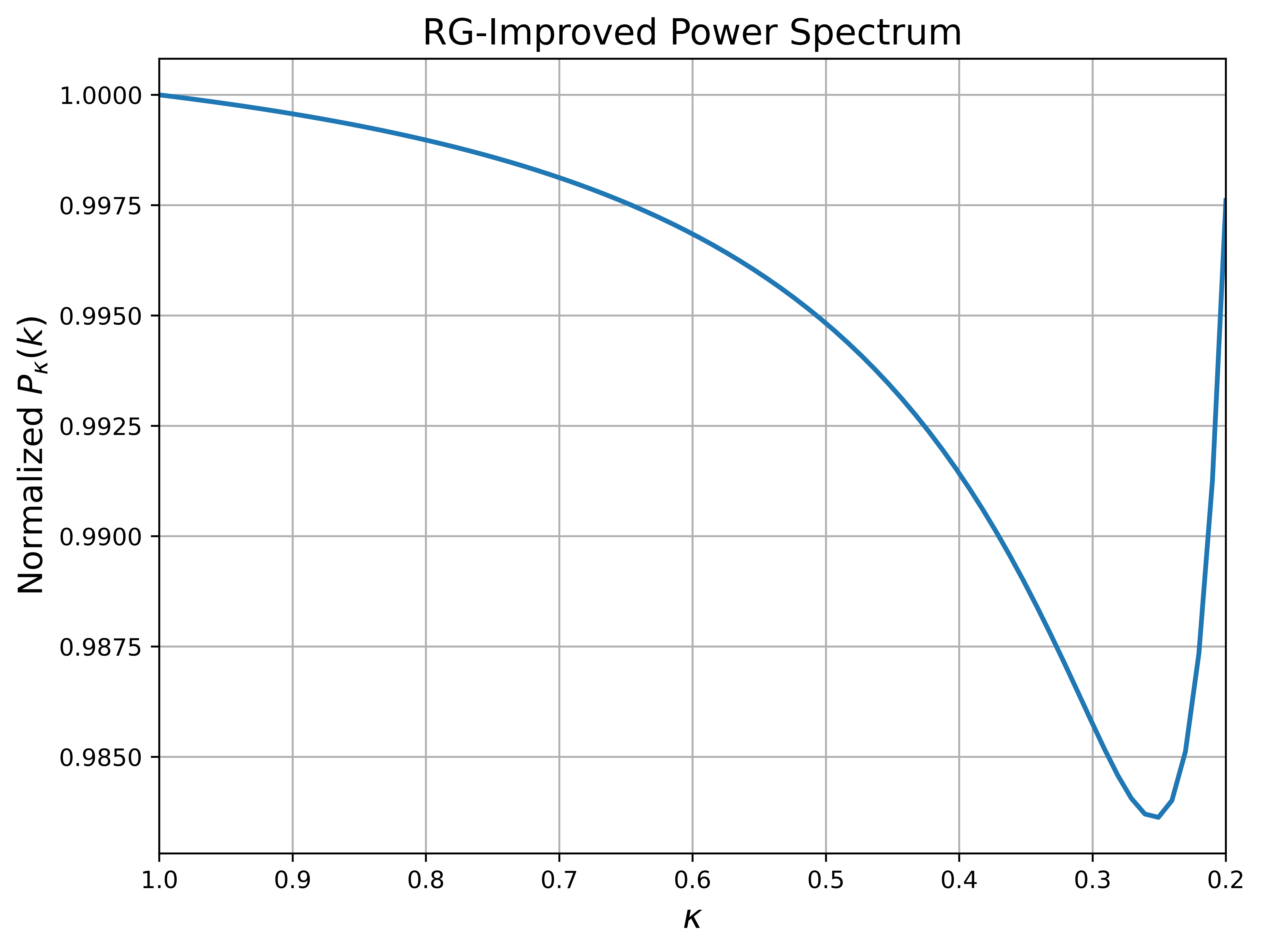}
\includegraphics[scale=0.23]{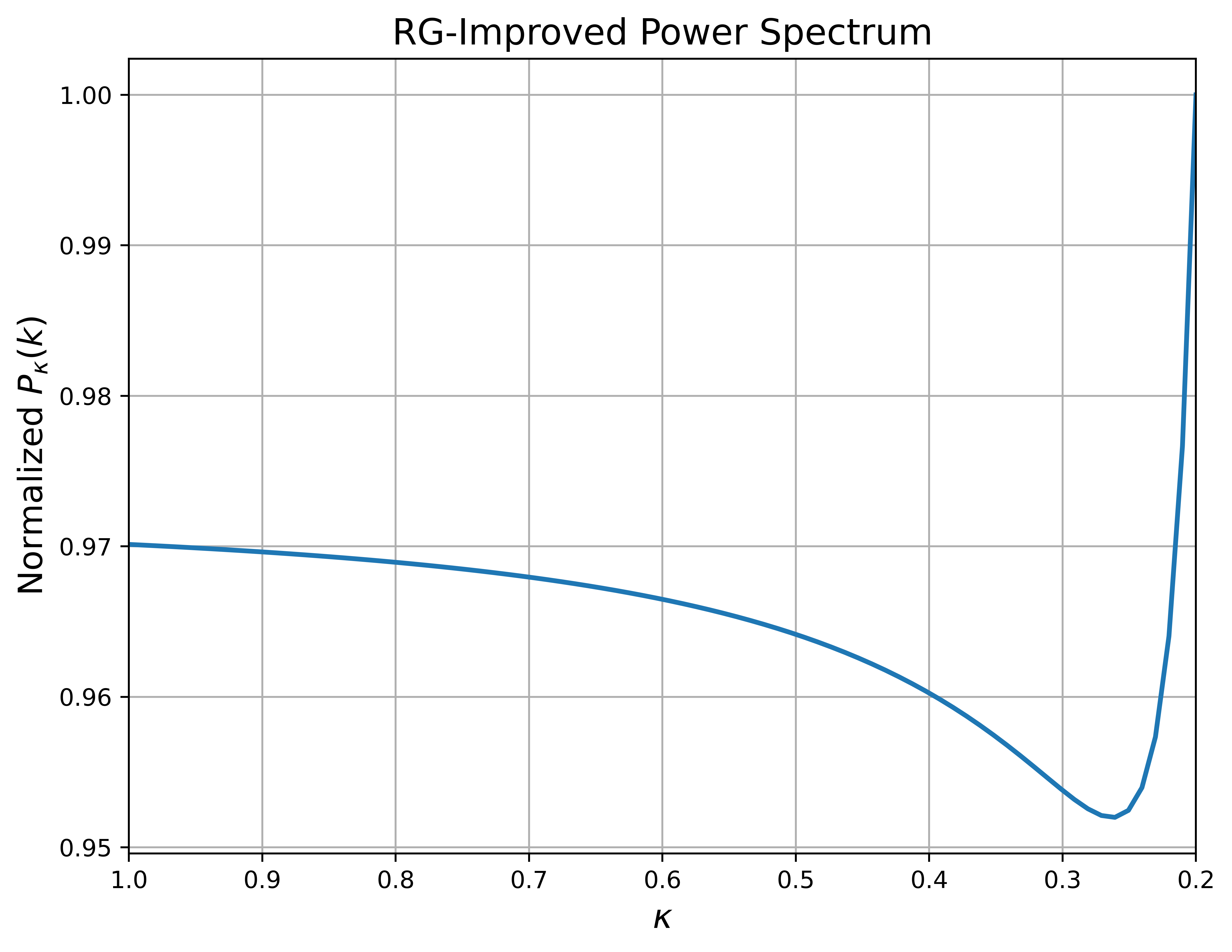}
\includegraphics[scale=0.23]{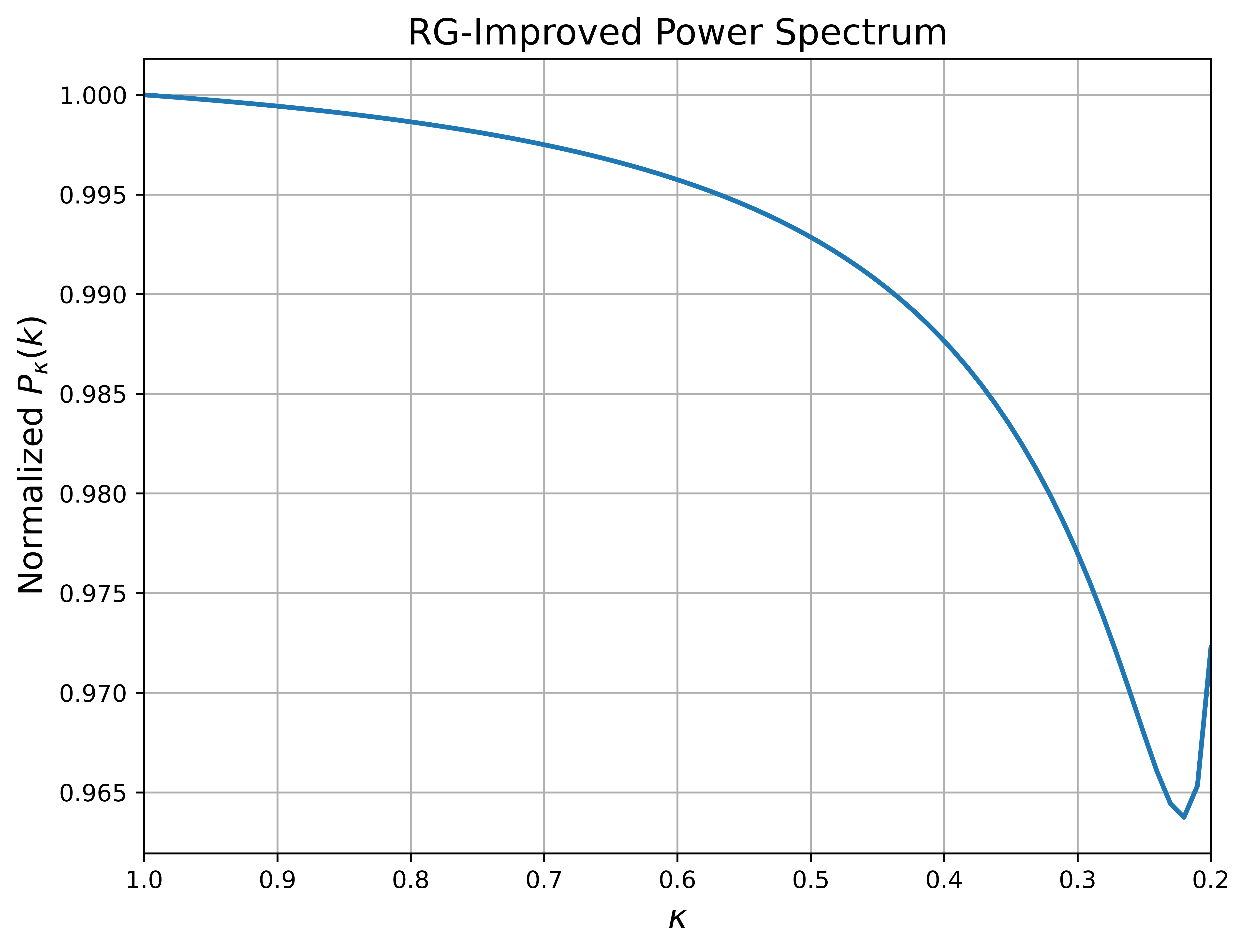}
\includegraphics[scale=0.23]{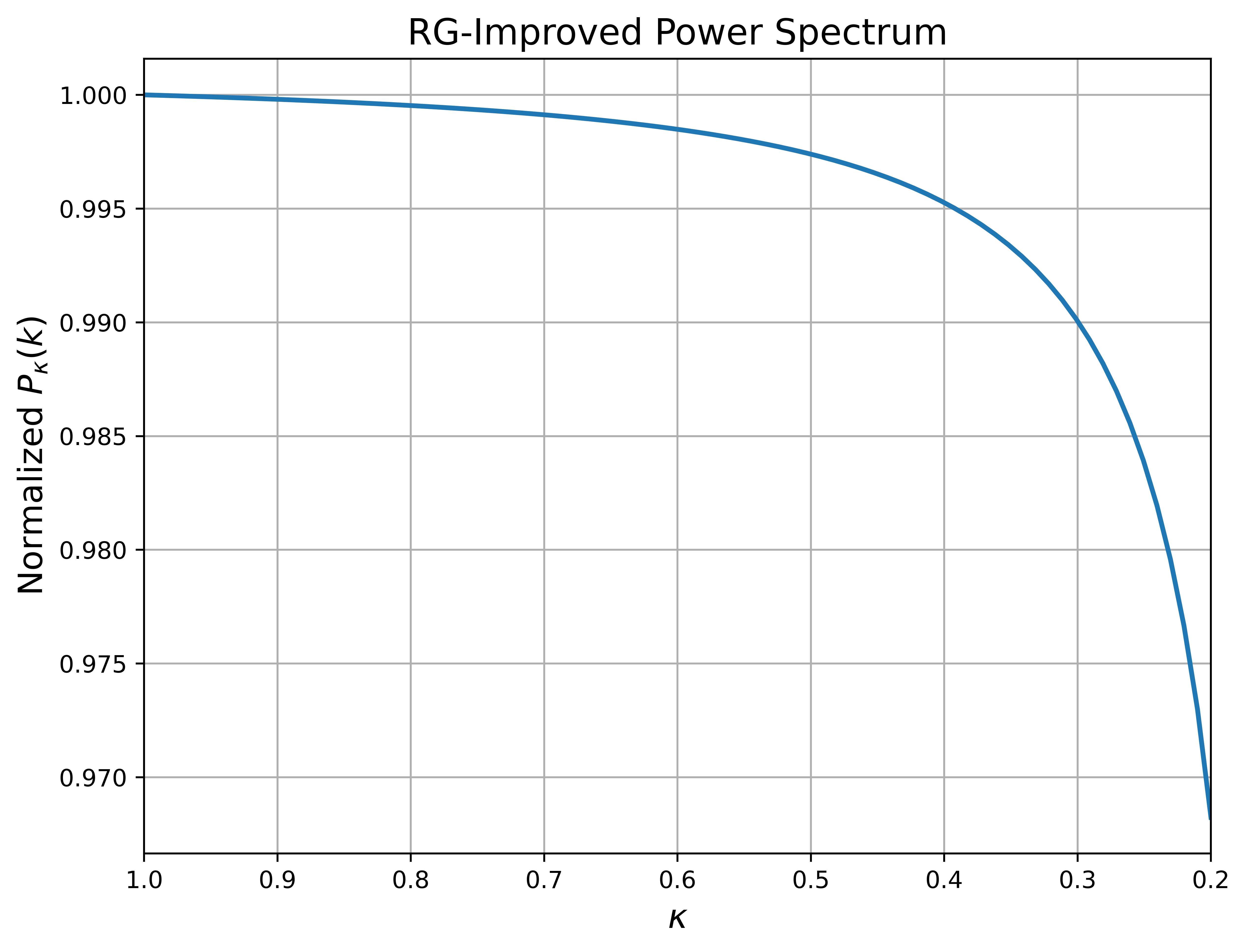}
\includegraphics[scale=0.23]{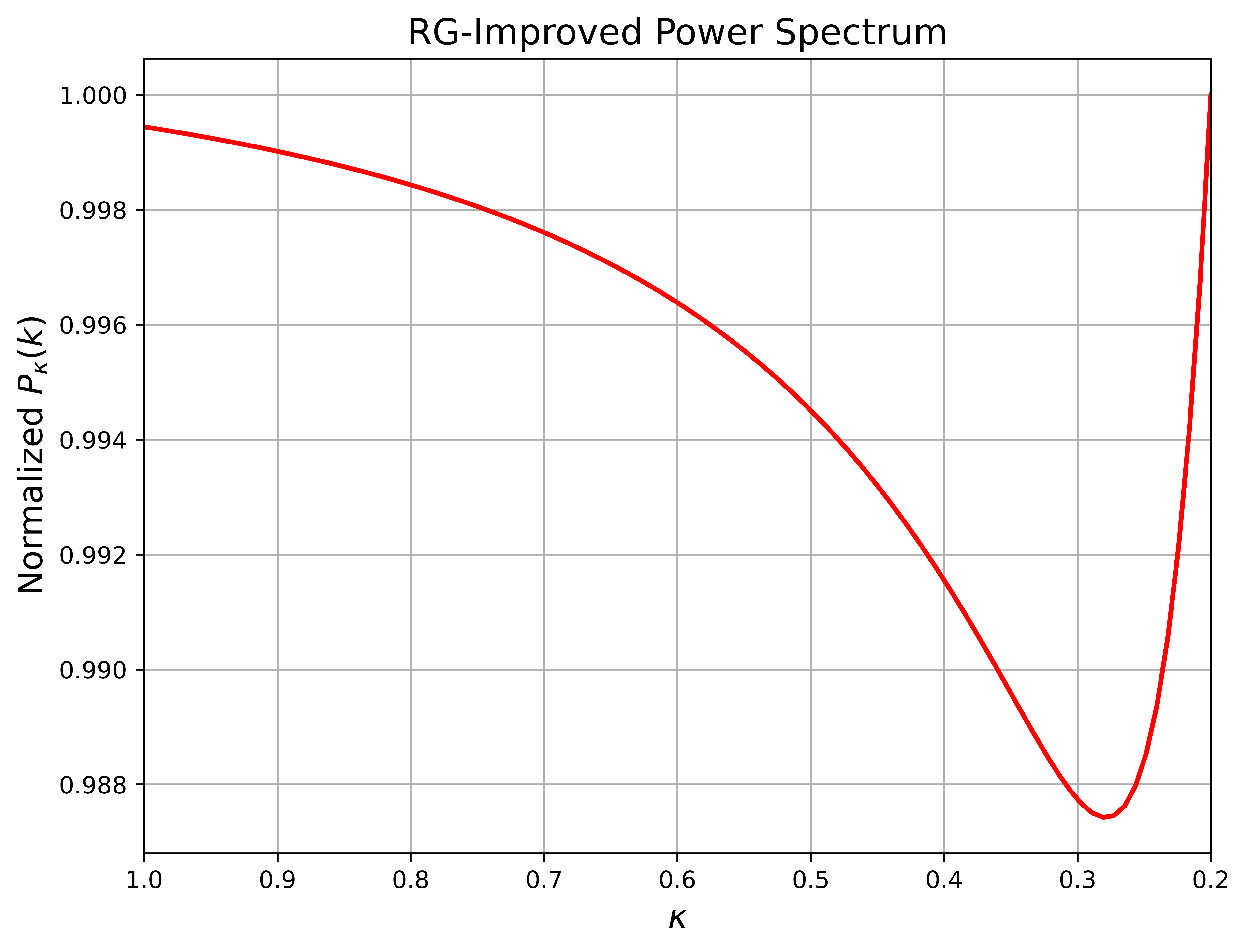}
\includegraphics[scale=0.23]{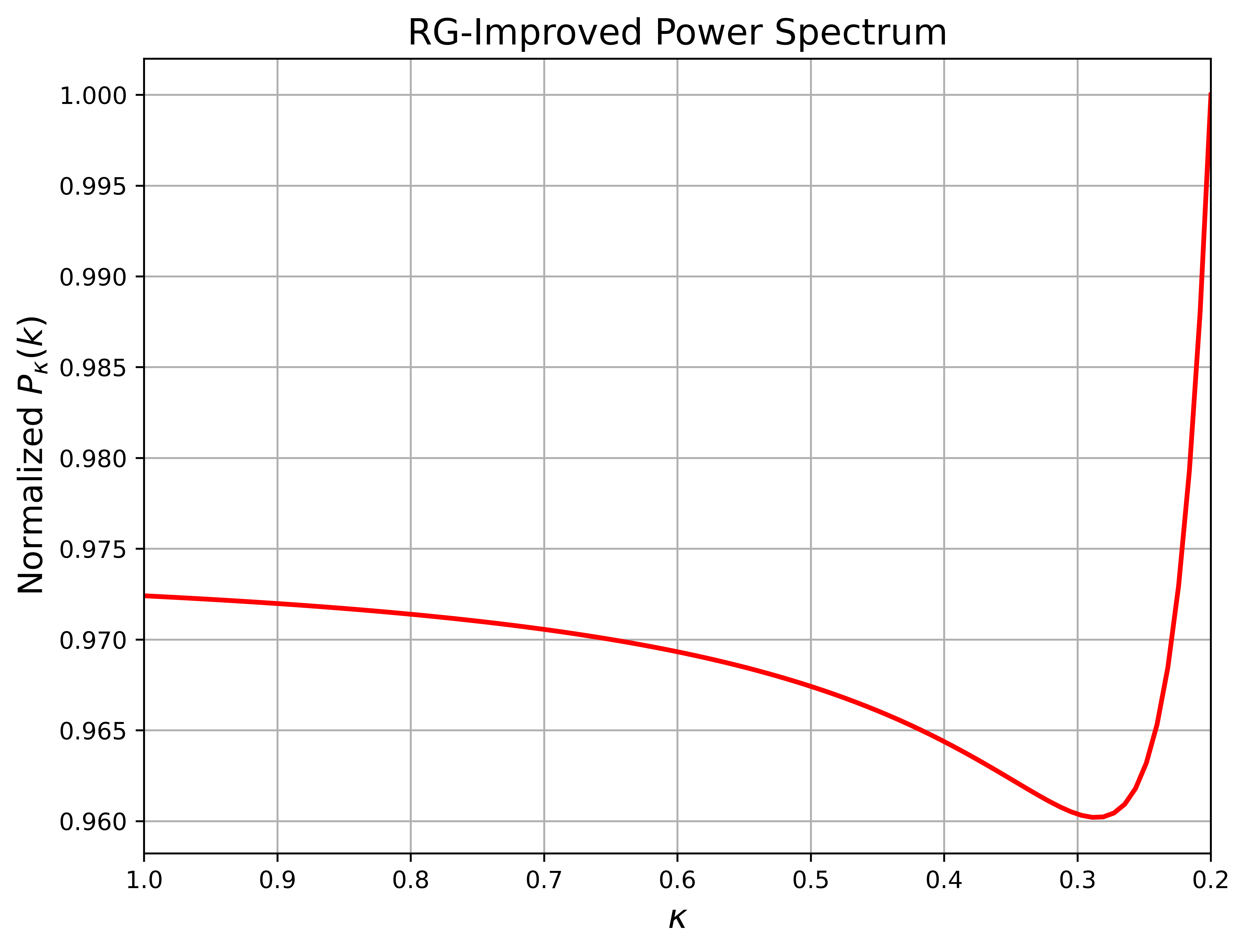}
\includegraphics[scale=0.23]{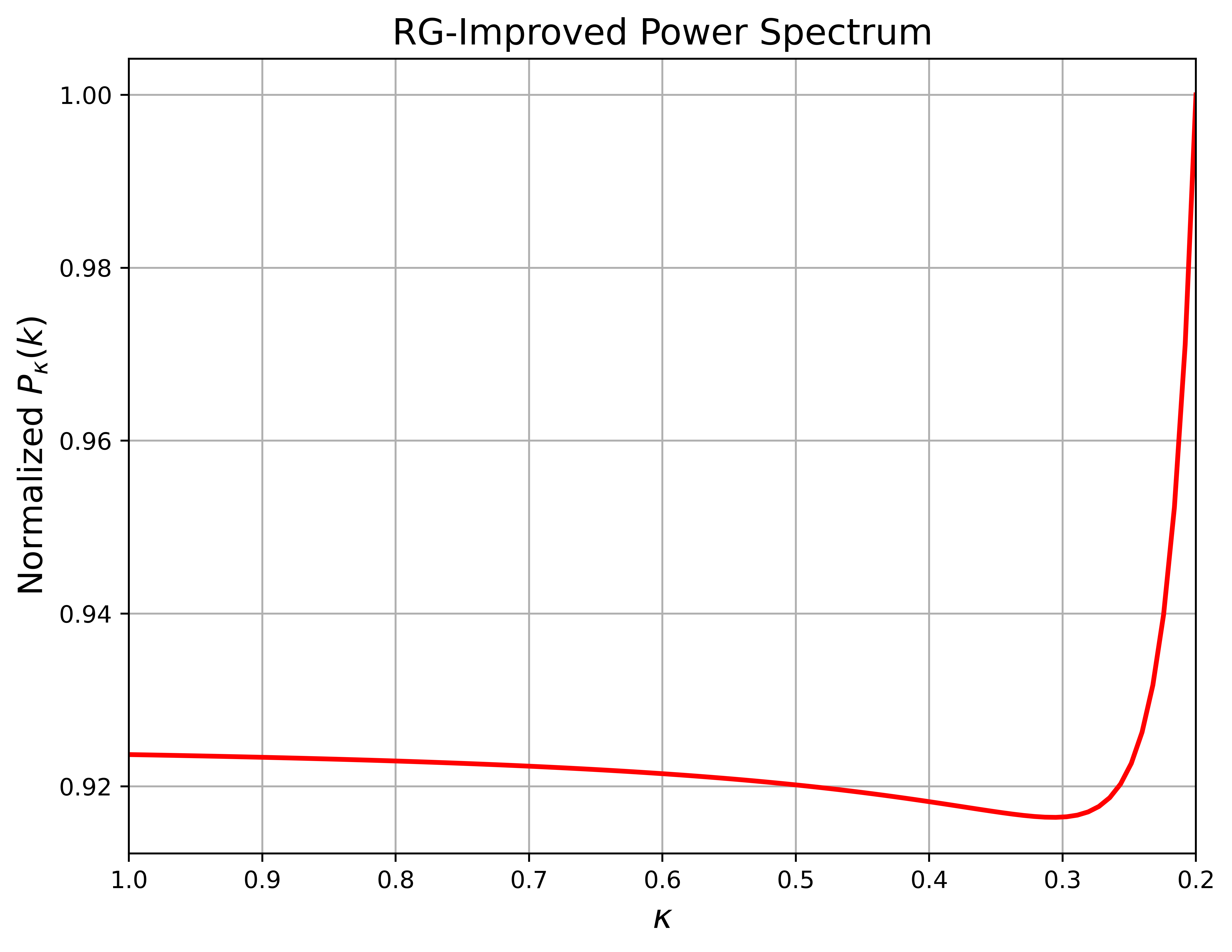}
\includegraphics[scale=0.23]{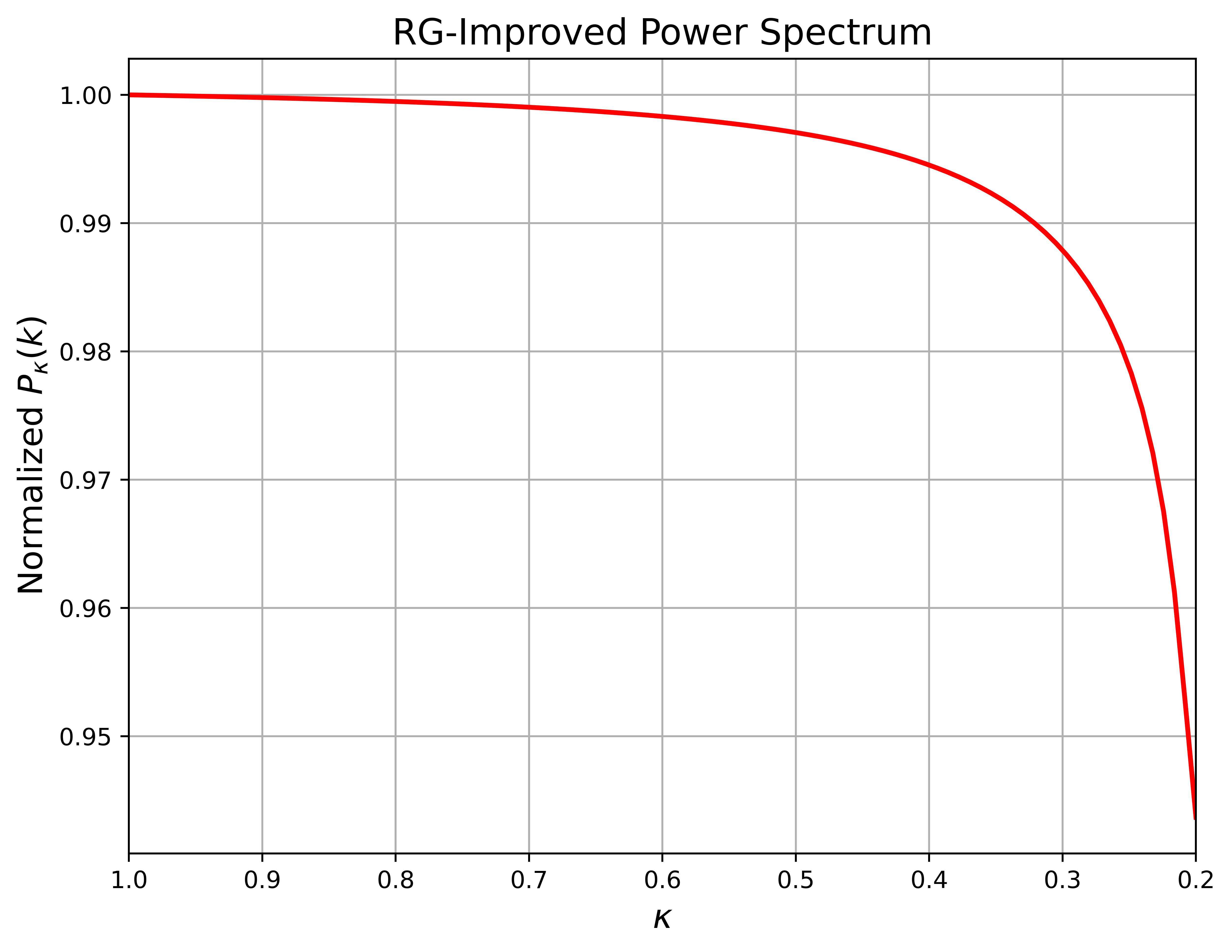}
\caption{  \it \small IR flow of the power spectra corresponding to  $O(N)$ effective potentials, for both Schwinger-DeWitt  (top row) and exact de Sitter cases (bottom row). The  numerical values have been borrowed from~\ref{fig3}, with  $N=3,7,20,50 $ respectively.  with the same numerical values as of~\ref{fig3} for Schwinger-DeWitt case.  Note the complete change for large $N$. See main text for discussion.  }
\label{fig7}
\end{center}
\end{figure}
\begin{figure}[H]
\begin{center}
\includegraphics[scale=0.23]{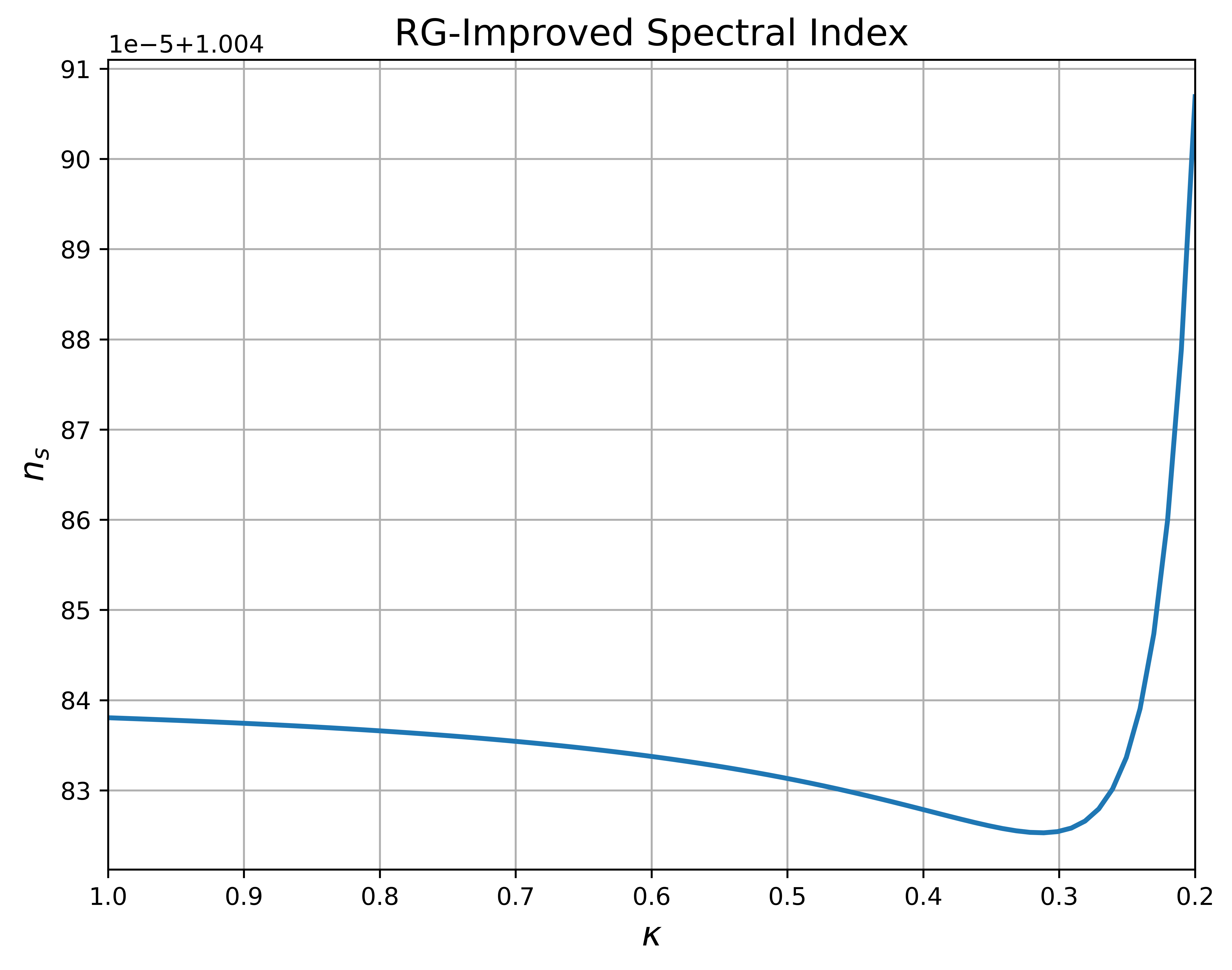}
\includegraphics[scale=0.23]{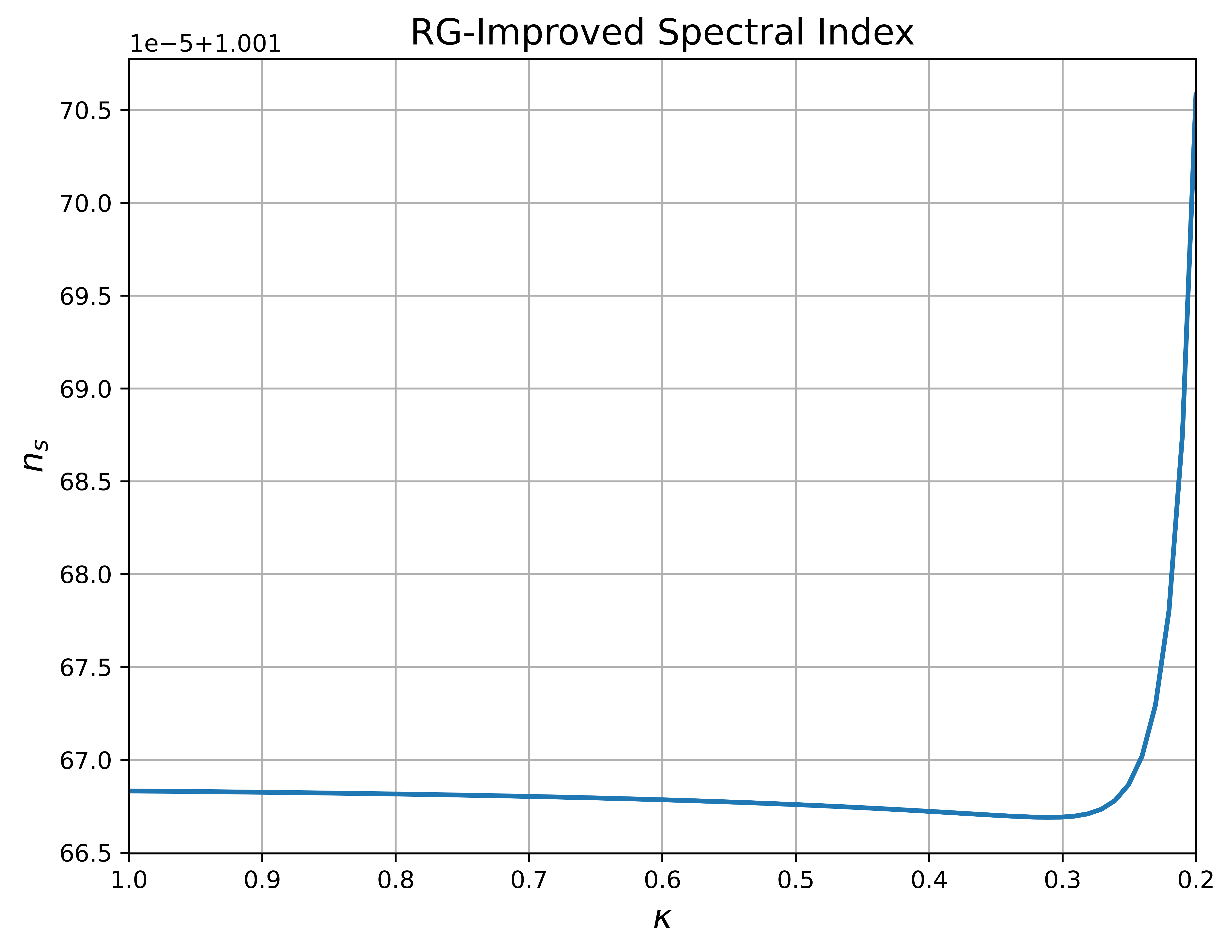}
\includegraphics[scale=0.23]{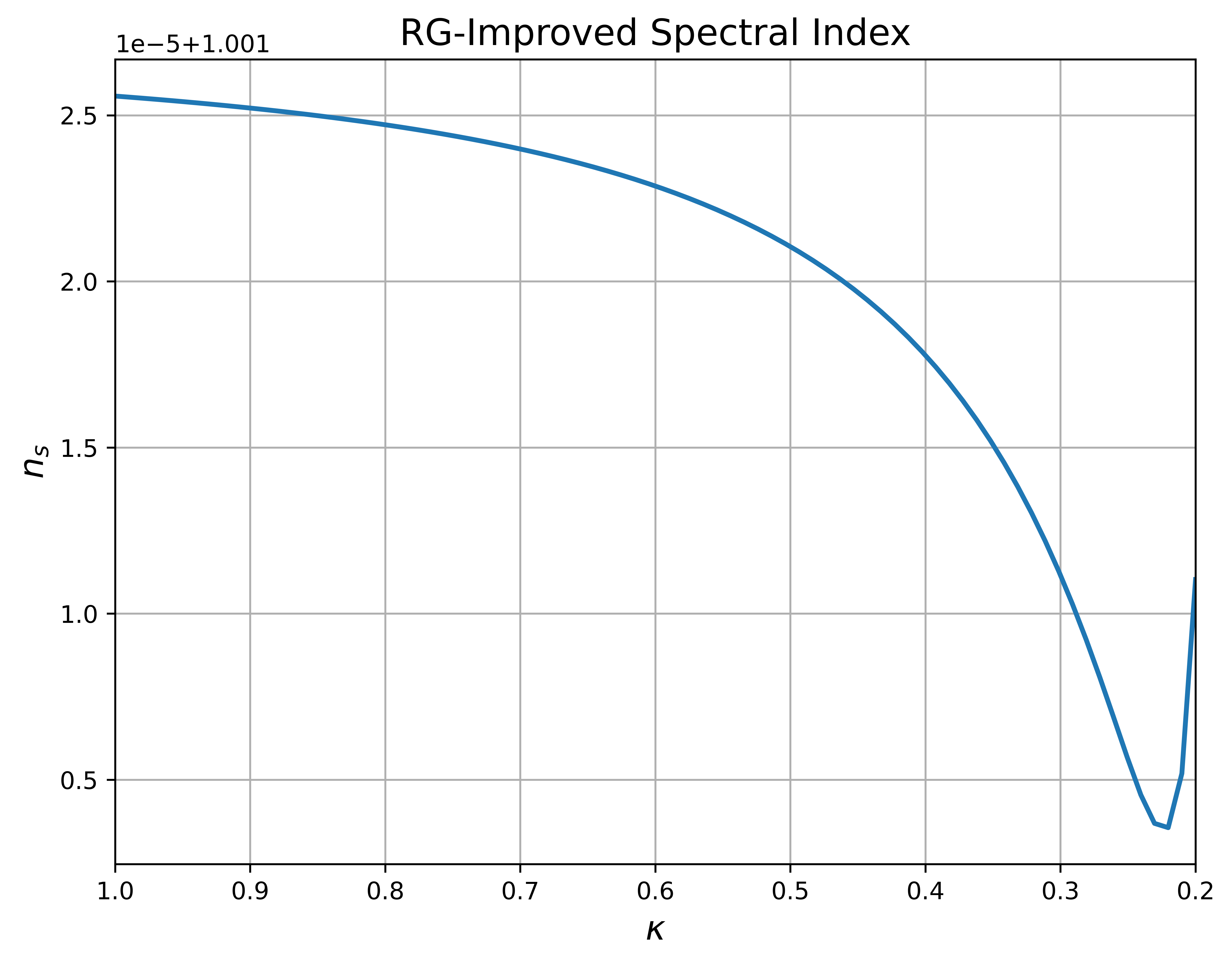}
\includegraphics[scale=0.23]{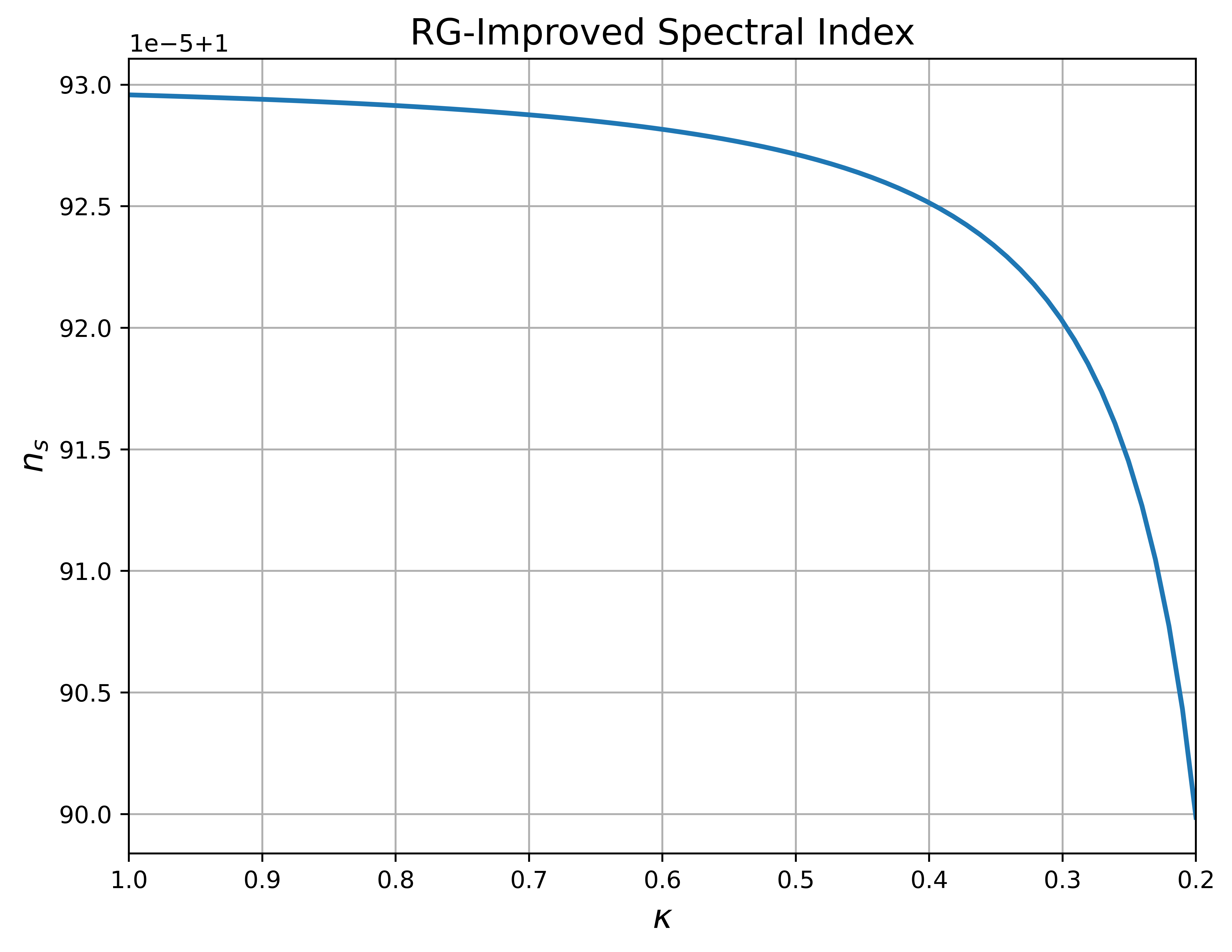}
\includegraphics[scale=0.23]{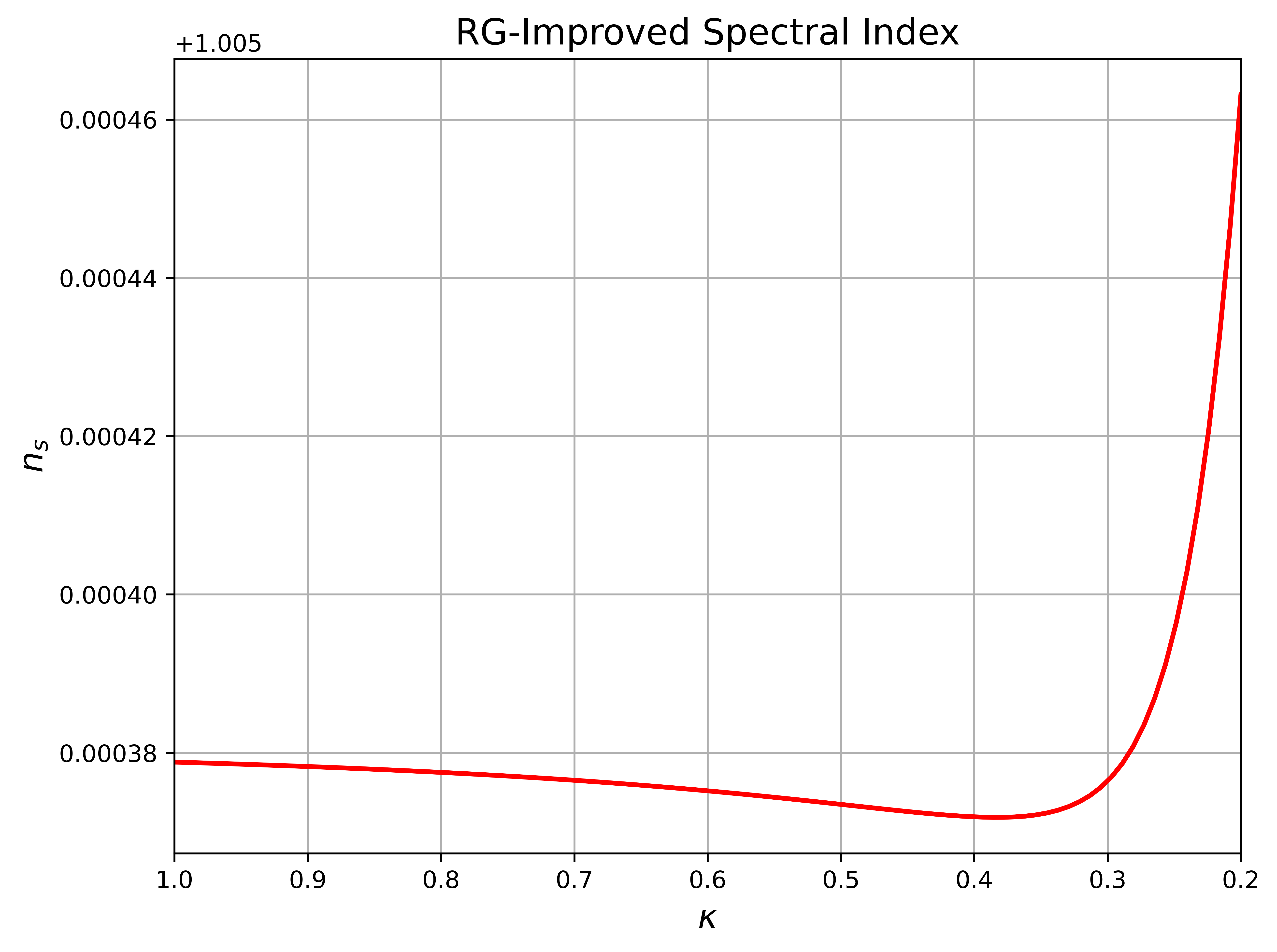}
\includegraphics[scale=0.23]{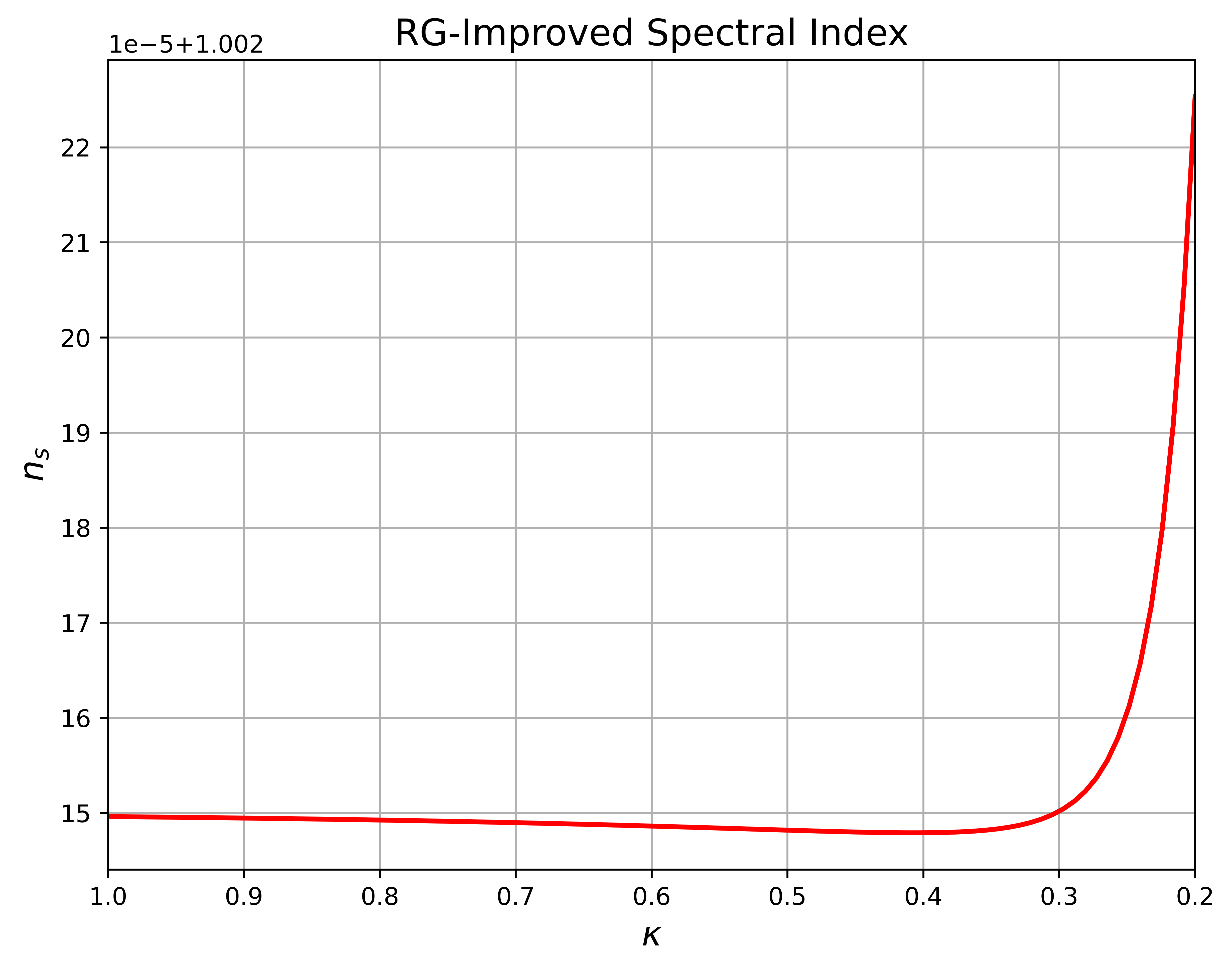}
\includegraphics[scale=0.23]{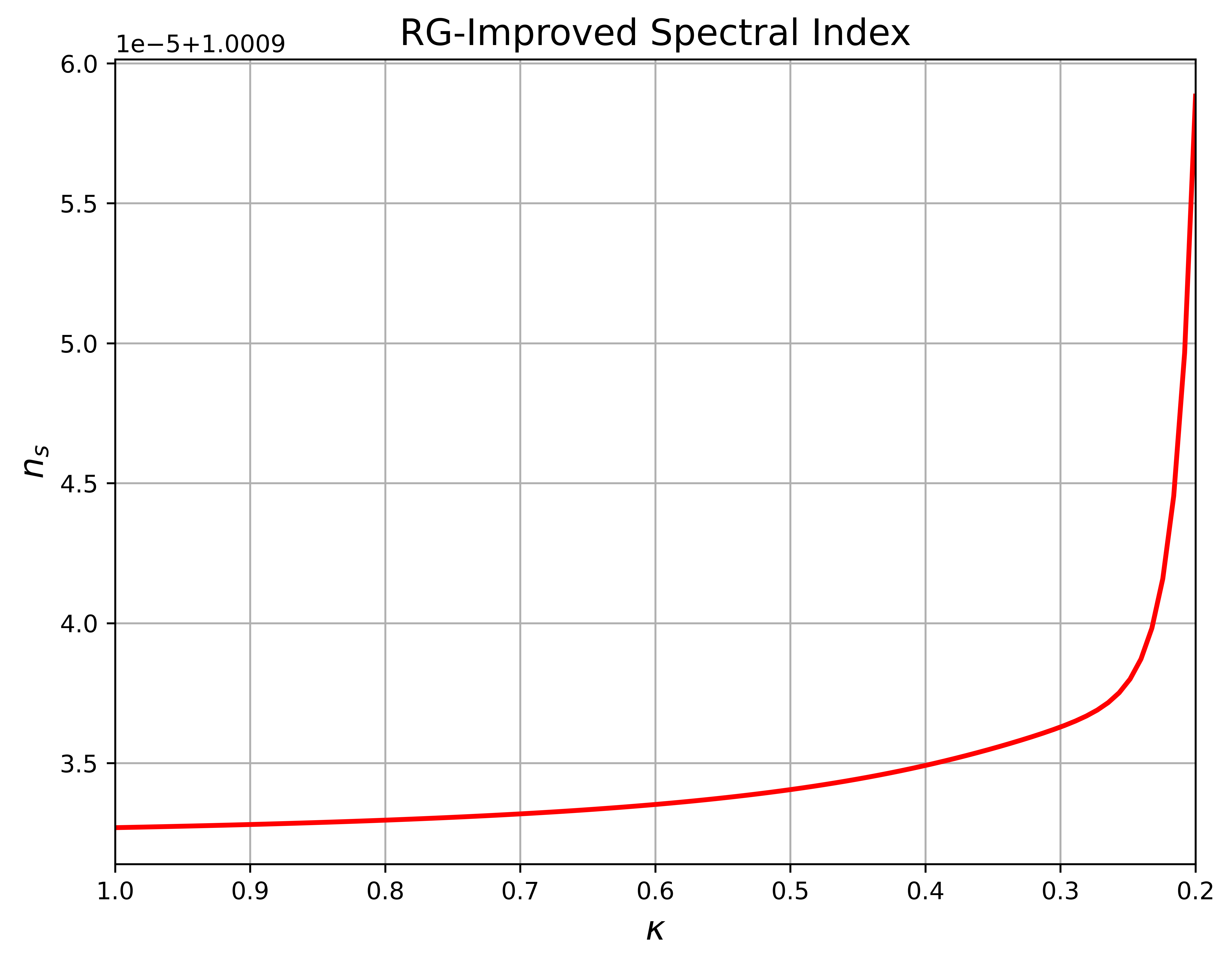}
\includegraphics[scale=0.23]{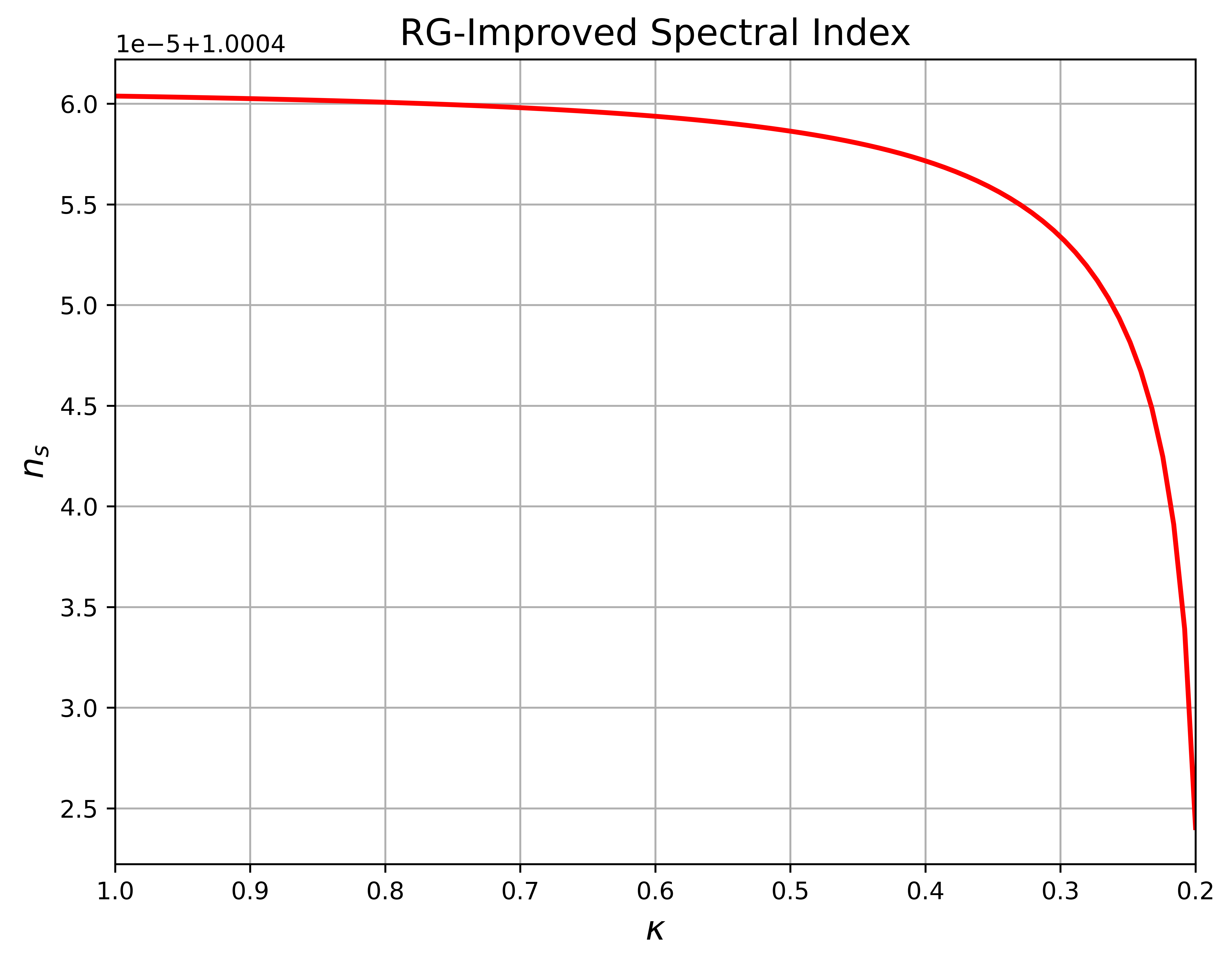}
\caption{  \it \small Spectral indices corresponding to the ${\cal O}(N)$ effective potentials. The numerical values are the same as of \ref{fig7}.}
\label{fig8}
\end{center}
\end{figure}
\noindent
From \ref{fig5}, \ref{fig6}, \ref{fig7} and \ref{fig8} it will be fair enough to conclude that  the IR-improved power spectrum and spectral index exhibit a nontrivial dependence on the IR scale, $\kappa$. For the $\phi^4$-theory, both quantities remain nearly constant throughout most of the flow, indicating that the IR stochastic dynamics is only weakly affected by the RG evolution at intermediate scales. As the flow approaches the deep IR regime, however, a significant enhancement develops, reflecting the increasing importance of long-wavelength fluctuations. The spectral index likewise exhibits a noticeable increase near the endpoint of the flow, suggesting that the lowest stochastic mode undergoes a substantial reorganisation deep in the IR. The inclusion of the Yukawa interaction modifies this behaviour. Although the power spectrum exhibits an even stronger IR enhancement, the spectral index evolves much more mildly and remains close to its ultraviolet value throughout most of the flow. This indicates that fermionic fluctuations affect the IR  dynamics differently from purely bosonic self-interactions.

For the ${\cal O}(N)$ scalar, the  IR-improved power spectrum $P_\kappa(k)$ and spectral index $n_s^{(\kappa)}$ reveal a clear evolution as the number of field components increases, \ref{fig7}, \ref{fig8}, as of the preceding section. For moderate values of $N$, both observables exhibit a weak but still visible dependence on the RG scale. The power spectrum remains nearly constant over most of the flow and develops only a modest IR enhancement near the endpoint, while the spectral index shows an extremely small variation and stays close to its ultraviolet value. This behaviour indicates that the IR stochastic dynamics is already significantly stabilised by the presence of multiple scalar degrees of freedom. The fluctuations are distributed among several field directions, reducing the impact of the RG flow on any individual mode, thereby suppressing large modifications of the IR spectrum. As $N$ becomes very large, the suppression of IR effects becomes even more pronounced. Both the power spectrum $P_\kappa(k)$ and the spectral index $n_s^{(\kappa)}$ become almost completely resistant to the IR evolution, exhibiting only minute variations throughout the entire flow. This behaviour is characteristic of the large-$N$ regime, where, as we have also emphasised it earlier, the dynamics is dominated by the collective effects rather than fluctuations of individual field components. The effective potential evolves very smoothly, the low-lying stochastic eigenvalues become highly stable, and the IR observables cease to show any significant sensitivity to the scale $\kappa$. The small residual variation that remains can be interpreted as a finite-$N$ correction to the ideal large-$N$ limit.

%%%%%%%%%%%%%%%%%%
\section{Conclusion}\label{S5'}
%%%%%%%%%%%%%%%
The chief objective of this paper was to investigate the effect of the Wetterich IR flow on the temporally coincident cosmological two point correlation function for a spectator scalar field in the inflationary de Sitter background. In order to derive the effective potentials, we have employed the Schwinger-DeWitt local expansion propagator, as well as the exact propagator in the de Sitter spacetime, \ref{S2}.  The Schwinger-DeWitt expansion is valid only in a sub-Hubble length scale. Such propagators furnish suitable framework to determine  the leading curvature corrections to high energy phenomena having  sub-Hubble Compton wavelengths. These effective potentials ($\phi^4$, $\phi^4$-Yukawa and broken $O(N)$ scalar) show spontaneous symmetry breaking feature, and are next used as the ultraviolet inputs into the IR flow equation, \ref{S3}.   The IR flow is shown to  drive the theory towards symmetry restoration. Using these IR-improved effective potentials, we  next compute the stochastic power spectra and spectral indices for the  spectator scalar, \ref{S4}. The variation of these quantities with respect to the IR scale is investigated. We show that  the results corresponding to the aforementioned two kind  of propagators are roughly in qualitative agreement with each other, but may differ quantitatively significantly. This indicates how our results might differ depending upon the energy scale associated with the process. The effect of fermionic interaction  on these quantities has been emphasised. In both cases we obtain enhancement of the spectral index in the deep IR. Also for the case of the $O(N)$ scalar, we demonstrate  that the effective potential's shape becomes more and more insensitive to the flow as $N$  becomes large. The power spectrum and the spectral index in this case  exhibit very non-trivial behaviour, as has been depicted towards the end of the preceding section. 

Note also that the tree level potential for the scalar field reads
$$V(\phi) \vert_{\rm Tree}= \frac12 \left( m_0^2+\xi R\right)\phi^2 +\frac{\lambda \phi^4}{4!}$$
Since $R=12 H^2$, from the chosen numerical values of various parameters, e.g.~\ref{T1}, we have  $\left( m_0^2+\xi R\right) \ensuremath{<} 0$, i.e. the above tree level potential has an SSB pattern as well. We have seen that this persists at one loop level too. However, washing away of  such patterns  as $\kappa$ decreases emphasises strong IR fluctuations in de Sitter. Since the physics towards the end of the inflation is essentially dominated by the IR, we  believe the results found here to be interesting in their own right.

An important extension of this investigation will be to compute the IR flow of the non-Gaussianity associated with various loop corrected higher point scalar and tensor correlation functions. Such computations might reveal important observable features associated with the CMB. We hope to address this issue in our future publication.

%%%%%%%%%%%
\section*{Acknowledgement} The work of VN is supported by the junior research fellowship of University Grants Commission, Govt. of India (Ref. no.~211610068936). 
%%%%%%%%%%

\bigskip
\appendix
\labelformat{section}{Appendix #1} 
%%%%%%%%% \cite{Nath:2024doz}
\section{Some detail of the Schwinger-DeWitt formalism using local four momentum space}\label{A}
%%%%%%%%%%
In this Appendix  we outline some detail of the Schwinger-DeWitt formalism using the local four momentum space in a general curved background for the scalar field theory, referring to~\cite{Parker} and references therein for detail. In this formalism, one is interested in quantum field theory in a neighbourhood of a given spacetime point, using the local Poincarre\'  \  invariance around that point. Accordingly, we expand the metric in the Riemann normal coordinates ($y^{\mu}$) as 
\begin{eqnarray}
&& g_{\mu\nu} = \eta_{\mu\nu} -\frac13 R_{\mu\alpha\nu\beta} y^{\alpha} y^{\beta} - \frac16 R_{\mu\alpha\nu\beta;\gamma} y^{\alpha} y^{\beta}y^{\gamma} +\left(\frac{2}{45}R_{\alpha\mu\beta\lambda}R^{\lambda}{}_{\gamma\nu\delta}-\frac{1}{20}R_{\mu\alpha\nu\beta;\gamma\delta} \right)y^{\alpha} y^{\beta}y^{\gamma}y^{\delta}\cdots  \nonumber\\
&& |g| = 1- \frac13 R_{\alpha\beta}y^\alpha y^\beta -\frac16 R_{\alpha\beta;\gamma}y^{\alpha} y^{\beta}y^{\gamma} +\left(\frac{1}{18} R_{\alpha\beta} R_{\gamma\delta} -\frac{1}{90} R^{\lambda}{}_{\alpha\beta}\,^{\kappa} R_{\lambda\gamma\delta\kappa} - \frac{1}{20} R_{\alpha\beta ; \gamma\delta}  \right)y^{\alpha} y^{\beta}y^{\gamma}y^{\delta}\cdots 
\label{v2}
\end{eqnarray}
The components of curvature  appearing above serve as expansion coefficients evaluated at  $y=0$. The above expansion holds good in a length scale small compared to the characteristic scale associated with the spacetime $\sim R^{-1/2}$, within which the variation of the curvature may be ignored. We will restrict ourselves to quadratic order of spacetime curvature in this work.

Let $G(x,x')$ be the Feynman propagator. One defines  $\bar{G}(x,x^{\prime})$ as
\begin{eqnarray}
G(x,x^{\prime}) = g^{-{\frac{1}{4}}}(x)\bar{G}(x,x^{\prime})g^{-{\frac{1}{4}}}(x^{\prime}).
\label{v3}
\end{eqnarray}
One can next use  the local flatness of the spacetime, \ref{v2}, in order to define a local momentum space ($k^{\mu}$), leading to a Fourier decomposition~\cite{Toms, Parker}
\begin{eqnarray}
\bar{G}(x,x^{\prime}) = \int \frac{d^d k}{(2\pi)^d}e^{ik\cdot y} \bar{G}(k).
\label{v4}
\end{eqnarray}
where  $\bar{G}(k)$ has an expansion 
\begin{eqnarray}
\bar{G}(k) = \bar{G_0}(k) +\bar{G_1}(k) +\bar{G_2}(k) +\bar{G_3}(k) +\cdots,
\label{v5}
\end{eqnarray}
with $\bar{G_i} (k)$ containing the $i^{\rm th}$ derivative of the metric at $y = 0$. We have at the leading order
\begin{eqnarray}
\bar{G_0}(k) =\frac{1}{k^2 +m^2}
\label{v6}
\end{eqnarray}
Since there is no first order derivative of the metric surviving at $y = 0$, we have $\bar{G_1}(k) =0$. We also have
\begin{eqnarray}
\bar{G_2}(k) =-\frac{\left(\xi-1/6\right)R}{(k^2 +m^2)^2},  \qquad 
\bar{G_3}(k) =-\frac{i\left(\xi- 1/6\right)(\nabla_{\alpha} R)}{k^2 +m^2} \p^{\alpha}_{k}\frac{1}{k^2 +m^2}, %\nonumber\\
\qquad
\bar{G_4}(k) =\frac{\left(\xi-1/6 \right)^2 {R^2}}{(k^2 +m^2)^3} +\frac{a_{\alpha\beta}}{k^2 +m^2}\p^{\alpha}_k\p^{\beta}_k\frac{1}{k^2 +m^2},\nonumber\\
\label{v7}
\end{eqnarray}
where we have abbreviated 
\begin{eqnarray}
a_{\alpha\beta} =\frac12\left(\xi -\frac{3}{20}\right)\nabla_\alpha \nabla_\beta R  -\frac{1}{40} \square R_{\alpha\beta} +\frac{1}{30} R_{\alpha}\,^{\lambda} R_{\lambda\beta} -\frac{1}{60} {R^\rho\,_\alpha}^{\lambda}\,_{\beta} R_{\rho\lambda} -\frac{1}{60} R_{\lambda\mu\rho\alpha} R^{\lambda\mu\rho}{}_{\beta}.
\label{v8}
\end{eqnarray}
We find after a little algebra
\begin{eqnarray}
\bar{G}(k) =\frac{1}{k^2 +m^2} +\frac{\left(\frac16 -\xi\right)R}{(k^2 +m^2)^2} +\frac{i}{2} {\left(\frac16 -\xi\right)(\nabla_\alpha R}){\p^{\alpha}_k}(k^2 +m^2)^{-2} +\frac13 a_{\alpha\beta}\p^{\alpha}_k\p^{\beta}_k(k^2 +m^2)^{-2} \nonumber\\ +\left[\left(\frac16 -\xi\right)^2 {R^2} -\frac23 a^{\lambda}\,_{\lambda}\right]\frac{1}{(k^2 +m^2)^3} +{\cal O}(R^3).
\label{v9}
\end{eqnarray}
Making now the usual decomposition, $\Phi=\phi+\varphi$, where $\phi$ is a classical background, we obtain  the one loop effective potential 
\begin{eqnarray}
&&V_{\rm{eff}}^{1-{\rm loop}} = \frac{1}{2}\int \frac{d^dk}{(2\pi)^d} \left[\ln (k^2 +m^{2}) +\frac{(\xi-1/6)R}{(k^2 +m^{2})} -\frac{2f_{2}}{(k^2 +m^{2})^2} +\frac12\frac{(\xi-1/6)^2 R^2}{(k^2 +m^{2})^2} \right] +{\cal O}(R^3),
\label{v10}
\end{eqnarray}
where  we have abbreviated
$$f_2 = \frac12\left(\frac16 -\xi\right)^2 {R^2} -\frac13 a^{\lambda}\,_{\lambda}.$$
After renormalisation, the result of \ref{v12} in the main text follows.

%%%%%%%%%%%%%%%%%%%%%%%%%%%%5

\end{document}